\documentclass[aps,pra,twocolumn,amsmath,amssymb,nofootinbib,showpacs,superscriptaddress]{revtex4-1}
\usepackage[english]{babel}
\usepackage{latexsym}
\usepackage{graphics}
\usepackage{graphicx}
\usepackage{epsfig}
\usepackage{color}
\usepackage{bm}
\usepackage{amsmath}
\usepackage{amssymb}
\usepackage{amsthm}
\usepackage{dcolumn}
\usepackage{bm}
\usepackage{float}
\usepackage{hyperref}
\usepackage{color}
\usepackage{epstopdf}
\usepackage{cleveref}
\usepackage[svgnames]{xcolor}
\usepackage{newfloat}
\usepackage{framed}
\usepackage[document]{ragged2e}

\usepackage{tikz}
\usetikzlibrary{quantikz}

\newcommand{\CX}{\textsf{C}$X$}

\newcommand{\Z}{\textsf{Z}}
\newcommand{\Y}{\textsf{Y}}
\newcommand{\X}{\textsf{X}}
\newcommand{\T}{\textsf{T}}
\newcommand{\Hg}{\textsf{H}}
\newcommand{\Sg}{\textsf{S}}
\newcommand{\NOT}{\textsf{NOT\,}}
\newcommand{\CNOT}{\textsf{CNOT\,}}
\newcommand{\CCNOT}{\textsf{CCNOT\,}}

\hypersetup{hidelinks,colorlinks=true,allcolors=DarkBlue}

\newcommand{\boxbegin}[2]
{\begin{story}
\begin{framed}
\begin{justify}
\vspace{-1em}
\caption{\label{#1} \bf #2}\vspace{1em}
}
\newcommand{\wideboxbegin}[2]
{\begin{story*}
\begin{framed}
\begin{justify}
\vspace{-1em}
\caption{\label{#1} \bf #2}\vspace{1em}
}
\newcommand{\boxend}[0]
{\vspace{-1em}
\end{justify}
\end{framed}
\end{story}
}
\newcommand{\wideboxend}[0]
{\vspace{-1em}
\end{justify}
\end{framed}
\end{story*}
}

\theoremstyle{remark}

\DeclareFloatingEnvironment[listname=Box,name=Box]{story}

\begin{document}
\justifying
\preprint{APS/123-QED}

\title{Quantum computing at the quantum advantage threshold: \\ a down-to-business review}

\author{A.K. Fedorov}
\affiliation{Schaffhausen Institute of Technology, Schaffhausen 8200, Switzerland}
\affiliation{Russian Quantum Center, Skolkovo, Moscow 143025, Russia}
\affiliation{National University of Science and Technology ``MISIS”, Moscow 119049, Russia}
\email{akf@rqc.ru}

\author{N. Gisin}
\affiliation{Schaffhausen Institute of Technology, Geneva, Switzerland}
\affiliation{Group of Applied Physics, University of Geneva, 1211 Geneva 4, Switzerland}

\author{S.M. Beloussov}
\affiliation{Schaffhausen Institute of Technology, Schaffhausen 8200, Switzerland}

\author{A.I. Lvovsky}
\affiliation{Russian Quantum Center, Skolkovo, Moscow 143025, Russia}
\affiliation{Department of Physics, University of Oxford, Oxford OX1 3PG, UK}
\email{Alex.Lvovsky@physics.ox.ac.uk}

\date{\today}
\begin{abstract}
\justifying
It is expected that quantum computers would enable solving various problems that are beyond the capabilities of the most powerful current supercomputers, which are based on classical technologies. 
In the last three decades, advances in quantum computing stimulated significant interest in this field from industry, investors, media, executives, and general public. 
However, the understanding of this technology, its current capabilities and its potential impact in these communities is still lacking. 
Closing this gap requires a complete picture of how to assess quantum computing devices' performance and estimate their potential, a task made harder by the variety of quantum computing models and physical platforms. 
Here we review the state of the art in quantum computing, promising computational models and the most developed physical platforms. We also discuss potential applications, the requirements posed by these applications and technological pathways towards addressing these requirements. 
Finally, we summarize and analyze the arguments for the quantum computing market's further exponential growth. 
The review is written in a simple language without equations, and should be accessible to readers with no advanced background in mathematics and physics.

\end{abstract}

\maketitle
\tableofcontents

\section{Introduction}\label{sec:Introduction}

\wideboxbegin{Box:BasicQ}{Basic quantum concepts.}


\textbf{States.} The primary notion of quantum physics is that of the \emph{state} of a given quantum object. The state is often defined by a certain value of an experimentally observable quantity. For example, a statement that the coordinate of a certain particle is $x=5$ meters constitutes a valid description of a state (denoted as $\ket{x=5\,\mathrm{m}}$). Note that, according to the uncertainty principle, the momentum of the particle in that state is fundamentally uncertain.

A variety of quantum systems are suitable for technological applications. A paradigmatic example is the \emph{spin} of a particle such as the electron. Spin can be visualized as rotation of a particle around its own axis --- akin to the diurnal spinning of the Earth. The quantum state is defined by the axis of rotation in some reference frame. 

\textbf{Superposition.} A primary  postulate of quantum mechanics is that one can perform mathematical operations with states: one can multiply states by numbers and/or add them together; the result of this operation is also a valid state (known as a \emph{superposition} state). In the language of mathematics, this means that quantum states form a vector space, or, more precisely, a Hilbert space. For example, the sum of the states $\ket{\uparrow}$ and $\ket{\downarrow}$ (spin axis directed along the positive and negative $z$ directions, respectively) gives the state $\ket{\rightarrow}$ with the spin axis along the positive $x$ direction, whereas their difference, $\ket{\uparrow}-\ket{\downarrow}$ corresponds to the state $\ket{\leftarrow}$ with the axis along the negative $x$ direction. 

\textbf{Qubit.} Conversely, any spin state can be represented as a weighted sum of two spin states corresponding to an arbitrary pair of opposite directions. Mathematically, this means that the Hilbert  space describing the electron spin  is \emph{two-dimensional}. This is an example of a \emph{qubit} --- the primary unit of quantum information: for example, the spin-up state $\ket{\uparrow}$ can be assigned a Boolean value of $0$ and  the spin-down state $\ket{\downarrow}$ a logical $1$. There exist many other physical incarnations of the qubit aside from the spin, many of which can be used for quantum information processing. 

\textbf{Measurements.} In classical physics, if we are given a classical object, we can precisely measure all properties of its state. For example, if we have a projectile, we can measure its position, velocity, acceleration, etc., at any moment in time, without disturbing its motion. This is not the case with quantum systems: if we have e.g.~an electron, we cannot measure the direction of its spin. The best we can do is to choose a certain direction and perform an experiment (known as the Stern-Gerlach measurement) that determines whether the spin is in the state oriented along or opposite to that direction. If the electron has been  prepared in one of the corresponding states (we will denote this pair of states, known as the \emph{measurement basis}, as $\ket{\nearrow}$ and $\ket{\swarrow}$), the measurement result will be certain. For an arbitrary state $\ket\psi$, on the other hand, the measurement will yield one of these two outcomes with some probability. In order to predict these probabilities for a known state $\ket\psi$, we need to write it as a weighted sum of the basis elements: $\ket\psi=\alpha\ket\nearrow+\beta\ket\swarrow$. The measurement probabilities are then ${\rm pr}_\nearrow=|\alpha|^2$ and  ${\rm pr}_\swarrow=|\beta|^2$ (we assume that $|\alpha|^2+|\beta|^2=1$).

This probabilistic, uncertain nature of measurement is a defining feature of quantum physics. Remarkably, a measurement will transform the state of a quantum system into whatever state that measurement has detected. For example, if we measure the $\ket{\rightarrow}$ state in the $\{\ket{\uparrow},\ket{\downarrow}\}$ basis and happen to detect $\ket{\uparrow}$, all subsequent measurements of that spin in the same basis will yield the same result: $\ket{\uparrow}$. 

\textbf{Entanglement.} Superposition states of multiple quantum objects, treated as a single system, are known as \emph{entangled}. Consider, for example, two electrons in the state $\ket{\uparrow\uparrow}+\ket{\downarrow\downarrow}$. This expression means that, whenever each of the electrons are measured in the $\{\ket{\uparrow},\ket{\downarrow}\}$ basis, they will be found in the same state. 

A simple calculation can show that the same is true for \emph{any} measurement basis: whenever the first electron is detected in a particular state, the second one is certain to be in the same state \cite{Lvovsky2018}. This property is remarkable. Suppose, for example, that one of the electrons is with a fictitious observer Alice on Venus and the other one with Bob on Venus. By choosing a  basis and measuring her electron, Alice can remotely prepare Bob's electron in the same state she has detected. For example, she can choose to measure in the $\{\ket{\rightarrow},\ket{\leftarrow}\}$ basis, in which case, dependent on the result, Bob's photon will become either $\ket{\rightarrow}$ or $\ket{\leftarrow}$. This \emph{remote state preparation}~\cite{RemPrep}, which occurs instantly and without interaction, and once called ``spooky action at a distance" by Einstein, has puzzled generations of physicists. We do not delve into this topic as it is tangential to our review; however an important fact that must be mentioned is that it is impossible to use quantum entanglement for superluminal communication of classical information.

\textbf{Decoherence.} The above argument also implies that if Alice happens to lose her electron, the state of Bob's electron becomes completely unknown. This gives rise to the phenomenon of \emph{decoherence}, which is a primary hurdle in quantum computation technology. In the process of quantum computation, qubits may undergo spurious interaction with the environment which will result in their entanglement with its quantum state. Since we have poor control of the environment, we cannot keep track of its state; essentially, it is lost as far a the quantum qubit register is concerned. As a result, the state of the register loses its superposition nature (decoheres) and becomes useless. 

The main quantitative benchmark describing decoherence is its characteristic time. 
The more controlled operations can be performed on a quantum register before it decoheres, the better.

\wideboxend

The progress in understanding our world at the smallest scales has culminated in a physical theory that is both the most controversial and, at the same time, most extensively tested of all physical theories: quantum physics.
Originally devised to explain the black-body radiation problem, one of the outstanding unsolved problems in physics at the end of the nineteenth century, 
quantum physics extended its reach in the subsequent decades to cover a great variety of microscopic systems. 
The theory's power~\cite{Planck1901} to predict collective phenomena in ensembles of quantum particles led to the development of many widely used practical devices,
the most impactful of which are transistors~\cite{Grumbling2019} 
and lasers~\cite{Milonni2010}. 
These inventions give rise to the development of semiconductor computing and microprocessors, optical communications and the Internet based on first wired and then mobile technologies 
--- in other words, made the world the way we know it today. 
This is known as the {\it first quantum revolution}.

\begin{figure}[b]
\center{\includegraphics[width=1\linewidth]{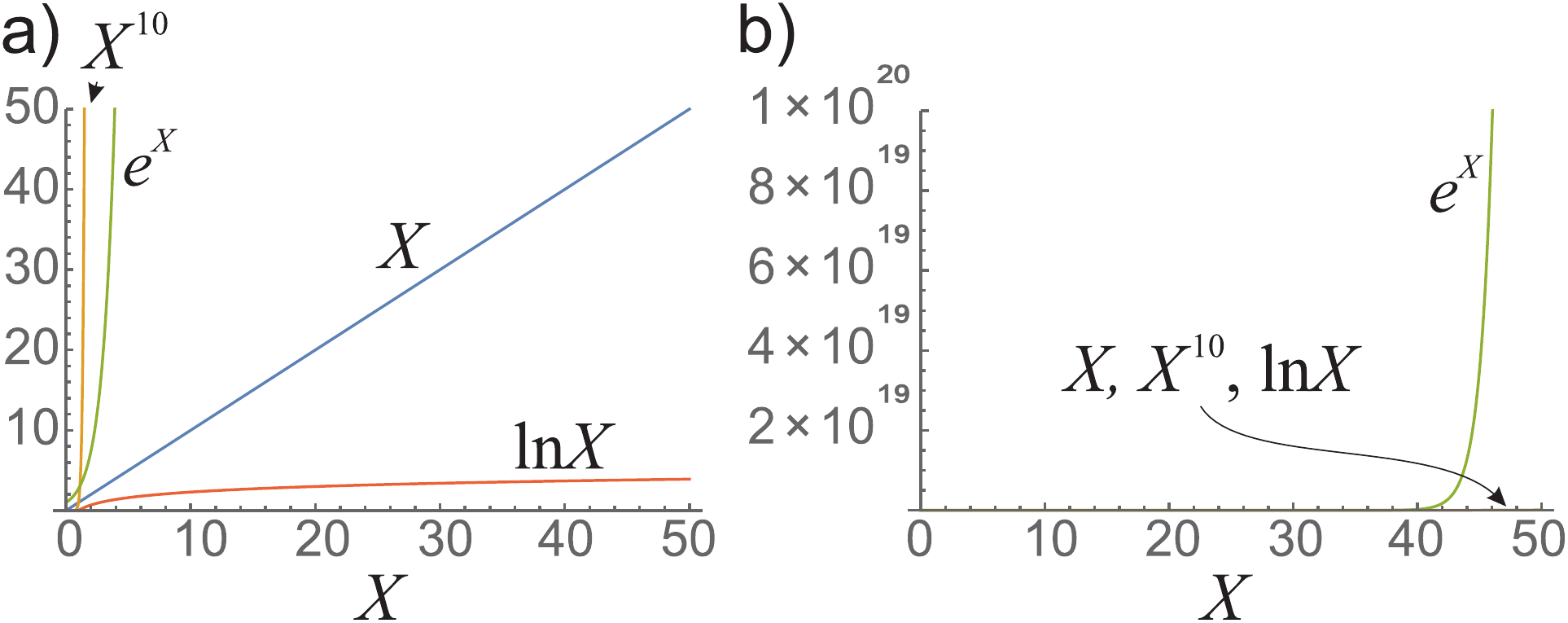}}
\caption{Linear, polynomial, and exponential scalings. Plots in (a) and (b) are different by the vertical axis scale.}
\label{fig:time}
\end{figure}

A characteristic feature of the technologies of the first quantum revolution is that they do not require handling individual quantum objects, such as atoms, photons (light quanta), or electrons.
Rather, they rely on their collective behavior in large (macroscopic) ensembles. 
For example, a gain medium of a gas laser consists of multiple atoms, but in order to design a laser one does need to precisely control the quantum physics of each of these atoms. 
It is enough to study the quantum physics of an ``average" atom, and then calculate how this physics translates into the properties of a collective of these atoms contained inside the laser tube, such as the dependence of the gain on the length of the tube, discharge voltage, etc. 
Such large systems are relatively easy to handle, both theoretically and experimentally, because their reasonable interactions with an environment does not degrade their useful properties. 

The next technological frontier is to learn how to handle a collective of {\it individual} microscopic quantum objects in such a way that each such object plays a unique, clearly defined role as a part of a complex entangled state of the collective (see Box~\ref{Box:BasicQ}). 
This will open up a whole new horizon of technological opportunities, such as new problems amenable to computational analysis, perfectly secure communications, and sensors with unprecedented precision. 
This is known as the {\it second quantum revolution} or {\it quantum technologies}~\cite{Dowling2003}. 
In this paper, we concentrate on a particular aspect of quantum technologies: quantum computing. 

Any computer processes information encoded in a string of bits, each of which can take a value of $0$ or $1$. 
A quantum computer operates with \emph{quantum bits}, or qubits (Box~\ref{Box:BasicQ}) and can process massive entangled superpositions of their states at once as a single quantum system. 
That is, a string of $N$ qubits can encompass an entangled superposition of $2^N$ classical $N$-bit stings. 
In this sense, we say that a quantum computer may possess exponential advantage with respect to classical (Fig.~\ref{fig:time})\footnote{We say that something is ``classical'' when one can describe its properties without invoking quantum phenomena like entanglement} computational devices 
(Box~\ref{fig:time}). 

There is caveat, however: entanglement is both a blessing and a curse. 
Since quantum computers process information in a superposition state, the computation results will also be superposed with each other. 
However, a human user is a macroscopic, classical entity and cannot handle such an entangled state. 
We need a specific classical answer to a specific classical problem, which is of interest to us at a given moment of time (Box~\ref{Box:QPhoneBook}). 
As a result, quantum computers offer a significant advantage for a specific class of problems. 

While this class is potentially large, its boundaries are not precisely known at this time. However, many of its elements --- problems with expected quantum advantage (albeit not always exponential) --- are now identified. 
This includes simulation of complex systems (fuels, drugs, biosystems, materials, etc.), optimization, data processing, and machine learning (see Sec.~\ref{sec:applications}). 
A problem  of particular practical importance, for which the advantage is exponential, is factorization (decomposing into prime factors) of large numbers with application in cryptanalysis (Sec.~\ref{sec:applicationcrypto}).

The development of a large-scale quantum computer capable of practical applications is a major challenge. 
Any interaction between the qubit register in a quantum computer and the environment will result in \emph{decoherence} --- uncontrolled entanglement of the two, bringing about the loss of the superposition nature of the register, which is fatal for the quantum information contained therein (Box~\ref{Box:BasicQ}). 
We are thus facing two antagonistic requirements: we must enable the qubits to be controllably affected by other (microscopic) qubits yet completely unaffected by the (macroscopic) environment. 
This is the main reason why we have not yet conclusively demonstrated quantum computational advantage in application to practical tasks, although 
the idea of quantum computing has been around for about 40 years and its primary working principles have been elaborated more than 25 years ago. 
\boxbegin{box:PhoneBook}{Quantum phone book.}

Some intuition regarding the quantum advantage (and limitations thereof) can be gained by the following, somewhat allegorical, example. 
Consider a telephone book of a city with, say, a million inhabitants encoded as an ASCII text file. 
Each entry includes e.g.~10 bytes containing the name of a line subscriber, and 7 bytes with a phone number --- so the entire book occupies 17 megabytes.

\begin{center}
\begin{tabular}{ll}
	\texttt{Abbott} & \texttt{ 123-4567} \\
	\texttt{Adams} & \texttt{ 765-4321} \\
	\texttt{Ahmed} & \texttt{ 222-3333} \\
	\texttt{Albrecht} & \texttt{ 456-7890} \\
	\dots&\dots
\end{tabular}    
\end{center}

A quantum analog of such phone book would place all these entries in an entangled superposition, involving only 17 bytes: 
\begin{align*}
    &|\texttt{Abbott  123-4567} \rangle + |\texttt{Adams 765-4321} \rangle 
    \\+& 
	|\texttt{Ahmed \quad 222-3333} \rangle + |\texttt{Albrecht 456-7890} \rangle +\dots
\end{align*}

This offers a massive advantage both in the storage capacity and processing of the information. 
Suppose, for example, that we need to add 1 to one of the digits or add an area code to all numbers. 
With a classical phonebook, this operation would need to be performed individually on each entry. 
But in the quantum case, only a single operation on the entangled system would suffice.

A complication arises, however, when we try to use the phone book according to its purpose --- read out a number associated with a specific name. 
Extraction of a single entry from this massive entangled superposition is not an easy task --- rendering such a quantum phonebook largely useless in a household. 
This illustrates that quantum computation is useful only for a limited class of problems, many of which are quite distant from those encountered by lay users.
\label{Box:QPhoneBook}
\boxend

\begin{figure}[b]
\center{\includegraphics[width=1\linewidth]{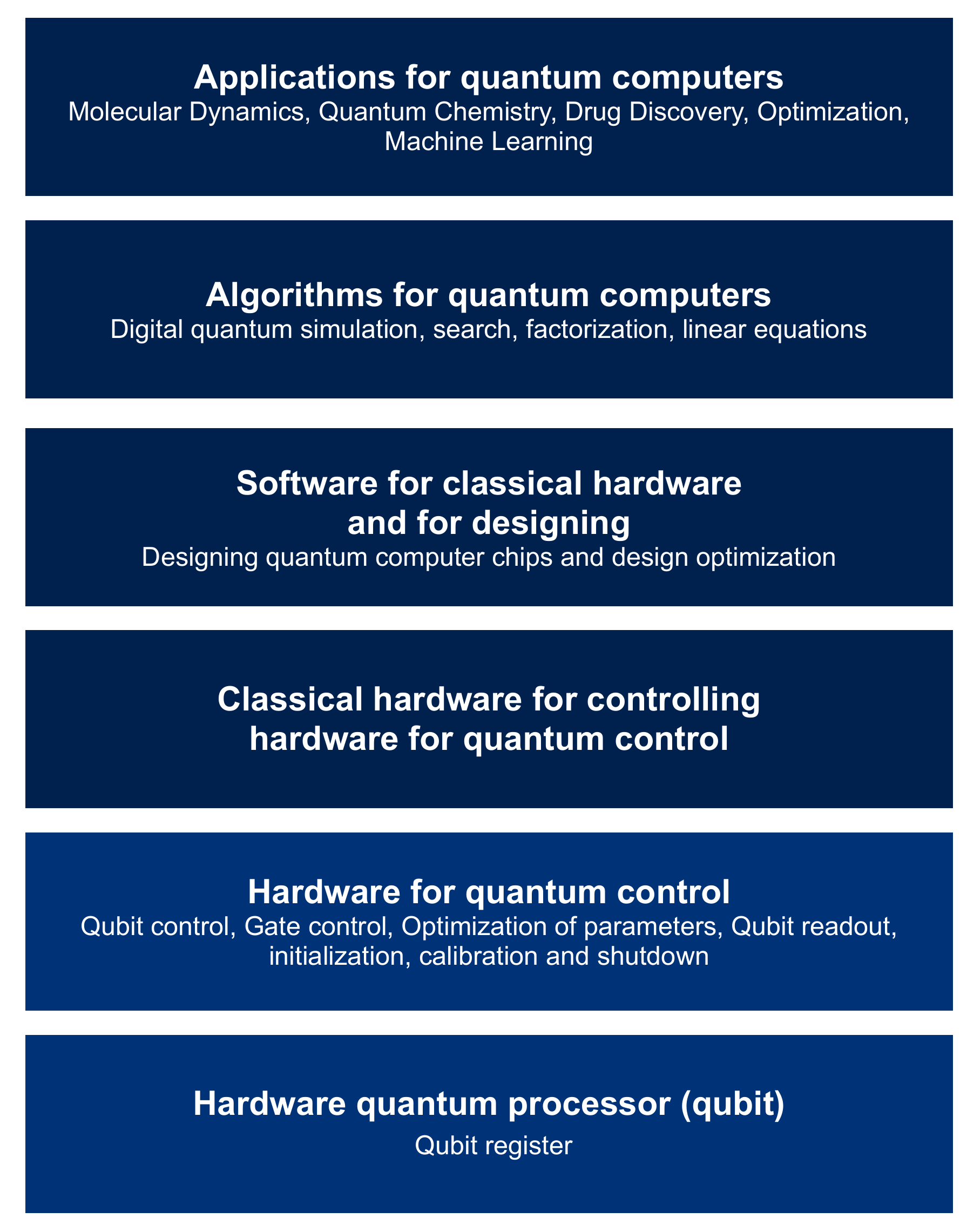}}
\vskip-4mm
\caption{Stack of quantum computing technology.}
\label{fig:stack}
\end{figure}

Current quantum computing devices operate with on the order of a hundred qubits with approximately 20--30 gate operations and are not capable of error correction. 
They are sometimes referred to as \emph{noisy intermediate-scale quantum (NISQ)}~\cite{Preskill2018}. 
NISQ devices have demonstrated advantage (supremacy) of quantum computing with respect to classical, albeit on tasks that have limited practical value.  
Applications of NISQ devices to practical problems, like optimization and chemistry, have also been attempted, however no quantum advantage has been achieved. 

Sadly, any progress in quantum computing, however minor, is a likely subject  for mass media coverage --- 
often in exaggerated and hyperbolized fashion, which might create an illusion that quantum computer technology is far beyond the point where it actually is. 
The true state of the art, in our opinion, in that the technology is at the threshold of quantum advantage, but not yet significantly beyond it. 

The quantum computing challenge is being approached by many research teams and many different ways, which gave rise to a large variety of devices. 
This includes both the physical platforms (superconducting circuits, trapped ions, neutral atoms, light, etc.; see Sec.~\ref{sec:platforms}) and computing models (for example, whether qubits are addressed one-by-one or all at once; see Sec.~\ref{sec:models}). 
This diversity complicates defining a metric for their comparison. 
While various metrics have been proposed, they do not give a complete and straightforward picture of how different approaches are related to each other or how close quantum computing devices are to solving real-world problems.

This lack of universal metrics, diversity of platforms, models and purposes, overselling in literature and media as well as general mysterious nature of quantum physics make quantum computation a challenging environment. 
The purpose of the present review is to demystify quantum computing not only for the broad scientific community, but also for decision makers, investors, media, industrial executives, and members of the public.
We attempt to cover the full stack of the quantum computing technology ranging from hardware for making individual qubits to software and potential applications (Fig.~\ref{fig:stack}). 
We can roughly identify three layers composing this stack.  
\begin{itemize}
    \item Quantum computational \emph{model (type)}  defines the general approach  to organizing the encoding and processing information in a quantum computer (Sec.~\ref{sec:models}).
    \item The \emph{platform} is the specific physical system (e.g.~superconducting circuits or trapped ions), whose quantum properties are used for calculation (Sec.~\ref{sec:platforms}).
    \item Within each platform, different \emph{architectures} --- suites of hardware and software solutions to implement programming, control, input and output,~--- are possible. Architecture details are highly technical are generally outside the scope of our work.
\end{itemize}
Additionally, in Sec.~\ref{sec:applications} we analyze \emph{applications} of quantum computing and discuss how far in the future we can expect to see quantum advantage in the context of this applications. 

We also summarize a set of parameters that could be used to estimate the readiness and potential of a specific quantum system, and formulate a two-pronged framework for their analysis. 
On the one hand, we discuss technical benchmarks, such as the system size, length of the operation sequence that can be implemented, and the degree of programmability. 
One the other hand, we discuss user-oriented criteria such as cost, speed, and range of tasks that can be solved. 
Based upon this analysis, we summarize the arguments predicting exponential growth for the quantum computing market (see Box~\ref{Box:MarketGlimpse}).


\section{Has classical computing approached its fundamental limits?}\label{sec:classical}

\boxbegin{box:QCmarket}{First quantum computing unicorn}
	On March 8, 2021 it was announced that IonQ (USA) would become the first publicly traded pure-play quantum computing company, with a pro forma implied market capitalization of approximately \$3 billion. 
	The company forecasts an increase in the number of algorithmic qubits (see Sec.~\ref{sec:comparison} for definition) from 22 in 2021 up to 1024 in 2028.
\label{Box:MarketGlimpse}
\boxend


Intel co-founder G. Moore published a paper~\cite{Moore1965} in 1965 (reprinted in 2006~\cite{Moore2006}), 
where he made a number of important observations. 
Two are particularly relevant for our review. 
The first one reads: ``Silicon is likely to remain the basic material, although others will be of use in specific applications. 
For example, gallium arsenide will be important in integrated microwave functions. 
But silicon will predominate at lower frequencies because of the technology which has already evolved around it and its oxide, and because it is an abundant and relatively inexpensive starting material''. 
The second one, today known as Moore’s law, 
is as follows: ``The complexity for minimum component costs has increased at a rate of roughly a factor of two per year''. 

Let us examine both observations in more details. 
Modern developments in computing technologies were a result of efforts to increase the quality of silicon-based transistors (basic elements of computing devices) and decrease their costs by improving manufacturing. 
The central element of this progress was the ``planar process'', a method of fabricating transistors on the surface of a flat silicon wafer. 
These transistors are connected via a metal layer to create a complete circuit. 
Such ``integrated circuits'', a consequence of the first quantum revolution, have for the last sixty years been the dominant venue for classical digital computation~\cite{Grumbling2019}. 

This second observation can be reformulated as follows: thanks to miniaturization of transistors, the number of transistors on a chip doubles every 24 months (the original prediction that the period is 12 months has been corrected)~\cite{Moore2006}. 
Remarkably, this trend has continued with the same zeal ever since it has been first observed. 
As an illustration, IBM has recently announced the production of a chip based on the 2 nm technology~\cite{Johnson2021}. 
In a scientific lab, transistors as small as 1 nm have been developed in 2016~\cite{Desai2016}. 
For comparison, the nearest-neighbor distance in the silicon crystal lattice is 0.235 nm. 

It is evident from the above figures that the transistor size is approaching fundamental physical limits. 
Other fast (exponential) scaling laws for classical computing technology, such as the growth of performance per watt (so-called Dennard scaling) or clock speeds, are already no longer valid~\cite{Markov2014,Waldrop2016}.  
This motivates many analysts to conclude that ``Moore's law is nearing its end''~\cite{Waldrop2016} and became a common argument to motivate the future of quantum computation. 

While we too advocate the future of quantum computing, we cannot agree with the above thesis. 
In addition to the existing reserve of at least a factor of three in the linear size, a further room for developing integrated circuits consists in 3D layering. 
Circuits with 96 layers have been demonstrated by Toshiba in 2018\footnote{https://www.businesswire.com/news/home/20180719005280/en/Toshiba-Memory-Corporation-Develops-96-layer-BiCS-FLASH-with-QLC-Technology}. 
Combining these figures, at least three orders of magnitude increase in the transistor density is possible with implies at least three more decades of Moore's law. 
New architectures, optimization, and application-driven specialization of processors open up even more opportunities for the development of classical computers. 
Quantum computing is therefore motivated not by the upcoming end of such developments, but rather by fundamental limitations of classical computers as Turing machines in solving certain classes of problems, as discussed above in Introduction. Quantum computers should be seen not as competitors to classical machines, but rather as a supplementary class of devices aimed to tackle a distinct class of problems.

The continued transistor  miniaturization implies that laws of microscopic world --- i.e., quantum laws --- play an increasingly significant role in the operation of even classical integrated circuits. 
For example, classical  bit values 0 or 1 correspond to a transistor switched ``on'' or ``off'', respectively. 
However, when the transistor is of microscopic size, the effect of quantum tunneling\footnote{The phenomenon of quantum tunneling allows a quantum particle to traverse energy barriers that are higher than the energy of the state itself, which is prohibited by classical physics. Quantum tunneling plays an essential role in various applications, including tunnel diodes in computing, flash memory, and in the scanning tunneling microscope.} results in the electric current flowing through the transistor even when it is in the ``off'' state. 
This results in excessive heating of integrated circuits becoming more and more of an issue in  computer design. 
Hence, even if we do not make directed efforts to harnessing the power of quantum phenomena, these phenomena will still enter computational technology, but as a hindrance rather that an opportunity. 

\wideboxbegin{box:EarlyAlgorithms}{Early quantum algorithms.}

\begin{itemize}
	\item In 1992, Deutsch and Jozsa presented an algorithm~\cite{Deutsch1992}, which finds whether a ``black-box" function of string of bits is balanced or constant. 
	A constant function returns the same result for any input, while a balanced function returns 0 for exactly half for all possible inputs and 1 for the other half. 
	The simple example of 2-bit functions was considered by Deutsch in Ref.~\cite{Deutsch1985}, whereas the $n$-bit function case was analyzed by Deutsch and Jozsa~\cite{Deutsch1992}. 
	Some improvements were proposed by Cleve, Ekert, Macchiavello, and Mosca in 1998~\cite{Mosca1998}. 
	Although currently lacking a practical use, it is one of the first examples of a quantum algorithm that is exponentially faster than any possible deterministic classical counterpart.
	\item The Fourier transform occurs in many different versions and is widely applicable in various tasks, such as signal processing, in which it allows transitions between representation of signals in time and frequency domains. 
	The quantum Fourier transform (QFT) --- the  discrete Fourier transform applied to the vector of amplitudes of a quantum state --- was proposed by Coppersmith in 1994 (published as Ref.~\cite{Coppersmith2002} in 2002). 
	QFT can be considered as a transformation between two bases of quantum states (the computational ($Z$) basis and the Fourier basis). 
	It is a subroutine of many quantum algorithms, most notably Shor's factoring algorithm and quantum phase estimation.
	\item In 1994, Simon invented~\cite{Simon1994,Simon1997} an algorithm for verifying whether an unknown black box function is one-to-one (maps exactly one unique output for every input) or two-to-one (maps exactly two inputs to every unique output). 
	Simon's algorithm uses the quantum Fourier transform. 
	\item In 1994, Shor proposed~\cite{Shor1994,Shor1996} a polynomial-time quantum computer algorithm for integer factorization, which is based on Simon’s algorithm. 
	Shor's quantum algorithm can be used in the cryptanalysis of currently widely deployed public-key cryptography algorithms, such as RSA and Diffie-Hellman (see Sec.~\ref{sec:applicationcrypto}). 
	This is one of the first quantum algorithms that solves a practically relevant problem.
	\item In 1995, Kitaev proposed a quantum phase estimation algorithm, which allows estimating the phase (or eigenvalue) of an eigenvector of a unitary operator. 
	This is an important part for many quantum algorithms including the current version of Shor's algorithm and quantum algorithm for solving linear systems of equations (see Secs.~\ref{sec:application-simulation} and \ref{sec:application-ML}). 
	\item In 1996, Grover considered~\cite{Grover1996,Grover1997} a black box function, which yields 0 for all bit strings expect one, for which the output is 1. Grover’s algorithm helps finding this unique bit string. 
	It takes an input a superposition of all possible bit strings and yields a superposition in which the string of interest has a high probability amplitude. 
	This algorithm gives quadratic speedup, however it has been proven~\cite{Bennett1997} that no higher speedup is possible for this problem. 
	\item In 1997, Bernstein and Vazirani~\cite{Bernstein1997} considered the class of functions, which is known to be the dot product between its argument and a secret bit string. The task of the algorithm is to find this bit sting. 
\end{itemize}
\label{Box:EarlyAlgorithms}
\wideboxend

\section{A historic excursion}\label{sec:history}


The roots of quantum computing can be traced back Szilard and von Neumann, who connected the thermodynamic concept of entropy (a measure of disorder introduced by Boltzmann in 1877) 
with information theory developed by Nyquist and Hartley in the 1920s, as well as Shannon in the 1940s. 
Based on these results, Landauer~\cite{Landauer1961,Landauer1970} in 1970 found a fundamental lower bound imposed on any irreversible operation on a bit. 
According to the Landauer limit, such an operation consumes energy in the amount of $kT\log2$, where $k$ is the Boltzmann constant $1.38\times10^{-23}$ Joules per Kelvin and $T$ is the absolute temperature. 
These results have been further analyzed by Bennett in 1973-1982~\cite{Bennett1973,Bennett1982}, who proposed the concept of reversible computation that does not involve erasing information, and hence overcomes the Landauer limit.
Subsequently, Bennett discussed his finding with Feynman and asked whether they are affected by the quantum nature of the world. 
Inspired by this question, Feynman in 1982-1986~\cite{Feynman1982,Feynman1986} published a series of papers showing not only that the answer to the above question is negative, 
but also that quantum properties of matter can be used to enhance power of computational devices. 
These papers~\cite{Feynman1982,Feynman1986} are universally considered as the origin of quantum computation. 
Note that a frequently quoted work by Benioff from 1980~\cite{Benioff1980} contains an analysis of the possible implementation of the (classical) Turing machine using quantum systems, but it does not use the feature of quantum entanglement.

Feynman’s results have influenced the work of Deutsch, who had at that time been working on similar ideas in the context of the Everett (multiworld) interpretation of quantum physics~\cite{Deutsch1985} 
(paper~\cite{Deutsch1985} was completed in 1978, but published only in 1985; see also Ref.~\cite{Deutsch2021}). 
Synthesizing Feynman’s results with his own thoughts, Deutsch came up with a mathematical study of capabilities of quantum computing machines titled 
``Quantum theory, the Church-Turing principle and the universal quantum computer'', published in 1985~\cite{Deutsch1985-2}. Lloyd in 1996 has rigorously proven Feynman's conjecture that a digital quantum computer can efficiently simulate an arbitrary quantum system \cite{Lloyd1996}.
These developments led to the start of research explicitly devoted to quantum computing in the mid-1990s.

In parallel, research on salient concepts of quantum information and computation took place in the Soviet Union. 
In 1970-1980 Holevo published in a series of papers~\cite{Holevo1973,Holevo1975,Holevo1977} 
establishing an upper bound to the amount of information that can be known about a quantum state (its ``accessible information'', which is now known as the {\it Holevo bound}).
In 1975, Poplavskiy~\cite{Poplavski1975} observed that quantum systems cannot efficiently simulated on a classical computer. 
Finally, in 1980 Manin published a book ``Computable and non-computable''~\cite{Manin1980}, in which he described the concept of quantum logic, i.e. logical operations on qubits. 
After becoming known worldwide in 1990s, these results significantly influenced the development of the field. 

\section{Why is quantum computing powerful?}\label{sec:advantage}

It is not yet strictly proven that quantum computers can be more powerful than classical counterparts. 
This is only a conjecture akin to many fundamental believes in computer science, such as P and NP complexity classes are not equivalent. 
However, there exists strong reasons to believe that this conjecture is valid as summarized by Preskill~\cite{Preskill2018}. 

A few dozens quantum algorithms~\cite{Montanaro2016,Montanaro2017} have been developed in the last 30 years that are significantly (sometimes exponentially) faster than the best known classical algorithms. Refs.~\cite{Montanaro2016,Aspuru-Guzik2021,Babbush2021-4} provide reviews of quantum algorithms, a more detailed guide can be found on a website ``quantum algorithm zoo''~\cite{zoo} and some important examples are listed in Box~\ref{Box:EarlyAlgorithms}.

Of particular relevance for the current state of the art are algorithms that provide samples from known probability distributions, 
which in certain cases cannot be efficiently obtained by classical means~\cite{Aaronson2013,Montanaro2016-2,Aspuru-Guzik2016-2,Neven2018}. 
These algorithms are the basis for current demonstrations of quantum advantage (see Sec.~\ref{sec:applications} below)~\cite{Montanaro2016,Montanaro2017},
which Preskill defines as ``computational tasks performable by quantum devices, 
where one could argue persuasively that no existing (or easily foreseeable) classical device could perform the same task, 
disregarding whether the task is useful in any other respect''~\cite{Preskill2018}.  

\begin{table}[h]
\caption{\textbf{Simulating quantum computers using classical devices}. 
The classical memory required to hold an equivalent amount of information, as well as the classical compute time required to implement a single operation on the given number of qubits are listed.
The quoted values are based on a lecture by Troyer titled ``High Performance Quantum Computing'' (\url{https://youtu.be/Hkz_Sn5qYWg}). The number $10^{80}$ bits quoted in the last line 
is on the order of the size of the universe, whereas the corresponding time of  $10^{50}$ years is unimaginably longer than the age of the universe (15 billion years).}
\begin{tabular}{c|c|c}
\hline \hline Qubits &Memory & Equivalent single-operation time \\
\hline
\hline $10$ & $16$ kByte & microseconds on a watch \\
\hline $20$ & $16$ MByte & milliseconds on smartphone  \\
\hline $30$ & $16$ GByte & seconds on laptop \\
\hline $40$ & $16$ TByte & seconds on computer cluster\\
\hline $50$ & $16$ TByte & hours on top supercomputer\\ 
\hline $60$ & $16$ EByte & minutes on next decade’s supercomputer? \\
\hline $70$ & $16$ ZByte & hours on potential future supercomputer? \\
\hline $80$ & $16$ YByte & $10^{50}$ years \\
\hline \hline
\end{tabular}
\label{table:simulation}
\end{table}

One may object that, even though existing classical algorithms are inferior to their quantum counterparts, perhaps in the future classical algorithms can be invented that close this gap.  
This can be countered by the observation that quantum matter, particularly the quantum computer, 
require exponentially growing amount of time and resources to simulate classically~\cite{Preskill2018} (so-called ``curse of dimensionality''). 
For example, for a modest-size quantum computer with 50 qubits, it would take 16 PByte of memory and hours on a top supercomputer in order to implement a single operation (see Table~\ref{table:simulation}). 
Simulations to this effect have recently been presented  by IBM (USA)~\cite{Pednault2017,Pednault2019} and Alibaba (China)~\cite{Huang2020}. 

An important challenge, which was extensively discussed at the early stage of quantum computing, is related to validation of the results that quantum devices produce~\cite{Cirac2012,Lewenstein2012}. 
Interestingly, issues of this nature are not unique to quantum computers, but relevant to any new generation of computing devices. 
They can be addressed in a variety of ways. 
First, the inherent nature of many problems, which are subject to quantum advantage, is that their solutions can be checked on a classical computer in polynomial time. 
Second, many problems, while being generally challenging, have certain instances or configurations that are amenable to classical computation. 
Third, one can test if the quantum computer provides a correct solution to problems of smaller sizes, which can be simulated numerically. 
After these multiple checks, calculations in classically inaccessible regimes can be considered reliable. 

\section{Quantum computing can be based on various models}\label{sec:models}

\begin{table*}\caption{Examples of common quantum gates.} 
\label{Table:Gates}
		\begin{tabular}[t]{p{0.15\textwidth} p{0.15\textwidth}p{0.3\textwidth}p{0.4\textwidth}}
			\hline\hline
			{\bf Gate name} & {\bf Notation in  diagrams} & {\bf Description (for spin qubits)} & {\bf Example action}\\  
			\hline\hline
			Pauli-Z (\Z, \NOT) &
			\begin{tikzcd}
			& \gate{Z} & \qw    		 
			\end{tikzcd}& Rotation of the spin axis by $\pi$ radians ($180^\circ$) around the $z$ axis. &Leaves the $\ket\uparrow$ and $\ket\downarrow$  unchanged, but transforms $\ket\rightarrow$ and $\ket\leftarrow$ into each other.
\\\hline
			Pauli-X (\X)& 		
			\begin{tikzcd}
			& \gate{X} & \qw
			\end{tikzcd}& Rotation of the spin axis by $\pi$ radians ($180^\circ$) around the $x$ axis. &Leaves the states $\ket\rightarrow$ and $\ket\leftarrow$ unchanged, but transforms $\ket\uparrow$ and $\ket\downarrow$ into each other.	 \\\hline
			Pauli-Y (\Y) & 
			\begin{tikzcd}
			& \gate{Y} & \qw    		 
			\end{tikzcd} & Rotation of the spin axis by $\pi$ radians ($180^\circ$) around the $y$ axis. & Swaps the states within both pairs $\{\ket\rightarrow,\ket\rightarrow\}$ and $\{\ket\uparrow,\ket\downarrow\}$.
 \\\hline
			Hadamard (\Hg) &
			\begin{tikzcd}
			& \gate{H} & \qw    		 
			\end{tikzcd} & Rotation of the spin axis by $\pi$ radians ($180^\circ$) around the vector halfway between the $x$ and $z$ axes. & Swaps the states within both pairs $\{\ket\uparrow,\ket\rightarrow\}$ and $\{\ket\leftarrow,\ket\downarrow\}$.
 \\	\hline
			\Sg &
			\begin{tikzcd}
			 \qw & \gate{S} & \qw    		 
			\end{tikzcd}&Rotation of the spin axis by $\pi/2$ radians ($90^\circ$) around the $z$ axis	 & Leaves the $\ket\uparrow$ and $\ket\downarrow$  unchanged, but transforms $\ket\rightarrow$ into the state with the spin along the $y$ axes; 
\\	\hline
			\T &
			\begin{tikzcd}
			& \gate{T} & \qw    		 
			\end{tikzcd}&Rotation of the spin axis by $\pi/4$ radians ($45^\circ$) around the $z$ axis & Transforms $\ket\rightarrow$ into the state with the spin halfway between the $x$ and $y$ axes; 
\\	\hline
			Controlled Not (\CNOT, \CX) &
			\begin{tikzcd}[baseline=-1em, anchor=north]
                        & \ctrl{1} & \qw &  \\
                        & \targ{} & \qw  & \\
                         \end{tikzcd}& X gate applied to the target (bottom) qubit if the control (top) qubit is in the logical 1 (spin-down) state. & Leaves the states with the control qubit in the state $\ket\uparrow$ unchanged; transforms $\ket\downarrow\ket\uparrow$ into $\ket\downarrow\ket\downarrow$ and vice versa.  The state $(\ket\uparrow+\ket\downarrow)\ket\uparrow=\ket\uparrow\ket\uparrow+\ket\downarrow\ket\uparrow$ will transform into the entangled superposition $\ket\uparrow\ket\uparrow+\ket\downarrow\ket\downarrow$.
\\	\hline
			Controlled Z (CZ, Controlled Phase) &
			\begin{tikzcd}[baseline=-1.5em, anchor=north]
                        & \ctrl{1} & \qw &  \\
                        & \gate{Z} & \qw  & \\
                         \end{tikzcd}& X gate applied to the target (bottom) qubit if the control (top) qubit is in the logical 1 (spin-down) state. & Leaves the states with the control qubit in the state $\ket\uparrow$ unchanged; transforms $\ket\downarrow\ket\rightarrow$ into $\ket\downarrow\ket\leftarrow$ and vice versa.  
\\	\hline		
			SWAP &
			\begin{tikzcd}
 			& \swap{1} & \qw \\
			& \targX{} & \qw
			\end{tikzcd}& 2-qubit swap &
\\	\hline		
			Toffoli (\CCNOT, CCX, TOFF) &
			\begin{tikzcd}[baseline=-1em, anchor=north]
                         & \ctrl{1} & \qw &  \\
                         & \ctrl{1} & \qw &  \\
                         & \targ{} & \qw &  \\
                         \end{tikzcd}& X gate applied to the target (bottom) qubit if both control qubits (top and middle) are in the logical 1 (spin-down) state.  &Transforms $\ket\downarrow\ket\downarrow\ket\uparrow$ into $\ket\downarrow\ket\downarrow\ket\downarrow$ and vice versa.
\\
		\hline	\hline
		\end{tabular}
\end{table*}

During the four decades of quantum computing history, a great variety of quantum computational models  have been proposed. 
We review these models in this section. 

First, we classify quantum computing devices into \emph{universal} and \emph{non-universal}. 
In classical digital computers, universality means an ability perform an arbitrary sequence of operations on a bit string. 
The quantum counterpart of this definition is the ability to perform an arbitrary transformation of the quantum state of a set of qubits. 
In other words, the universal quantum computer can be understood as extension of the Turing machine into the quantum domain, as formalized by Deutsch in 1985~\cite{Deutsch1985-2}. 

In contrast, non-universal quantum computers aim to solve a specific problem or a specific class of problems. 
There are two important subclasses of non-universal quantum computers. 
The first one is \emph{analog quantum simulators}~\cite{Nori2014,Cirac2012}: devices that simulate a process in a complex quantum system, 
e.g.~solid matter, by another quantum system with well-known and controllable properties, e.g.~an ensemble of cold atoms trapped in an optical field. 
The second class is \emph{special-purpose quantum computers} to solve a specific restricted class of abstract mathematical problems, 
e.g., quantum annealing devices, which implement discrete optimization.  
These two classes overlap as we detail below.

We consider the following quantum computing models.
\begin{itemize} 
	\item Universal:
		\begin{enumerate} 
			\item gate-based, also known as digital or circuit~\cite{DiVincenzo1995,Brassard1998,Ladd2010};
			\item adiabatic~\cite{Aharonov2007,Lidar2018};
			\item one-way, also known as cluster-state~\cite{Briegel2001,Nielsen2006};
			\item variational~\cite{Aspuru-Guzik2016,Babbush2021-4}, also known as hybrid.
		\end{enumerate}
	\item Non-universal:
		\begin{enumerate} 
			\item quantum simulators; 
			\item special-purpose. 
		\end{enumerate}
\end{itemize}

\subsection{Gate-based (circuit-based, digital) quantum computing}\label{Sec:gate-based}

The most intuitive and popular approach to universal quantum computing is generalize classical digital computing to the quantum case. 
This is the basis of the gate-based model of quantum computing \cite{Brassard1998,Ladd2010,NielsenChuang2000}. 

In the digital quantum computing model, we start by preparing our qubits (\emph{quantum register}) in a certain initial state: normally, all logical zeros. 
These qubits are then subjected to a sequence of logical gates, known as the \emph{quantum circuit} (or \emph{quantum network}), 
and subsequently measured yielding the computation result in the form of classical string of bits. 

Akin to classical digital computing, quantum gates are simple operations on qubits. 
One can prove that any arbitrary transformation of multiqubit states can be decomposed into a sequence of 1- and 2-qubit gates. 
Although the length of such a sequence generally scales exponentially with the number of qubits, this scaling is polynomial, or even linear, for many practical situations. 
Finding such situations and composing the corresponding efficient gate sequences constitutes the art of quantum algorithmics~\cite{NielsenChuang2000}.

Generally, a 1- or 2- qubit gate is parametrized by several real numbers, 
such as the angle of rotation around a particular axis. 
For practical purposes, however, it would be convenient if all gates used in an algorithm belonged to a fixed small set. 
Fortunately, there do exist gate sets (known as universal sets, see Box~\ref{Table:Gates}), that can be used to efficiently represent any sequence of arbitrary 1- or 2- qubit gates. 

Specifically, a quantum circuit of $m$ arbitrary 1- or 2- qubit gates can be approximated to $\varepsilon$ error (in so-called operator norm) by a quantum circuit of $O(m\log^c (m/\varepsilon))$ gates from a universal gate set, 
where $c$ is a constant approximately equal to 2 (this is known as a consequence of the \emph{Solovay-Kitaev theorem}~\cite{Kitaev1997}). 
This result is remarkable because a general gate parameter set is an arbitrary point in a multidimensional continuum, whereas  sequences of universal set gates are represented by a set of discrete points in this continuum. 
To construct an intuitive analogy, one can think of decimal fractions, in which any real number can be arbitrarily well approximated by a short sequence of discrete numbers. 

While there are many ways to define a universal set, the most common one consists of \Hg, \Sg, \T  and \CNOT. Three elements of this set, \Hg, \Sg\, and \CNOT, form the so-called Clifford group, which enables formation of highly-entangled states: 
 the Hadamard gate makes superposition states, the \CNOT gate creates entanglement between qubits, and the \Sg gate introduces complex amplitudes. However, according to the Gottesman-Knill theorem~\cite{Gottesman1998,NielsenChuang2000,Gottesman2004}, a circuit consisting of Clifford gates can be efficiently simulated classically. 
 To achieve quantum advantage, the \T gate must be added to this group\footnote{One may notice that the \Sg gate can be obtained by repeating the \T gate twice. 
However, the \Sg gate is still traditionally included in the universal set for the historical reasons described above. }. 

The choice of gates can be influenced by what is easier to implement on the physical platform at hand. Such platform-specific gates are known as \emph{native gates}.
For example, some platforms use the controlled phase gate instead of \CNOT.
The \CNOT gate can then be obtained by applying the \Hg gate to the target qubit before and after the controlled phase gate. 

During last decades, basic working principles of gate-based quantum computing have been demonstrated using various physical platforms with quantum registers reaching 50--100 qubits. 
However, noise and decoherence prevent sustainable sequences of more than 20--30 logical operations. 
Hence the primary task with today’s gate-based NISQ computers is the implementation of error correction allowing long-lived logical qubits (Sec.~\ref{sec:errors}). 
In the meanwhile, gate-based quantum devices are mostly useful for variational quantum computation (see Sec.~\ref{sec:variational}). 

\subsection{One-way quantum computing}\label{sec:one-way}

Two-qubit gates require controlled interaction of two specific qubits while keeping other qubits completely free of interaction with both their counterparts and the environment. 
This is a primary challenge of the gate-based model. 
The one-way (cluster-state) model addresses this challenge by preparing a complex state, in which all qubits are entangled with each other in a known way, in advance of the computation. 
The computation consists in measuring these qubits sequentially and applying single-qubit operations to other qubits depending on the result of the measurement (so-called {\it feedforward processing}). 
This process takes advantage of the remote state preparation effect (Box \ref{Box:BasicQ}), i.e.~when one part of an entangled state is measured, other part change. 
Any gate-based algorithm can be reformulated in the quantum one-way computation framework as demonstrated by Raussendorf and Briegel in 2001~\cite{Briegel2001}. 

One-way quantum computing is particularly relevant in the context of the optical platform because of lacking tools for deterministic implementation of two-qubit gates~\cite{Zeilinger2005} and quantum memory for light \cite{Lvovsky2009}. 
Cluster states and operation therewith have been demonstrated in free-space optics up to 12 qubits~\cite{Pan2018}. 
Integrated-optics implementation of optical one-way quantum computing is being pursued commercially (see Sec.~\ref{sec:optical}). 

\subsection{Adiabatic quantum computing}\label{sec:adiabatic}

Both gate-based and one-way quantum computing models rely on manipulating the quantum state of a multiqubit register. 
This manipulation results in a complex entangled state that constitutes the result of the calculation. 
Adiabatic quantum computing also works with a multiqubit system as the carrier of quantum information, but manipulates it according to a very difficult paradigm. 
Rather than implementing the algorithm as a sequence of operations on individual qubits (or pairs thereof), 
it puts the system into the physical conditions (\emph{Hamiltonian}) such that the state, which is desired as the output of the calculation, has the lowest energy among all possible states of the system (i.e.~is the \emph{ground state} of the Hamiltonian). 
It might appear that, in order to device such conditions, one would need to know the desired state. 
However, remarkably, this is not the case. 
The Hamiltonian can be calculated efficiently using a classical computer just from the corresponding circuit in the gate-based model~\cite{Kitaev2006,Aharonov2007,Lidar2018}
(therefore making this approach universal quantum computing~\cite{Aharonov2007}). 
Alternatively, the Hamiltonian can be calculated for the computational problem at hand, e.g. factorization, directly, bypassing the intermediate step of circuit design~\cite{Aspuru-Guzik2019}. 

Once the required Hamiltonian is calculated and its physical implementation is devised, a question arises is how to bring the quantum system to its ground state. 
In principle, this can be achieved by cooling it to very low temperatures close to absolute zero. 
However, these temperatures may not not achievable in practice. 
Therefore, one instead takes advantage of the so-called \emph{adiabatic theorem} of quantum mechanics. 
This theorem states that, if a system is prepared in the ground state of a Hamiltonian, and this Hamiltonian evolves slowly enough, the system will always remain in its instantaneous ground state. 
We therefore can start from some simple Hamiltonian, in which the ground state is easy to reach. 
Then we gradually (adiabatically) evolve the conditions towards the Hamiltonian that encodes the problem while keeping the system in the ground state. 
At the end of the evolution, the ground state is measured yielding the result of the computation. 

A subtle question is how slowly the Hamiltonian must be evolved to prevent the system from leaving the ground state. 
According to the adiabatic theorem, the ``speed limit'' is inversely proportional to the energy gap between the ground and second lowest energy state. 
This gap decreases with the number of qubits in the system, but fortunately not at an exponential rate~\cite{Lidar2018} --- hence making adiabatic quantum computation feasible. 

An important advantage of the adiabatic model is that it exhibits inherent robustness against certain types of quantum errors~\cite{Aharonov2007}. 

Universal adiabatic quantum processors have not yet been implemented.
The quantum annealer manufactured by D-Wave Systems\footnote{https://www.dwavesys.com} 
can be seen as the first step toward adiabatic quantum computing; however, it does not enable encoding Hamiltonian corresponding to an arbitrary computational problem and is therefore not universal. 
Furthermore, it is a subject of research whether the adiabatic theorem is satisfied in this machine (see Box~\ref{Box:DWave}). 

\wideboxbegin{Box:QUBO}{Quadratic unconstrained binary optimization (QUBO) problems.}

A particularly important class of optimization problems is \emph{quadratic unconstrained binary optimization (QUBO)}. 
It consists in finding the bit string ($\sigma_i=\pm1$) that minimizes the objective function $H = \sum_{i}R_i\sigma_i - \frac{1}{2}\sum_{i,j}J_{ij}\sigma_i \sigma_j$, 
where $R_i$ is the given ``bias vector'' and $J_{ij} = J_{ji}$ is a given ``coupling matrix''. 
This objective function emerges in solid state physics under the name of the \emph{Ising model}, hence QUBO is sometimes referred to as Ising problem. 

An equivalent formulation of QUBO is the \emph{maximum cut (MaxCut)} problem. 
It considers a graph with each edge (connecting vertices $i$ and $j$) associated with a real number $J_{ij}$. 
The problem consists in dividing the set of graph vertices into two subsets such that the total of ($J_{ij}$)’s connecting the vertices in these two subsets is maximized. 

The applications of QUBO range from basic science to problems of everyday practical nature. 
An example is portfolio optimization. 
Suppose a number of discrete assets are available for purchase. 
The expected investment return ($R_i$) is known for each asset. 
Also, the correlation $ J_{ij}$ between the expected returns is known for each pair of assets. 
This correlation is a measure of risk associated with buying these two assets; the risk is minimized if no correlation is present, which corresponds to the highest diversification of the portfolio. 
The value $\sigma_i$ of $1$ or $-1$ corresponds to buying on not buying the asset. 
The task is to choose the subset of assets with the desired balance between the expected return and the risk. 
This corresponds to the point on the efficient frontier in the framework of the 1952 Markowitz model~\cite{Markowitz1952}, for which a Nobel prize in Economics was awarded in 1990. 
This task of selecting the optimal asset set is exponentially hard because the number of possible bit strings $\sigma_1\ldots\sigma_N$ grows exponentially with the number $N$ of available assets. 

An important particular case of QUBO is the maximum independent set problem. 
It consists in finding the largest set of vertices on a graph that are not connected by edges. 
Mathematically, this problem corresponds to QUBO with all $J_{ij}$ being equal to either 1 (if an edge between a given pair of nodes is present) or 0 (if it is absent). 

While a number of classical heuristic solutions for the Ising/QUBO/MaxCut/maximum independent set problem have been proposed, none of them is efficient for large problem sizes.
 
\label{Box:QUBO}
\wideboxend

\subsection{Variational (hybrid) quantum computing}\label{sec:variational}

\begin{figure}
\center{\includegraphics[width=1\linewidth]{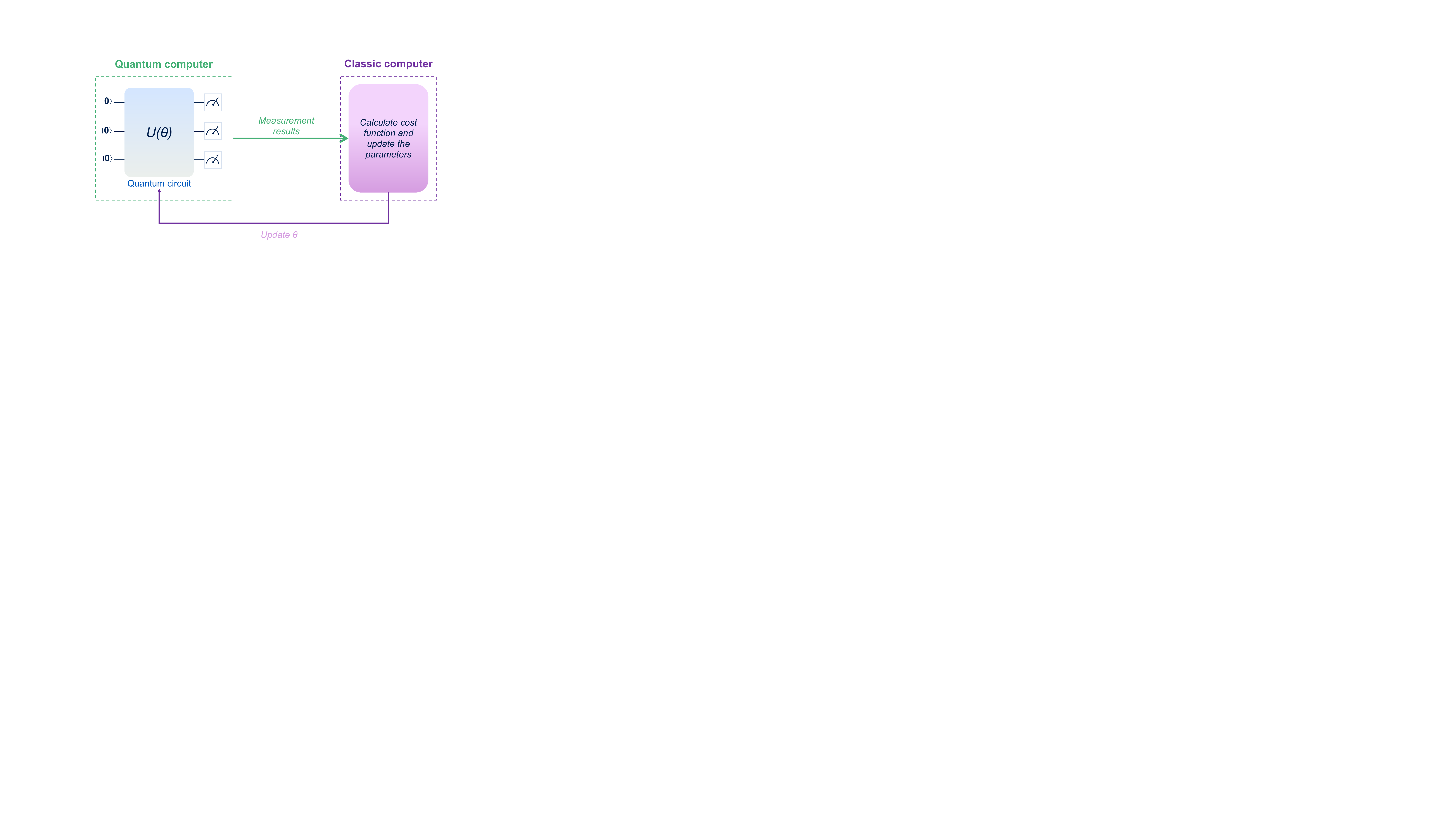}}
\vskip-4mm
\caption{An iteration of variational quantum computing. 
At the first stage of each iteration, the circuit is run multiple times and its output is measured. 
At the second stage, based on the measurement results, the energy value associated with objective Hamiltonian is evaluated classically. 
A classical optimization algorithm then provides feedback to the parameters $\theta$ of the quantum circuit in order to minimize the energy.}
\label{fig:variational}
\end{figure}

This model combines the features of the gate-based and adiabatic models. Similarly to the adiabatic model, variational quantum computing utilizes the observation that the final state of a quantum computation can be seen as the ground state of a certain Hamiltonian efficiently calculable on a classical computer. 
On the other hand, like the gate-based model, the variational quantum computer does use a quantum circuit. 
However, the gates in the circuit are not fixed, but described by continuous parameters (for example, the angle by which a qubit is rotated around a certain axis). 
At each iteration of the algorithm, the circuit output is measured and the energy value corresponding to the Hamiltonian of interest is calculated. 
Small adjustments to the parameters of the gates are then calculated  using a classical optimization algorithm with the aim to produce the state with a lower energy (Fig.~\ref{fig:variational}). Iterations continue until the energy of the output state no longer reduces.

An important advantage of variational algorithms is that the optimization cost function --- the energy --- may not only be  computed from a gate circuit, but can also represent the actual energy of a real physical object, such as a molecule. 
Then the quantum variational optimization will result in the output state representing the ground state of the electrons in this molecule with the corresponding energy. 
This idea gives rise to an algorithm known as the variational quantum eigensolver (VQE)~\cite{Bharti2021,Aspuru-Guzik2016,Aspuru-Guzik2014}, which was developed in 2014. Moreover, 
the scope of VQE can be extended to arbitrary cost functions beyond energy, leading to the quantum approximate optimization algorithm (QAOA), proposed in 2014 to solve combinatorial optimization problems~\cite{Farhi2014,Farhi2019}. 

Historically, these two algorithms have been invented outside of the context of universal  quantum computation. Formal universality proofs have been presented 
by the groups of Lloyd and Biamonte in 2018--2021~\cite{Lloyd2018-2,Biamonte2020,Biamonte2021}. Notably, these proofs relied on an unproved assumption that the optimization algorithm is capable of converging to the lowest energy state~\cite{Bittel2021}. Closing this gap is an open problem.

The variational model is particularly relevant at present as  current NISQ devices have limited recourses (number of qubits, fidelity, number of operations, etc.) and furthermore the parameters of each gate cannot be precisely controlled. 
The variational model appears to be more forgiving to these shortcomings because it does not require precise knowledge and control of the absolute values of each circuit parameters, but only needs small relative adjustments thereof. 
A shortcoming, on the other hand, is the need to have much more complete information about the output quantum state in order to calculate the energy value. 
This means at each iteration the circuit must be run multiple times and the output state measurements must be performed in different bases. 
For example, the estimation of the energy of a relatively simple molecule Fe$_2$S$_2$ would require as many as $10^{13}$ measurements~\cite{Aspuru-Guzik2020}. 
Assuming (optimistically) that each quantum circuit run takes 10 ns, 
the single iteration would require about 24 hours. 

Variational quantum algorithms have been demonstrated in the context of  optimization~\cite{Monroe2020,Babbush2021,Lukin2020} (for example, QAOA was implemented experimentally by IonQ [USA] and Google [USA]), 
machine learning~\cite{Biamonte2017}, 
quantum chemistry~\cite{Aspuru-Guzik2020},
linear algerba~\cite{Yuan2021},
and quantum simulation~\cite{Yuan2019}. 
A detailed review of variational quantum algorithms can be found in Bharti {\it {\it et al.}}~\cite{Bharti2021} and Cerezo {\it {\it et al.}}~\cite{Babbush2021-4}.

\subsection{Quantum simulators}\label{sec:quantumsimulators}

Complete and precise theoretical descriptions of complex quantum systems, such as solid state, which involves interaction of multiple microscopic quantum objects, is beyond the reach of current science and technology. 
There do exist simplified models that capture their salient properties. 
But even in the framework of these models, the curse of dimensionality makes the analysis exponentially hard for classical computers. 
As discussed above, this is one of main motivations behind quantum computing. 

Each of the above described universal quantum computing models can in principle be used to simulate arbitrary complex quantum systems~\cite{Lloyd1996}. 
For example, a digital superconducting quantum computer was used to simulate the interaction of two fermions, 
whose states are encoded in four qubits~\cite{Martinis2015}. 
This rather simple simulation required as many as 300 single-qubit and two-qubit gates. 
Another example is the aforementioned variational quantum eigensolver, which can find the lowest energy state of a quantum system of interest. 
In an experimental realization based on trapped ions, interaction within a system of multiple high-energy particles governed by the so-called Schwinger model was simulated~\cite{Blatt2019,Biamonte2021-2}. The ground state of the system was found as well as the phase transition as the function of the particle mass. 

However, the simulation of quantum systems also permits an entirely different approach~\cite{Cirac2012,Nori2014}: quantum machines with known and controllable properties imitating the quantum system of interest. 
This is known {\it as analog quantum simulation}. 
The advantage of this approach in comparison to digital quantum computing is that the simulation can be done at a much larger scale at the expense of loosening precise control over individual elements and lack of fault tolerance~\cite{Lewenstein2012}. 
In this context, an important benchmark of a quantum simulator is its \emph{programmability}, which is the degree of control that can be imposed on its elementary quantum units and their interaction. 

Progress over the last two decades has produced more than 300 quantum simulators in operation worldwide, using a wide variety of experimental platforms~\cite{Nori2014,Bloch2012,Blatt2012,Aspuru-Guzik2012,Altman2021}. 
They range from highly optimized special-purpose non-programmable simulators to flexible programmable devices. 
Physical platforms include solid-state superconducting circuits, quantum dot arrays, nitrogen-vacancy centers, atomic and molecular systems, such as Rydberg atoms and trapped ions, interacting photons, and others. 
We describe these platforms in detail in Sec.~\ref{sec:platforms}.

\wideboxbegin{Box:DWave}{The D-Wave quantum annealer.}
Good examples  of special-purpose quantum machines are the superconducting quantum annealers produced by  D-Wave Systems. These devices feature remarkably many qubits (5,000 in the latest model D-Wave Advantage), greatly exceeding that in other existing quantum processors. The D-Wave machines can be seen as a step towards the adiabatic model (Sec.~\ref{sec:adiabatic}): 
they gradually vary (``anneal”) the physical conditions into which the qubits are placed in order to drive them to the ground state of a particular Hamiltonian. 

However, the D-wave annealer is not yet a universal adiabatic quantum computer. 
The Hamiltonian it is capable of implementing is not arbitrary (as required by the adiabatic model), but limited to the Ising (QUBO) Hamiltonians  --- in other words, it can look for bit strings that minimize the QUBO objective function (Box~\ref{Box:QUBO}). 
But even within the framework of the QUBO problem, it is not able to realize any arbitrary coupling matrix $J_{ij}$. This is because every qubit is connected to only a small number of other qubits [7 in D-Wave 2000Q and 15 in D-Wave Advantage, see Fig.~\ref{fig:DWave}(a)]. 
This leads to a major overhead when an all-to-all coupled Ising problem needs to be solved. Physical qubits are then grouped into clusters such that the qubits within each cluster are forced to the same logical value and share their outside connections. The entire cluster then plays the role of a single ``logical qubit" for the purpose of the calculation; the number of such logical qubits is limited to a few dozen~\cite{Benedetti2017}. 
Moreover, there exist no confidence whether the D-Wave annealer properly fulfils the adiabatic theorem~\cite{Vazirani2014}. In practice, this means that the calculation output bit string may not be the true minimum of the QUBO objective function.
 
The question of quantum advantage of D-Wave machines has been widely discussed in the literature. 
Evidence of quantum effects in the annealing process was claimed in 2014~\cite{Troyer2014} by D-Wave One (108 qubits), 
but these claims have been disputed by other groups presenting classical models that efficiently simulate the annealer's behavior~\cite{Vazirani2014}. 
Subsequent D-Wave annealer models featured significantly higher qubit numbers, and have been used to attempt demonstration of quantum advantage on specially tailored problems~\cite{Katzgraber2020}. 
Existing studies of this matter compared the performance of the D-Wave annealer 2X with 2000 qubits and classical algorithms in application to various problems and produced controversial results~\cite{Roy2014,Troyer2014-2,Martinis2016,Lidar2018-3,Katzgraber2018}. 
One of the latest results by Google and D-Wave in 2021 is a claim of quantum advantage in the physically relevant problem of simulating geometrically frustrated magnets~\cite{Amin2021}. 
Thus, at this moment there is no universally accepted conclusive evidence of quantum advantage of the D-Wave machine. 

In spite the lack of such evidence, the D-Wave annealer is being extensively studied in application to various problems of practical significance (see Sec.~\ref{sec:application-optimization}). 
However, as discussed above a major issue is embedding the problem in the native structure of the annealer~\cite{Benedetti2017,Troyer2014,Troyer2014-2}, which limits applicability of the device to problems of very small sizes only. 

\label{Box:DWave}
\wideboxend

\wideboxbegin{Box:CIM}{Coherent Ising Machine for annealing.}

The coherent Ising machine (CIM) is another example of a hardware annealer. 
CIMs store the information about the optimization variables in optical pulses and use optoelectronic feedback to implement the couplings between them. 
In contrast to D-Wave processors, CIMs have no restrictions on connectivity between variables. 
In 2016 CIMs with the capacity of 2048 variables ~\cite{Takesue2016,Yamamoto2016}, and in 2021 with as many as 100512 variables, have been demonstrated~\cite{Honjo2021}. 
It was shown that CIMs significantly outperform D-Wave processors in dealing with dense QUBO functions~\cite{Yamamoto2019}, although this claim was disputed by the D-Wave team~\cite{McGeoch2018}.

However, because the coupling of the optical pulses in the CIM is implemented via measurements and optoelectronic feedforward, no entanglement between the pulses is possible. 
This means that any speedup observed cannot be of quantum nature. 
This was confirmed by classical simulations~\cite{Lvovsky2019}, which ran on graphic processors and achieved solution speed and quality that is comparable or superior to that of CIM. 
This shortcoming can be addressed, and quantum entanglement can be achieved by coupling the optical information units in the CIM by direct interference. This has proven to be challenging, but is an important vector for the future development of this technology~\cite{Marandi2014}. 

\label{Box:CIM}
\wideboxend

\subsection{Special-purpose quantum machines}\label{sec:special-purpose} 
An important example of special-purpose problems solvable by quantum machines is discrete optimization, which arises in various industries ranging from logistics to finance, such as QUBO (see Box~\ref{Box:QUBO}). 
Traditionally, such problems have been solved by a family of classical algorithm known as simulated annealing, 
in which the set of variables to be optimized is treated as a physical system with probability of different configurations given by the thermal distribution associated with some temperature. 
As the ``temperature'' is decreased, the probability of the optimal configuration increases to one. 
This process is reminiscent to annealing in metallurgy, giving rise to its name. This term is now also used for a variety of quantum and analog methods for combinatorial optimization, 
even though they may not involve any thermal distribution or temperature variations. 

\begin{figure*}[ht]
\center{\includegraphics[width=1\linewidth]{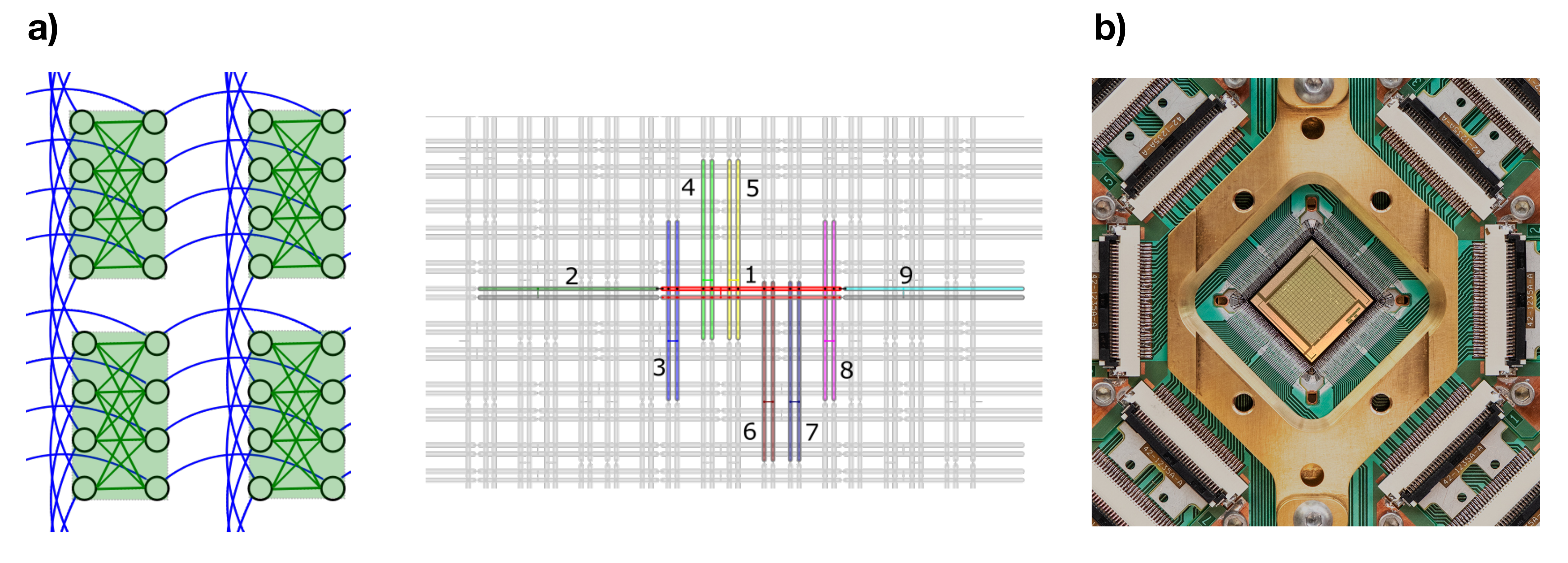}}
\vskip-3mm
\caption{D-Wave quantum annealer (reproduced from dwavesys.com): 
a) Qubit coupling topology. Left: 2000Q processor (chimera unit cell): qubits are shown as circles and couplers as lines. Right: 5000Q processor, also known as Advantage (Pegasus unit cell). 
Qubits are represented by horizontal and vertical loops. Qubits coupled to qubit \#1 (red) are colored.
b) Photograph of the D-Wave 5000Q Advantage processor. }
\label{fig:DWave}
\end{figure*}

A prominent example of a quantum annealer is the superconducting device for solving QUBO problems manufactured by D-Wave Systems (see Box~\ref{Box:DWave} and Fig.~\ref{fig:DWave}). 
Another solution that is frequently measurement in the context of hardware annealing is the optical coherent Ising machine (see Box~\ref{Box:CIM}); 
however, current realizations of this approach do not feature entanglement between computational units and hence do not exhibit  quantum advantage. 

\begin{figure*}[t]
\center{\includegraphics[width=1\linewidth]{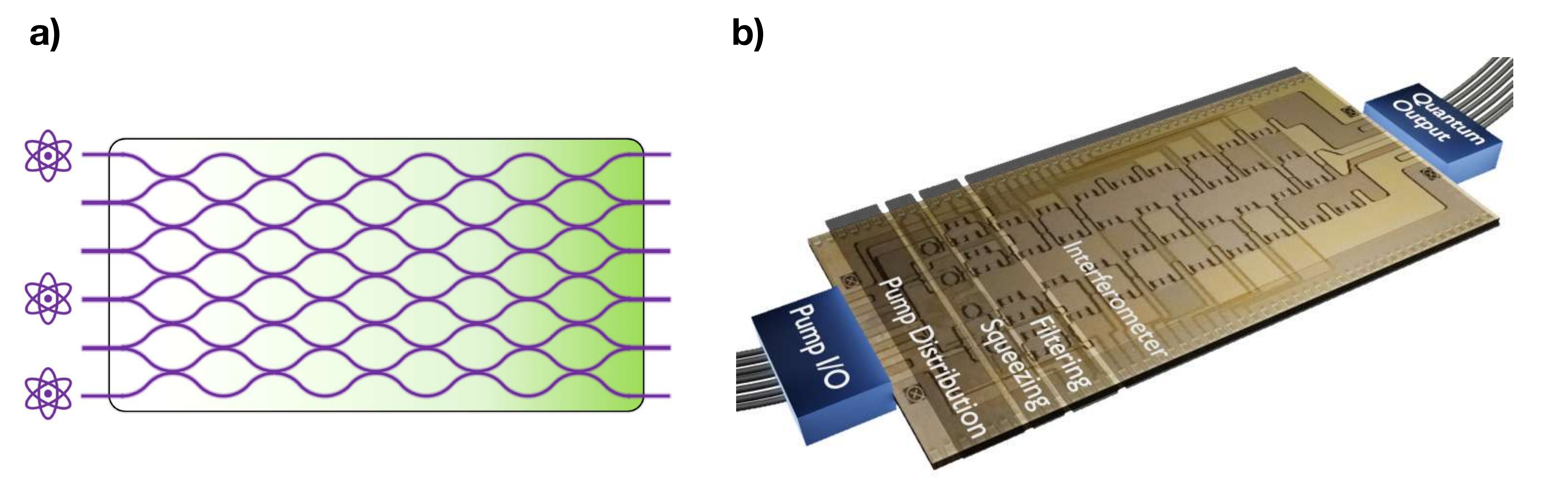}}
\vskip-3mm
\caption{Boson sampling: a) basic principles (reproduced from Chabaud {\it et al.}~\cite{Chabaud2021}; b) rendering of the chip (based on a micrograph of the actual device) for Gaussian boson sampling
(reproduced from Arrazola {\it et al.}~\cite{Schuld2021}).}
\label{fig:boson-sampling}
\end{figure*}

Beyond combinatorial optimization, an important class of special-purpose quantum machines is the boson sampler~\cite{Aaronson2013}. 
This is a network of intersecting optical waveguides with $n$ input and $n$ output channels, where $n$ is large [Fig.~\ref{fig:boson-sampling}(a)]. 
Single photons are injected into  $m<n$ input channels and subsequently detected at the output. As the photons propagate through the network, 
they can jump between waveguides at their intersections or experience interference with each other if they arrive at an intersection together. 
As a result, the output state of the photon paths feature complex entanglement. 
Hence the probability, with which the photons will emerge in a particular combination of output channels, is conjectured to be exponentially difficult to calculate 
(as it involves calculating the so-called permanents of the matrix describing the network)~\cite{Aaronson2013}. 
The output photon detection produces a sample of such a distribution, thereby offering a solution to a classically hard problem. 
Boson sampling is therefore of interest as a setting, in which quantum superiority can be demonstrated. First attempts to realized boson sampling were implemented in 2013 with up to 4 photons in 6 modes~\cite{Tillmann2013,White2013,Walmsley2013}. 

The boson sampling scheme in its original form is difficult to scale up because no on-demand sources of high-quality single photon exist yet.
An important breakthrough is associated with the concept of Gaussian boson sampling [Fig.~\ref{fig:boson-sampling}(b)]. In this scheme, the states of light injected into the optical network are the so-called squeezed  vacuum --- 
a class of states of light, which, like single photons, exhibit quantum features, but can be produced  on-demand with relatively little effort~\cite{Lvovsky2015}. 
The idea of Gaussian boson sampling enabled experimental realization on a scale, at which quantum advantage is present (see Sec.~\ref{sec:sciapplications} for further detail).

Boson sampling was initially introduced purely as a problem for demonstrating quantum advantage~\cite{Aaronson2013} (see Sec.~\ref{sec:sciapplications}), abstract from any practical utility. 
However, it was later discovered to have applications in chemistry~\cite{Aspuru-Guzik2015} (calculating molecular vibronic spectra) and mathematics~\cite{Schuld2020} (graph similarity). 

The final example of special-purpose quantum machines is the previously mentioned neutral Rydberg atom simulator. 
This system is remarkable in that it occupies several positions in our classification. 
On the one hand, it can be used to solve the maximum independent set problem~\cite{Lukin2018,Lukin2018-2,Lukin2020,Browaeys2020-2,Ayral2020} (see Box~\ref{Box:QUBO}), 
which corresponds to finding the minimal energy configuration of an ensemble of Rydberg atoms. 
On the other hand, it is a programmable quantum simulator capable of probing exotic phase transitions in condensed matter, which is a classic quantum simulation problem. 
Furthermore, this platform can also be used as a fully digital quantum computer. 
We discuss this system in detail in Sec.~\ref{sec:ionsatoms}.  

\section{How can these models be compared?}\label{sec:comparison}

As seen from the previous discussions, various quantum computing models dramatically differ not only in their physical and technical implementation, but also in their fundamental computational paradigms. 
Hence it is difficult to define universal performance evaluation criteria. 
A first attempt at this task was made by DiVinchenzo in a classic paper of 2000~\cite{DiVincenzo2000}, who formulated five qualitative requirements that a physical setup must satisfy in order to support gate-based quantum computing. 
These requirements have since evolved in adaptation to emerging quantum computational models and a number of quantitative benchmark have been proposed, which we summarize below. 
The specific values for some of these parameters for various physical platforms are listed in Table~\ref{tab:parameters}.


\begin{enumerate}
	\item User-oriented criteria.
        \begin{enumerate}
		\item Accessible class of problems. 
		\item Speed. 
		\item Cost. 
        \end{enumerate}
	\item Technical criteria.
	\begin{enumerate}
		\item Size (number of elementary quantum units). 
		\item Quality.
		\begin{enumerate}
			\item Decoherence time.
			\item Fidelity of  operations (gates), state preparation and measurements. 
	\end{enumerate}
		\item Dimensionality of elementary quantum units. 
		\item Duration of an elementary operation.
		\item Connectivity. 
		\item Parallelism. 
		\item Programmability.
		\item Ability to realize error suppression / correction. 
	\end{enumerate}
\end{enumerate}

While the first group in the above list is rather intuitive, the second one requires explanation, which we provide below. 
Each of these requirements gives rise to a quantitative benchmark, which can be used to assess and compare different quantum computational platforms. 

The size of problem at hand dictates the number of elementary quantum units required for the quantum computing device to solve it. 
However, the notion of such a number is in itself ambiguous. 
For example, a quantum algorithm in the gate model operates with idealized quantum logical variables ({\it logical qubits}) that are assumed to be perfectly isolated from the environment and able to store information for infinitely long time. 
The practical {\it physical qubits} are, however, subject to a variety of imperfections and errors, in particular, decoherence. 
To address these imperfections, redundant encoding is used: multiple physical units encode a single logical qubit compensating each other’s errors by means of quantum error correction (see Sec.~\ref{sec:errors}). 
The overhead ratio of physical and logical qubits depends on another major benchmark: the error rate $\varepsilon$, which is the ratio between the duration of a single operation (gate) and the characteristic decoherence time. 
For example, factoring a 2048-bit RSA key requires approximately 6000 logical qubits, but 20 million physical qubits, assuming that the error rate of $\varepsilon=10^{-3}$ (i.e., a single error occurs once in every 1000 gate operations)~\cite{Gidney2021}. 

When the error rate in a given platform exceeds a certain threshold, error correction cannot be implemented at all, no matter how high the the overhead. 
Moreover, error correction techniques have to day  been developed mainly for the gate-based model, and much less so for other models. 
Thus the ability of a quantum computer to implement error correction depends not only on the error rate, but also on the quantum computational model as well as other factors. 
That being said, for some models, such as analogue quantum simulators, error correction is not a requirement at all. 


\begin{table*}[]
\caption{Performance benchmarks of primary quantum computing platforms, represented by the record values achieved to date. The data for the neutral-atom, trapped-ion and superconducting platforms are taken from Ref.~\cite{Shi2021}, which also contains the corresponding bibliography references.}
\label{tab:parameters}
\begin{tabular}{|l|l|lllll|lll|}
 \hline & \multicolumn{1}{l|}{\begin{tabular}[c]{@{}l@{}}Quantum  \\ units (qubits)\end{tabular}} & \multicolumn{1}{l|}{\begin{tabular}[c]{@{}l@{}}1-qubit \\ fidelity\end{tabular}} & \multicolumn{1}{l|}{\begin{tabular}[c]{@{}l@{}}2-qubit \\ fidelity\end{tabular}} & \multicolumn{1}{l|}{\begin{tabular}[c]{@{}l@{}}Coherence \\ time\end{tabular}} & \multicolumn{1}{l|}{\begin{tabular}[c]{@{}l@{}}Gate \\ time\end{tabular}} &\multicolumn{1}{l|}{\begin{tabular}[c]{@{}l@{}}Error \\rate\end{tabular}}         &\multicolumn{1}{l|} {\begin{tabular}[c]{@{}l@{}}Achievable\\dimension\end{tabular}}                                                        & \multicolumn{1}{l|}{Connectivity}                                                        & Parallelism                                                               \\ \hline
Neutral atoms                                                                & 256                                                                                                       & \multicolumn{1}{l|}{99.986}                                                      & \multicolumn{1}{l|}{99.5}                                                               & \multicolumn{1}{l|}{1-10 s}                                  & \multicolumn{1}{l|}{100 ns}                                                   & $\sim$10$^{-7}$ & \multicolumn{1}{l|}{\begin{tabular}[c]{@{}l@{}}Qudits, \\ 3-5 levels\end{tabular}}  & \multicolumn{1}{l|}{All-to-all}                                                           & Yes                                                                       \\ \hline
Trapped ions                                                                 & 30                                                                                                        & \multicolumn{1}{l|}{99.999}                                                      & \multicolumn{1}{l|}{99.9}                                                              & \multicolumn{1}{l|}{600 s}                                    & \multicolumn{1}{l|}{100 $\mu$s}                                               & $\sim$10$^{-9}$ & \multicolumn{1}{l|}{\begin{tabular}[c]{@{}l@{}}Qudits, \\ 3-8 levels\end{tabular}}  & \multicolumn{1}{l|}{All-to-all}                                                          & \begin{tabular}[c]{@{}l@{}}Hard for \\ 2-qubit \\ operations\end{tabular} \\ \hline
\begin{tabular}[c]{@{}l@{}}Superconducting\\ junctions\end{tabular} & 127                                                                                                       & \multicolumn{1}{l|}{99.92}                                                       & \multicolumn{1}{l|}{99.7}                                                                  & \multicolumn{1}{l|}{0.5 ms}                           & \multicolumn{1}{l|}{10 ns}                                                    & $\sim$10$^{-4}$ & \multicolumn{1}{l|}{\begin{tabular}[c]{@{}l@{}}Qudits, \\ 3-8 levels\end{tabular}}  & \multicolumn{1}{l|}{Neighbors}                                                           & Yes                                                                       \\ \hline
Photons                                                                    & 76                                                                                                        & \multicolumn{1}{l|}{99.998}                                                      & \multicolumn{1}{l|}{99.86}                                                            & \multicolumn{1}{l|}{n/a}                                       & \multicolumn{1}{l|}{$<1$ ps} & n/a            & \multicolumn{1}{l|}{\begin{tabular}[c]{@{}l@{}}Qudits, \\ 3-10 levels\end{tabular}} & \multicolumn{1}{l|}{\begin{tabular}[c]{@{}l@{}}All-to-all \\ (potentially)\end{tabular}} & Yes                                                                       \\ \hline
\begin{tabular}[c]{@{}l@{}}Silicon spins \\ (quantum dots)\end{tabular}    & 4                                                             & \multicolumn{1}{l|}{99.9}                                                        & \multicolumn{1}{l|}{99.8}               & \multicolumn{1}{l|}{10 ms}                                                                                   & \multicolumn{1}{l|}{0.8 ns}                                                   & $\sim$10$^{-6}$ & \multicolumn{1}{l|}{Mostly qubits}                                                  & \multicolumn{1}{l|}{Neighbors}                                                           & Yes                                                                       \\ \hline
\end{tabular}
\end{table*}

In addition to the effect of passive interactions with the environment, the quality of the state of the quantum register is influenced by imperfect control, i.e., state preparation and measurement (SPAM) errors and gate inaccuracies. 
The quantitative measure of these imperfection is the \emph{fidelity}, i.e., how close the prepared quantum state (gate) is to theoretically desired. 
There exists an arsenal of experimental tools for estimating the fidelity. In the gate-based model, this criterion is further specialized in terms of single- and two-qubit gate fidelities. 

Even if the qubits are of perfect quality, a quantum computing device may not enable their arbitrary pairwise interaction. 
This capability, known as the {\it connectivity}, is another important requirement for a quantum computing platform. 
For example, in the trapped-ion platform, in which all ions are situated in the same trap and gates are implemented through their mechanical interaction, all-to-all connectivity is possible. 
In contrast, connectivity is a challenge in the superconducting model, in which a two-qubit gate is realized by a physical junction between these qubits. 
For example, the publicly accessible IBM superconducting quantum computer has only two to three connections per qubit, whereas the state-of-the-art Google Sycamore unit constitutes a 2D square grid of qubits~\cite{Martinis2019}, 
so each qubit is connected to four others. 
A 2D architecture is also used in Rydberg atom quantum simulators~\cite{Lukin2021,Browaeys2021}, although the connectivity issue in this setting can be addressed by physically repositioning the atoms with respect to each other. 
In the absence of all-to-all connectivity, a two-qubit gate between arbitrary qubits can be realized by swapping (applying the SWAP gate, see Box \ref{Table:Gates}) the quantum state through a chain of neighboring qubits. 
The price to pay is the decoherence resulting from the need to include additional operations in the algorithm. 

A related important performance criterion is the ability to implement operations in parallel. 
For some quantum computational settings, such as the atomic simulator, the parallel interaction of multiple pairs of units is inherent in their nature and essential for proper operation. 
In other models, such as gate-based, the parallelism is optional, but desired for faster implementation of quantum algorithms. 
The achievability of parallelism depends on the specific physical platform and is often complementary to connectivity. 
For example, it is relatively straightforward in the superconducting model, but more challenging in the trapped-ion model. 

The time required to solve a computational problem is directly proportional to the duration of an individual quantum gate operation. 
The aforementioned factorization of 2048-bit RSA key would require 8 hours with the average gate time of 10 $\mu$s~\cite{Gidney2021}. 
As seen in Table~\ref{tab:parameters}, the gate time strongly depends on the physical platform of a quantum computer.  

Many practical quantum devices operate with units that naturally have more than two independent quantum states, i.e., are multidimensional (for example, multiple energy levels in atoms). 
Using such multidimensional units, known as qudits, instead of qubits to encode quantum information helps to reduce the number of gates required for the realization of quantum algorithms~\cite{Sanders2020}. Thus, the dimensionality of the elementary information unit is an important parameter of a quantum computing device. Quantum operations with qudits have been demonstrated with various physical systems. Most progress has been achieved with the  superconducting, trapped-ion, and optical platforms, on which qudit processors have been reported (Rigetti Computing~\cite{Galda2021}, AQT with collaborators~\cite{Blatt2021}, and the Peking University team with collaborators~\cite{Wang2022}, respectively). 

As we see from the above discussion, assessing a quantum computing platform involves relatively large number of complementary and sometimes conflicting criteria~\cite{Blume-Kohout2021}. 
One is therefore tempted to simplify the task and introduce a single-number metric to express a quantum computer’s power. 
One such metric is the {\it quantum volume} introduced by IBM in 2017--2019~\cite{Gambetta2018,Gambetta2019-2} for gate-based quantum computers. 
Quantum volume is defined as $2^{\rm AQ}$, where ${\rm AQ}$ --- number of algorithmic qubits --- 
is the maximum size of a ``square" quantum circuit that can be successfully implemented with this platform (Box~\ref{Box:QV}). 
Historically, the number of algorithmic qubits as a figure of merit for a quantum computational platform has been introduced \emph{after} quantum volume, namely by IonQ in 2020\footnote{\url{https://ionq.com/algorithmic-qubit-estimator}}. 
To date, the largest quantum volume of 2048 (${\rm AQ}=11$) has been demonstrated by Quantinuum (previously Honeywell) in 2021 in a trapped ion machine\footnote{https://www.quantinuum.com/pressrelease/demonstrating-benefits-of-quantum-upgradable-design-strategy-system-model-h1-2-first-to-prove-2-048-quantum-volume}. 

\section{Quantum computers can deal with errors}\label{sec:errors}

In the early stage of quantum computing development, the accumulation of error caused by environmental noise (decoherence) was widely used to argue that it is infeasible to build a large-scale quantum computer~\cite{Unruh1995,Landauer1995}
As a result of this controversy, the development of the field in last few decades followed several vectors. 
First, experimental efforts were made towards investigating what quantum computers are capable of in the presence of decoherence, giving rise to the current NISQ technologies. 
Second, concepts of quantum gate-based computing devices with {\it digital error correction}, which use redundant qubits, have been developed. 
Third, it was found that certain structures of quantum matter are robust to decoherence, leading to topologically-protected quantum computation. 
We discuss each of these below.

\wideboxbegin{Box:QV}{Quantum volume.}

The value of quantum volume is formally defined as $2^{\rm AQ}$, where $$
	\mathrm{AQ}=\arg\max_{n<N}(\min{[n,d(n)]}
$$ 
is the number of algorithmic qubits with the {\it quantum depth} $d(n)=1/n\varepsilon$ being the number of operations before a single error has occurred. 
To understand this expression consider a quantum computer with $N=10^6$ qubits and the error rate $\varepsilon=10^{-3}$. 
To determine the quantum volume, we need to find the optimal $n$, for which $\min[n,d(n)]$ is maximized. 
For example, if we choose $n=1$ we have $d(n)$ equals 1000, so $\min[n,d(n)]=1$. 
On the other hand, choosing $n=10^6$ will results in $d(n)=10^{-3}$, so $\min[n,d(n)]=10^{-3}$. 
Both these cases are suboptimal: in the first case one can perform many operations with only one qubit, 
whereas in the second case, if we take too many qubits, there is a high probability that at least one of them will decohere before even a single operation takes place. 
The optimal value of $n$ in our situation is about $30$, in which case $d(n)\approx{30}$ as well, meaning that the circuit is of square shape. 
So the quantum volume is $2^{30}\approx{10^9}$. 

\label{Box:QV}
\wideboxend

\subsection{Noisy intermediate-scale quantum devices}

As mentioned above, NISQ devices have 50-100 physical qubits and do not implement any tools for error correction. 
In spite of these limitations, these machines have been used to implement basic quantum algorithms~\cite{Montanaro2016,Bharti2021} and demonstrate quantum advantage~\cite{Montanaro2017}. 
In some cases, such as analog quantum simulation, decoherence forms a natural part of finding the solution, because simulated objects themselves experience decoherence. 
This is reason for successful application of such devices to simulate phases and transition between them in condensed matter~\cite{Lukin2017}. 
A further promising model of quantum computation within the NISQ framework appears 
to be the variational model~\cite{Babbush2021-4,Bharti2021} because it uses relatively short operation sequences.

In the absence of digital quantum error correction, there exist techniques for reducing the effect of errors at the level of individual qubits~\cite{Gambetta2017,Gambetta2019,Benjamin2017,Benjamin2018,Lidar2014,Lidar2017-2,Lidar2018-4}, 
such as error mitigation, error suppression, and fidelity amplification. 
As an example, computation accuracy can be enhanced through extrapolation of results from a collection of experiments with varying noise with no additional hardware modifications~\cite{Gambetta2019}. 
The potential of this family of approaches has not yet been systematically studied. 
More generally, it is not known at this time whether quantum devices without error correction can provide quantum advantage for practically relevant problems. 

\subsection{Devices with error correction: Fault-tolerant quantum computing}

The existence of computational errors is not limited to quantum domain. 
Classical digital computers use redundant bits to nondestructively detect and correct errors. 
Such a direct approach is, however, not applicable in quantum technology because any measurement results in the loss of coherence and entanglement. 
Moreover, no-cloning theorem of quantum physics~\cite{Zurek1982} precludes creating an independent and identical copy of an arbitrary unknown quantum state.
Thus, quantum error correction should be tackled in a subtler way. 
The idea is still to use many (imperfect) physical qubits to encode one (perfect) logical qubit and perform measurements in order to detect errors. 
However, these measurements must be specially constructed to reveal no information about the values of the qubits, but only indicate whether and at which position the error has occurred. 

\wideboxbegin{Box:Shorscode}{Shor’s error correction code.}

Examples of errors are bit flips (e.g., $|0\rangle\mapsto|1\rangle$ or $\alpha|0\rangle+\beta|1\rangle\mapsto\beta|0\rangle+\alpha|1\rangle$) 
or phase flips ($\alpha|0\rangle+\beta|1\rangle\mapsto\alpha|0\rangle-\beta|1\rangle$; there is no classical equivalent to the phase flip error).
Specializing to spin qubits, this is equivalent to the rotation of the qubit around the $z$ or $x$ axes, respectively. 
Suppose we wish to implement a code capable of correcting a bit flip. 
If we have only one qubit, the only thing we can do is to measure this qubit, which will destroy the superposition state, but not tell us whether the error has occurred. 

Shor in 1995 proposed instead to encode a single logical qubit into three physical qubits~\cite{Shor1995}, i.e. for the logical qubit superposition $\alpha|0\rangle+\beta|1\rangle$ 
to use an entangled state of the form $\alpha|\uparrow\uparrow\uparrow\rangle+\beta|\downarrow\downarrow\downarrow\rangle$, 
which is easy to prepare using standard gates. 
To detect errors, one needs to measure the \emph{product} of the $z$-components of any pair of these three qubits --- for certainty, let it be the first and second qubit. 
Quantum mechanics allows measuring such products without measuring each qubit individually and, moreover, such a measurement reveals no information about the qubit contents. 
Indeed, a healthy Shor state will have both qubits in the same state: either $\ket{\uparrow\uparrow}$ (corresponding to $z_1=z_2=1$) or $\ket{\downarrow\downarrow}$ (then $z_1=z_2=-1$). 
In both cases, the measurement will yield the same result: $z_1z_2=1$ --- and hence it will neither tell us anything about the values of $\alpha$ and $\beta$ nor destroy the superposition. 

Suppose now that the first of the physical qubits in Shor's code underwent a bit flip, so the state became $\alpha|100\rangle+\beta|011\rangle$. 
Measuring the three qubit pairs, we find $z_1z_2=-1$, $z_1z_3=-1$ and $z_2z_3=1$, which will tell us unequivocally that the error occurred in the first qubit. 
We can correct for this error by applying a Pauli X gate. 

The procedure of detecting phase flip errors is similar to the above, but the error can be detected by measuring the spins' $x$ components instead of $z$. 
Moreover, both single phase flips and single bit flips can be detected by a more complicated encoding of a single logical qubit in nine physical qubits. 

\label{Box:Shorscode}
\wideboxend

Shor's seminal error-correcting code uses nine physical qubits to encode one logical qubit (see Box~\ref{Box:Shorscode} and Ref.~\cite{Shor1995}). 
Soon after Shor’s code, new error-correction codes were developed that lowered the number of physical qubits in a logical qubit to five while maintaining the same level of protection~\cite{Steane1996,Bennett1996,Knill1997,Bacon2006}. 
The theory quickly evolved, producing more and more sophisticated classes of error correction codes~\cite{Devitt2013}. 

\begin{figure}
\center{\includegraphics[width=1\linewidth]{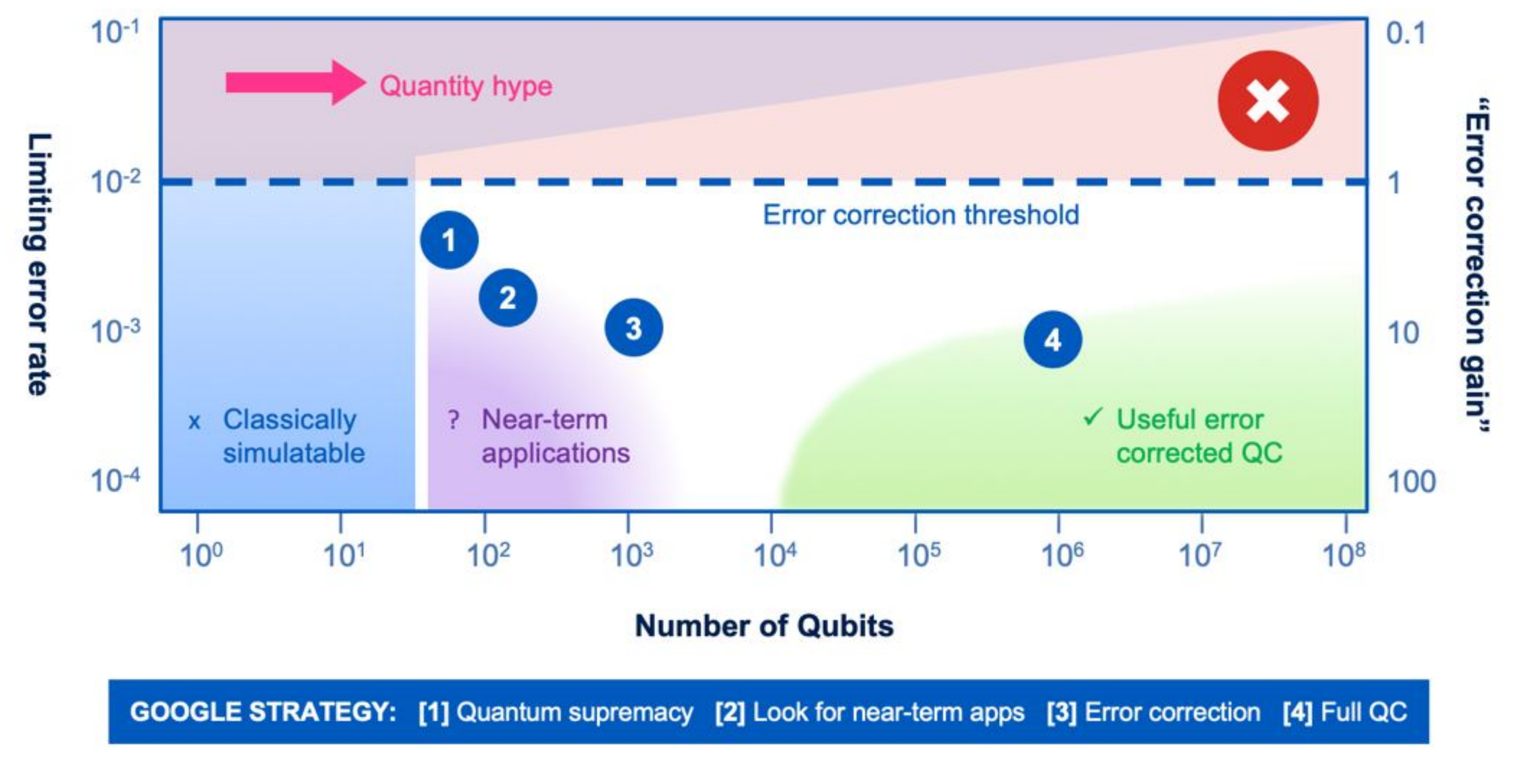}}
\vskip-4mm
\caption{Google's strategy on error correction (\url{https://quantumai.google/learn/map}).}
\label{fig:errors}
\end{figure}

One of the primary results of this theoretical research is the quantum threshold theorem (or quantum fault-tolerance theorem)~\cite{Shor1996-2,Knill1998,Kitaev2003,Aharonov2008}. 
It states that a quantum computer with a physical error rate below a certain threshold can, through application of quantum error correction schemes, suppress the logical error rate to arbitrarily low levels (see Fig.~\ref{fig:errors}). 
Roughly speaking, we need to be
 ``correcting errors faster than they are created"\footnote{Aaronson, Scott; Granade, Chris (Fall 2006). ``Lecture 14: Skepticism of Quantum Computing". PHYS771: Quantum Computing Since Democritus. Shtetl Optimized. Retrieved 2018-12-27.}. 
The optimal error correction method in a given quantum computational setting depends not only on the value of the threshold, but also on its architecture, in particular, the qubit connectivity.

The error threshold of Shor's code is relatively low. For example, in the widely studied square 2D lattice model, in which each qubit is connected to its four nearest neighbours, the error threshold for Shor’s code is as low as $2\times10^{-5}$~\cite{DiVincenzo2007}. 
This is because the implementation of Shor’s code requires operations on pairs of qubits that are not necessarily nearest neighbors. 
These operations must be implemented through a chain of connected qubits, resulting in a high likelihood of additional errors. 
A further shortcoming of Shor’s code is the difficulty to implement operations on logical qubits as such operations simultaneously involve multiple distant physical qubits, therefore requiring complex connectivity. 

These issues are remedied through the so-called surface code~\cite{Martinis2012}, which involves only nearest-neighbor qubit interactions for error correction and, furthermore, streamlines the logical gate operations to some extent. 
In this case the error threshold increases to 0.01~\cite{Fowler2011,Fowler2012,Martinis2012}.
This property makes the surface code and its descendants mainstream approaches to error correction. 
Practical realization of the surface code was expected to be the primary feature of the Bristlecone quantum chip with 72 qubits announced in 2018 by Google\footnote{\url{https://ai.googleblog.com/2018/03/a-preview-of-bristlecone-googles-new.html}}. 
However, no experimental results with this chip have been reported yet. 

The number of physical qubits needed to implement a logical qubit rapidly increases with the error rate and tends to infinity when the threshold is approached. 
For example, for the error rate of $10^{-3}$ in the surface code setting this number is about $3000$~\cite{Martinis2012}. 

In practice, the construction of error correction codes for a specific platform must consider its peculiarities, such as the dominating decoherence channels and temporal characteristics of the interaction with the environment. 
The latter property determines whether decoherence is Markovian or non-Markovian~\cite{Kitaev2006-2}, 
i.e. to which extend the effect of environment on the system at a given moment of time is correlated with past and future moments. 

In the last decade, error correcting codes have been reported in experiments with linear optics~\cite{Franson2005}, 
trapped ions~\cite{Wineland2004,Blatt2011,Blatt2020,Monroe2021,Blatt2021-2}, 
and superconducting circuits~\cite{Schoelkopf2012,Schoelkopf2016,Kelly2021}. 
Extension of the coherence time of a logical qubit using error correction was demonstrated with superconducting qubits~\cite{Schoelkopf2016} in 2016. 
In 2021 the Google team reached a breakthrough \cite{Kelly2021} with arbitrary single-qubit error correction on their Sycamore superconducting device. 
When increasing the number of physical qubits in a logical qubit from 5 to 21, the error rate per round of error correction reduced exponentially by more than 100 times: from $10^{-2}$ to $10^{-4}$. 
The experimental run lasted for 50 such rounds, each of which had a duration of about 1 $\mu$s. 
This performance, however, was achieved with 1D array of qubits, whereas for the 2D surface code only basic operations have realized. 
Proof-of-principle realization of the surface code on 17 qubits has been demonstrated by Zhao {\it {\it et al.}}~\cite{Pan2021-3} using the \textit{Zuchongzhi} 2.1 superconducting quantum processor. 

A drastically different approach to error correction was proposed by Gottesman, Kitaev and Preskill (GKP) in 2001~\cite{Kitaev2001}. 
The idea is to encode qubit in a quantum system of infinite dimension, such as a harmonic oscillator, which can be realized as an optical or superconducting platform. 
This infinite dimensionality provides the redundancy required for error correction, so a role of a logical qubit can be played by a single physical oscillator. 
This approach has been experimentally demonstrated in 2020 with a superconducting quantum circuit~\cite{Devoret2020}. 
The coherence lifetime has been increased by a factor of 2--3. 
The application of the GKP approach has been also proposed in the optical quantum computing setting~\cite{Fukui2018}.

\subsection{Topologically protected quantum information processing}\label{sec:error-topology}

Error correction can be implemented not only at the ``software level'', but also thanks to the physical properties of the platform. 
An example from classical IT is fault-tolerant information storage in a magnetic medium. In a ferromagnetic state, most of the atoms have their magnetic moments oriented in the same direction. 
While an individual atom may flip its direction due to thermal fluctuations, the interaction with neighboring atoms will quickly reverse it to the original orientation. In 2003, Kitaev extended this concept to the quantum domain by theoretically designing a model, in which errors are energetically unfavorable~\cite{Kitaev2003}. 
That is, whenever any physical qubit experiences an error, the energy of the system’s collective state increases above that of any error-free state. 

He envisioned interaction of qubits in a 2D lattice embedded on a torus surface and showed that, under a particular interaction model, the system has exactly four ground states. 
These four ground states are interpreted as the logical states of two qubits. 
These qubits are highly non-local and, therefore, unlikely to transform into each other as a result of a random local fluctuation. 
The existence of these four ground states is a consequence of the the boundaries of a toric surface being “stitched” together; 
for this reason, this approach to error correction is known as the {\it toric code}~\cite{Kitaev2003}. 

We should note that the aforementioned surface codes~\cite{Martinis2012} have been inspired by the idea of toric code and have a lot in common with the latter. More specifically, in the surface code, the boundary conditions are imposed only on two sides of the computational surface, which can be visualized as a surface wrapped into a tube. In the toric code, boundary conditions are additionally introduced to the remaining two sides, wrapping the tube into a torus.

Realizing the toric code experimentally is a challenging problem since it involves realizing a complex model of many-body interactions while implementing or simulating the torus topology. 
Following early experiments limited to small-scale systems~\cite{Weinfurter2009,Pan2012,Ioffe2009,Pan2017}, Google team with collaborators in 2021 realized a 31-qubit ground state of the toric code using the Sycamore superconducting quantum processor, 
suppressing the error probability for the protected state~\cite{Neven2021}. 
A qubit with basic topological properties inspired by toric code has also been demonstrated using neutral atoms~\cite{Lukin2021-2}. 

A shortcoming of both the original toric and the surface code  is that, while these codes provide topological protection to qubit states during storage, their active manipulation in fault-tolerant manner is difficult to realize. 
This problem has been later addressed by a family of methods known as {\it topologically-protected quantum computation}~\cite{Kitaev2003,Stern2013,Field2018}. 
We shall discuss these methods only briefly because of their relative complexity and many challenges arising in their experimental realization. 

The fundamental concept within this paradigm is the {\it anyon}~\cite{Wilczek1982,Field2018} --- a stable vortex with particles circulating around its center. 
These anyons can be compelled to move around each other by applying external fields, which leads to the change of their quantum state. 
For example, in a Bose-Einstein condensate such a vortex can be created by illuminating a certain location with a laser beam with a vortex-like spatial structure~\cite{Ueda2009,Simula2019}. 
In a superconducting plane, one can create a vortex by applying a localized  ``beam'' of the magnetic field in the direction perpendicular to the plane. 

The quantum state of an ensemble of anyons contains information about the history of their movement, more specifically, about the {\it topology} (or ``braiding'') of their trajectories, i.e.~how they ``wove'' around each other. 
A qubit is formed by several such anyons, for example, four in the widely considered Fibonacci model~\cite{Brennen2008,Field2018}.
Certain trajectories correspond to single- and two-qubit gates. 

The quantum information carried by such a system is protected because it is stored not in the local states of individual particles, but the history of how anyons have been moved around each other. 
A local perturbation, as long as it is small enough to keep the vortex intact, will not change that history. 
This is analogous to the physical error correction in classical information storage that we mentioned in the beginning of this section. 

As such, the topological approach is considered to be supremely promising as a path towards fault-tolerant quantum computing. 
It is sometimes even classified as an independent model of quantum computing~\cite{Field2018}, although, as we discussed above, it is also can be seen as the way to implement the gate-based model. 
A variety of physical systems and computational protocols for topologically-protected quantum computation have been theoretically proposed --- albeit so far without successful experimental demonstration. 

One such system is the so-called Majorana zero modes --- anyons, which are expected to emerge in superconducting nanowires~\cite{Kitaev2001-2,DasSarma2008,DasSarma2015}. 
Observation of signatures of Majorana zero modes has been reported~\cite{Kouwenhoven2018} by the Microsoft laboratory in the Netherlands, but subsequently retracted citing ``insufficient scientific rigor'' in the original data analysis \cite{Kouwenhoven2020}. 
Nevertheless, Microsoft is still committed to the topological approach to quantum computing~\cite{Castelvecchi2021}. 

Even though quantum braids are inherently more stable than quantum particles within the standard (non-topologically-protected) systems, they are not a universal panacea against all types of errors~\cite{Field2018}. 
Therefore, large-scale quantum computers based on these principles are not expected in the next 5--10 years. 

\wideboxbegin{box:Transmons}{Superconducting circuits: Transmon qubits.}

A circuit consisting of an inductor and a capacitor (so-called LC-circuit) is a harmonic oscillator, i.e.~a system that is capable of exhibiting simple harmonic motion associated with periodic charging / discharging of the capacitor through the inductance. 
This harmonic motion can be quantized, resulting in energy levels positioned at equal intervals from each other. 
One can select two lowest energy levels to comprise a qubit. 
Transitions between these levels, corresponding to single-qubit operations, can be implemented by applying a microwave field, whose frequency is resonant with the separation between levels, typically on a scale of a few GHz. 

In practice, such a qubit needs to be maintained at very low temperatures 
(on a scale of a hundredth of a kelvin above absolute zero).
This is necessary, first of all, to bring the circuit into the superconducting regime, so the conductors lose electric resistivity and energy dissipation is prevented. 
Second, this will preclude spurious excitation of the qubit due to thermal fluctuations 
(according the Boltzmann distribution. the probability of an excitation is given by $e^{\hbar\omega/kT}$, where $\hbar\omega$ is the transition energy, $k$ is the Boltzmann constant, 
and $T$ is the absolute temperature; in typical superconducting computing circuits this probability is on a scale of $10^{-9}$). 
Superconductivity has an additional important function. It helps to deal with the equidistant distribution of the energy levels in the harmonic oscillator. 
This equidistance is problematic if we use a resonant electromagnetic field to implement a transition between two qubit states. Higher energy levels, with which this field is also resonant, will also get excited,  taking the system out of the  qubit Hilbert space. To prevent this, the inductance is replaced by a so-called Josephson junction --- a superconducting circuit element, whose inductance depends on the current. 
This results in the LC oscillator losing its harmonic nature, eliminating, in turn, the equidistant level structure. 
This circuit combining a capacitor and a Josephson junction is known as the {\it transmon} qubit, and is currently most common in superconducting quantum computing. 
For other types of superconducting qubits we refer the reader to review papers~\cite{Martinis2004,Oliver2019,Oliver2020}. 

Two-qubit operations require coupling between qubits. 
To this end, another LC oscillator is used, whose frequency is also controlled by means of a Josephson junction. 
By tuning it with respect to the qubit resonance, the coupling can be switched on and off on demand. 
\label{Box:Transmons}
\wideboxend

\section{Quantum computing can be based on various physical platforms}\label{sec:platforms}

At the dawn of classical computing it was not known which physical platform is best suitable for its implementation. 
Various platforms have been tried, such as mechanical, electromechanical, vacuum tubes, etc., until the engineering community has converged on semiconductor microstructure as the optimal approach. 
The current situation in quantum computing resembles that of early days of classical computers: a number of platforms are under consideration, 
but the leader is not yet determined. 
In this section, we review existing platforms, their basic principles, advantages and shortcomings as well as achievements recorded to date with each of them. 

\subsection{Solid-state quantum computing}\label{sec:solid-state}

\begin{figure}
\center{\includegraphics[width=0.8\linewidth]{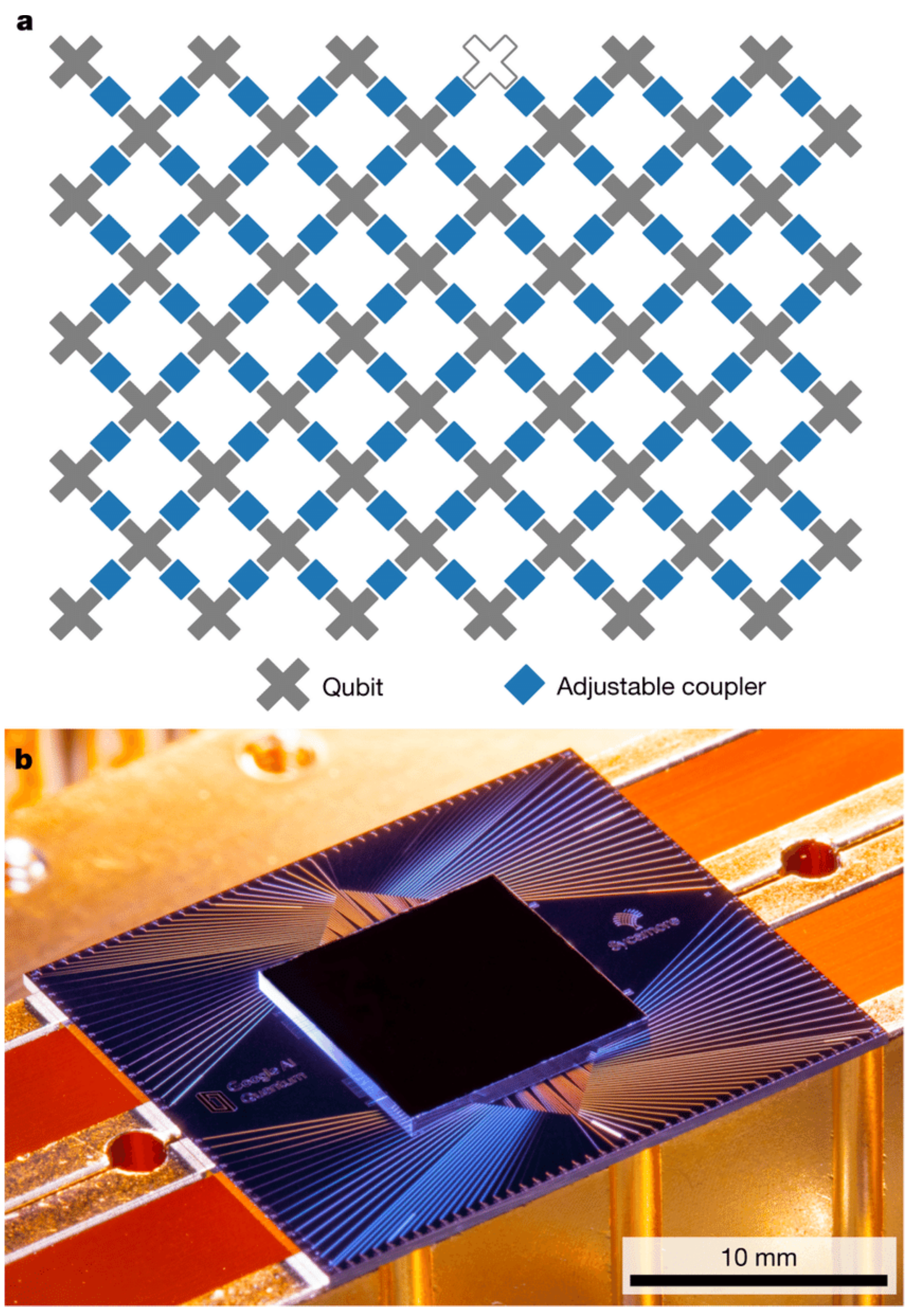}}
\vskip-4mm
\caption{Google Sycamore superconducting quantum processor, figure and caption are reproduced from Ref.~\cite{Martinis2021}: 
a) Layout of the processor, showing a rectangular array of 54 qubits (grey), each connected to its four nearest neighbours with couplers (blue). b) Photograph of the Sycamore chip.}
\label{fig:sycamore}
\end{figure}

Solid-state quantum circuits rely on nanotechnologies to construct qubits as artificial structures connected to each other in a hardwired fashion akin to classical electronic integrated circuits. 
The advantages of this approach include faster gates, possibility of industrial fabrication and broad availability of control equipment~\cite{Oliver2019,Oliver2020}. 
Additionally, solid-state systems can be used for designing topologically-protected qubits~\cite{Ioffe2009,Castelvecchi2021} (see Sec.~\ref{sec:error-topology}). 
These features play a role in attracting major industrial computing companies --- such as Google and IBM --- to this class of platforms. 
However, quantum solid-state devices also suffer from important shortcomings: 
their hardwired nature results in limited connectivity and potential loss of scalability as all qubits and their junctions must be individually controlled by an electrical connection. 
A further challenge is the fabrication of  circuit elements that are both defect-free and sufficiently identical. 
Furthermore, solid-state platforms suffer from occasional decoherence bursts associated high-frequency cosmic ray particles~\cite{Martinis2021-2,Neven2021-2}. 
An additional technological challenge is the requirement to maintain the quantum computing chip at temperatures on a scale of tens of millikelvins, which demands expensive dilution refrigerators. 
The leading solid-state quantum computing platform is superconducting circuits with the runner-up being arrays of semiconductor quantum dots. 

\paragraph{Superconducting circuits.} Currently, the most advanced superconducting quantum computers include
\begin{itemize}
    \item a 53-qubit Sycamore quantum processor presented by Google in 2019~\cite{Martinis2021} (Fig.~\ref{fig:sycamore}), 
which has been used for demonstrating quantum advantage (see Sec.~\ref{sec:sciapplications}), 
\item a family of publicly accessible quantum computers with sizes up to 127 qubits by IBM with the largest quantum volume of 64~\cite{Gambetta2021}, 
\item processors \textit{Zuchongzhi} 2.0~\cite{Pan2021-4} and \textit{Zuchongzhi} 2.1~\cite{Pan2021-5} developed by a group from University of Science and Technology of China and collaborators 
with up to 66 qubits, which have been also used to address the quantum advantage challenge~\cite{Pan2021-4,Pan2021-5}, 
\item a 5,000-qubit quantum annealer developed by D-Wave in 2020~\cite{Amin2021} (see Sec.~\ref{sec:special-purpose} and Box~\ref{Box:DWave}).
\end{itemize}
In August 2020, IBM published a roadmap targeting 1,121 qubits by the end of 2023\footnote{https://www.ibm.com/blogs/research/2020/08/quantum-research-centers/}. 
A similar roadmap by Google (December 2020) promises 1,000,000 qubits with error correction by 2029\footnote{https://quantumai.google/learn/map}.
On the other hand, an independent forecast review~\cite{Sevilla2020} estimates that ``proof-of-concept fault-tolerant computation based on superconductor technology is unlikely ($<5\%$ confidence) to be exhibited before 2026.''

All the above achievements have been reached with transmon qubits (see Box~\ref{Box:Transmons}). 
However, limitations of transmon qubits are currently becoming increasingly manifest~\cite{Neven2021-2,Ustinov2019,Ustinov2020,Martinis2021-2,DiVincenzo2021,McDermott2021}. 
First, their coherence times are relatively short (about $10^4$ gate times), which complicates error correction (see Sec.~\ref{sec:errors}).  The second cause for concern is relatively high qubit frequencies (on a scale of a few GHz), 
which requires expensive control electronics and complicates the wiring between these electronics and the cryostat where the quantum computing chip is located. 
Third, the transmon capacitor must be $\sim100$ femtofarads, which implies sizes on a scale of $\sim 100$ micrometers, making it a challenge to pack more than a few hundred transmons on a few-mm chip. 
This being said, progress has been reported in reducing the qubit area by up to a thousand  with the help of atomically thin heterostructures \cite{Fong2021,Oliver2022}. 
Fourth, nanofabricated transmons are non-identical and require special efforts to tune in resonance with each other. Other types of superconducting qubits are currently being developed that alleviate these issues, such as the fluxonium qubit~\cite{Manucharyan2019,Ustinov2019-2,Oliver2019-3}. 

\paragraph{Semiconductor quantum dots.} A further important solid-state platform is based on semiconductor quantum dots --- nanoscopic conglomerates of a semiconducting material deposited on a substrate. 
In such a quantum dot, single electrons can be isolated and their spins can be used as qubits. 
Individual qubit operations are implemented by applying magnetic fields. 
Qubits are coupled by direct magnetic interaction, which can be controlled by means of another quantum dot, which creates a variable potential barrier between the two electrons~\cite{Loss1998}. 
The typical size of a semiconductor qubit is hundreds of nanometers, which is two--three orders of magnitude smaller than that of superconducting qubits. 
This feature combined with the widely available semiconductor fabrication technology make the system promising in terms of scalability~\cite{Vinet2021,Clarke2021}. 
These promises motivated Intel to switch their quantum computing program to the semiconducting platform in spite of their impressive progress with superconductor technology 
(in 2018 Intel presented a 49-qubit superconducting quantum processor\footnote{https://www.intel.com/content/www/us/en/research/quantum-computing.html}). 

A challenge associated with the semiconducting platform is strong decoherence caused by impurities in a crystal structure. 
Note that defects are also present in superconducting circuits, however, their role is reduced due to larger sizes of the circuit elements~\cite{Kuemmeth2021}. The way to address this challenge is to use exceptionally pure materials for fabrication.

Currently, two primary materials for semiconducting quantum computations are silicon and germanium.
Four-qubit germanium quantum processors with fast high-fidelity gates have been demonstrated in 2021~\cite{Veldhorst2021}. 
Three groups in 2022 independently reported two-qubit gates with silicon quantum dots with fidelities over 99\%, which is sufficient to enable surface-code error correction \cite{Vandersypen2022,Morello2022,Tarucha2022}. 
A further important recent achievement has been to bring the temperature of a semiconduictor quantum logic setup up to $\sim 1$ Kelvin \cite{Veldhorst2020}, 
which is more than an order of magnitude warmer than typical solid-state quantum computing experiments. 
Increasing this temperature further to 4 Kelvin will obviate the need for dilution refrigerators, thereby drastically decreasing the cost and footprint of quantum processors. 

\paragraph{Other approaches.} We conclude this section by mentioning a few promising alternative solid-state platforms, on which no multiqubit processes have been demonstrated yet.
\begin{itemize}
	\item color centers, where qubits are realized by the electronic or nuclear spin of defects in the crystal lattice caused by impurities, for example, donors in silicon~\cite{Kane1998}, 
	vacancies near a nitrogen atom in diamond \cite{Lukin2007}, 
	or rare-earth ions in 
	crystals~\cite{kinos2021roadmap}\footnote{These systems are also promising for making quantum repeaters~\cite{Gisin2011} --- devices for increasing the quantum communication distances.}; 
	\item fullerene molecules with the qubits based on nitrogen or phosphorus atoms encapsulated therein~\cite{Harneit2002};
	\item qubits represented by spins of electrons positioned at the surface of a liquid helium film deposited on an insulator substrate~\cite{Dahm2001,Dahm2003};
	\item bound states of electrons localized in an array of nanowires~\cite{Bertoni2000};
	\item spins of itinerant electrons within metallic-like carbon nanospheres~\cite{Nafradi2016};
	\item point-defect spin qubits in engineered quantum wells~\cite{Ivady2019};
	\item Andreev spin qubit combining properties of superconducting circuits and semiconductor setting~\cite{Devoret2021};
	\item qubits based on split-ring polariton (light-matter superfluid) condensates~\cite{Kavokin2021};
	\item electron spin qubits in graphene quantum dots~\cite{Loss2007,Thomas2006,Tahan2019}, van der Waals heterostructures~\cite{Oliver2019-2}, 
	and quantum simulators for the Hubbard models based on twisted heterostructures~\cite{Basov2021}, such as twisted bilayer graphene~\cite{Mak2020}.
\end{itemize}

\begin{figure}[ht]
\center{\includegraphics[width=1\linewidth]{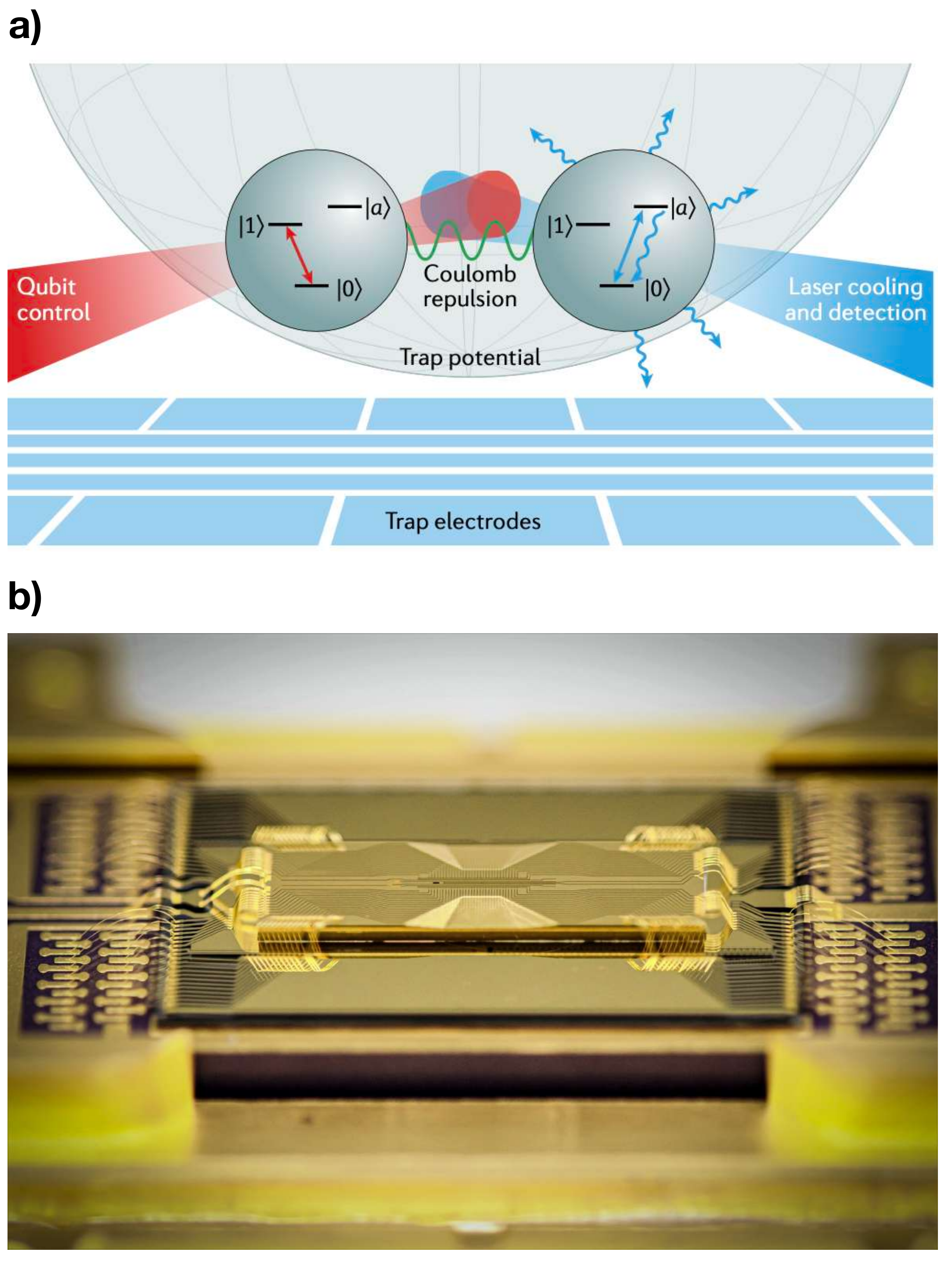}}
\vskip-3mm
\caption{Trapped ions quantum computing. a) The ions are held in
an electromagnetic trap. Lasers or microwaves are used to control the internal states of qubits, $|0\rangle$ and $|1\rangle$. The internal control and the Coulomb repulsion between ions combine to form conditional logic gates. Readout is performed by measuring laser- induced ion fluorescence using an auxiliary state $|a\rangle$. The laser-induced fluorescence is also used to cool the ions in preparation for quantum logic (figure and caption are reproduced from Brown {\it et al.}~\cite{Brown2021}.
b) Photograph of the IonQ's ion trap (reproduced from ionq.com).}
\label{fig:iontrap}
\end{figure}

\subsection{Atoms, ions, and molecules for quantum computing}\label{sec:ionsatoms}

Elementary units of solid-state quantum systems sometimes are referred to as artificial atoms because of their compact nature, reduced interaction with the environment, and well-defined, narrow-band energy spectrum. 
All these properties are essential in quantum computation. However, as discussed above, these artificial atoms are hard to make identical. 
Hence an alternative approach is to use actual atoms and molecules, which are identical by their nature, as elementary quantum units. 
The price to pay is the challenge associated with controlling and engineering interactions between them. 
This is achieved by means of {\it traps} --- arrangements of external force fields keeping the particles steady during the experiment (see Boxes~\ref{Box:TrappedIons}, \ref{Box:Rydberg}, and \ref{Box:AtomsLattice}). 
Special efforts need to be applied to prevent mechanical oscillations of particles within these trap --- that is, the particles must be {\it cooled} to temperatures on a microkelvin scale or even lower. 
Generally, this cooling is achieved using lasers, electric, and magnetic fields. 
This constitutes a significant cost advantage in contrast to the solid-state platform, in which cooling requires a dilution refrigerator. 
Cooling of atoms, ions, and molecules is a broad field in its own right~\cite{Letokhov1995,Schreck2021}, but is beyond the scope of our review. 

\wideboxbegin{box:TrappedIons}{Trapped ions.}

Trapping of ions relies on their charged nature and utilizes electric or magnetic fields oscillating at radio frequencies. 
Trapping force produces a single potential well, which pushes the ions towards its center. However, the ions keep away from each other due to their electrostatic (Coulomb) repulsion. 
A typical ion trap in quantum computing is a 1D array of ions separated by a few micrometer distance, so they can be individually resolved by optical means and addressed by lasers. 

Each ion carries a single qubit, typically encoded in the state of its electrons. 
The energy separation between qubit states can be as low as few GHz (hyperfine states) or as high as hundreds of THz (electronic states). 
Dependent on this magnitude, qubits are classified into radiofrequency or optical qubits. 
The advantage of radiofrequency qubits is that they are more robust to decoherence, however they require more than one laser for single-qubit gates.  

A critical feature that enables two-qubit gates is the presence of synchronized mechanical oscillations of ions within the trap. 
These oscillations form a part of the collective quantum state of the ion ensemble. 
They can be used to communicate quantum information between ions and entangle them. 
Specifically, by applying lasers to the control ion in a certain way, an oscillation can be exited dependent on the state of this qubit. 
When the target qubit is addressed by a subsequent laser pulse, the presence of this oscillation may determine whether  this qubit will change its state, thereby completing a \CNOT gate~\cite{Wineland1995}. 
This is the basis of the original idea of ion-based quantum computing proposed by Cirac and Zoller in 1995~\cite{CiracZoller1995}. 
While the specific procedure of multiqubit gates has varied over past few decades~\cite{Blatt2003-2,Molmer-Sorensen1999,Molmer-Sorensen1999-2,Molmer-Sorensen2000}, the collective oscillations have always remained their primary concept. 
This approach enables gates between any two ions arbitrary chosen within the trap. 

\label{Box:TrappedIons}
\wideboxend

\begin{figure*}[ht]
\center{\includegraphics[width=1\linewidth]{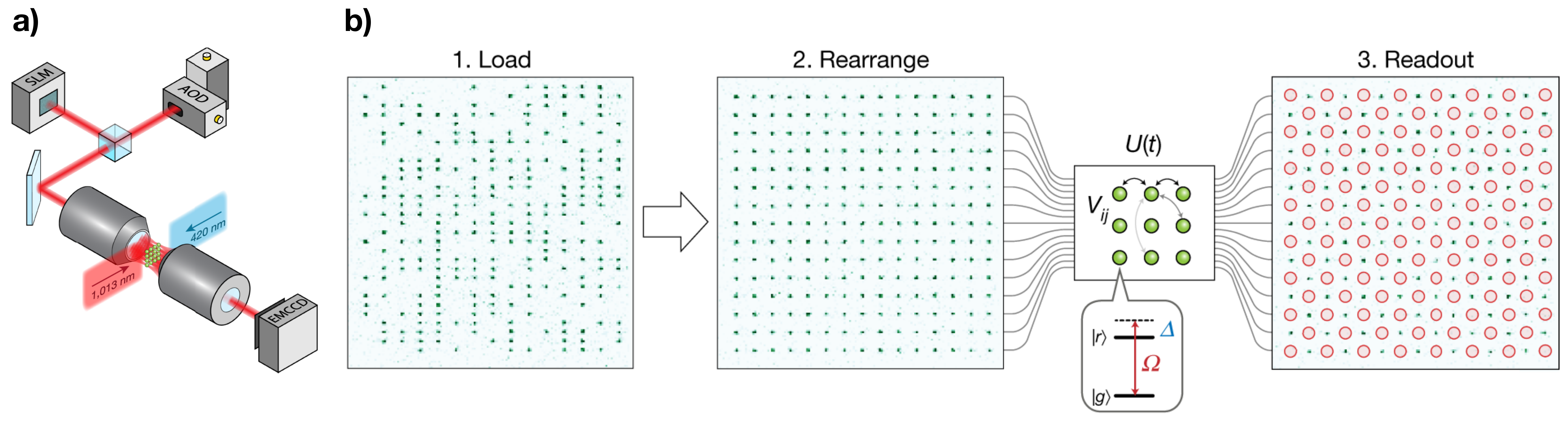}}
\vskip-3mm
\caption{Rydberg-atom quantum computing (reproduced from Ebadi {\it et al.}~\cite{Lukin2021}). a) Optical setup. The 2D array of atoms is placed between two powerful microscope objectives that enable their individial addressing and imaging. The trapping fields are created by the spatial light modulator (SLM) and the tweezers are implemented by means of an acousto-optical deflector (AOD). 
b) Initially loaded atoms are rearranged into defect-free patterns by a set of moving tweezers. Their states can be changed in the programmable manner through Rydberg blockade. The ground $\ket g$ and Rydberg $\ket r$ states of the atom constitute a qubit.}
\label{fig:rydberg}
\end{figure*}
\wideboxbegin{box:Rydberg}{Rydberg atoms.}

An atom (typically, of group 1 in the periodic table) is in the Rydberg state when one of its electrons is exited to a very high energy orbit (principal quantum number 50--100). 
Such orbits are characterized by large radii (fractions of micrometres) and strong interaction with neighbouring atoms (the interaction strength scales as the 11th power of the principal quantum number). 
One atom exited by a laser field into the Rydberg state may prevent its neighbours from achieving the same state, a phenomenon known as the {\it Rydberg blockade}. 
In other words, the behaviour of an atom can depend on the state of another atom. 

On the one hand, this can be interpreted as the CNOT gate~\cite{Deutsch1999,Lukin2000,Ahn2000} for two qubits (each of which is encoded in the ground and Rydberg states of an atom) and hence enabling digital quantum computation. 
On the other hand, this system of atoms can be seen as a graph, in which the atoms are nodes (whose value can be 0 or 1 dependent on whether the atom is in the Rydberg state) and edges connect neighbouring atoms. 
The Rydberg blockade prevents the connected nodes from simultaneously taking on the value of 1. 
At the same time, by choosing the detuning of exiting lasers, one can make the state with the largest number of exited Rydberg atoms to be the most energetically favourable. 
This sets the natural condition for solving the maximum independent set problem~\cite{Lukin2018,Lukin2018-2} as well as simulating the Ising model in condensed matter physics (see Box~\ref{Box:QUBO}).  

A primary control tool in the Rydberg atom technology is a strong, tightly-focused laser beam known as an optical tweezer~\cite{Browaeys2016,Lukin2016}. 
Laser light provides a force attracting the atoms towards the beam center, thereby creating a potential well, in which the atoms can be trapped.  
To make a Rydberg atom quantum computing device, a cloud of atoms (typically, rubidium or cesium) is cooled to submillikelvin temperatures. 
Then some of the atoms are individually trapped in optical tweezers; the remainder is released from the trap. 
These tweezers are used to arrange atoms into spatially order arrays, in which Rydberg gates are possible between neighbouring atoms (Fig.~\ref{fig:rydberg}). 

\label{Box:Rydberg}
\wideboxend

The leading platforms within this family are cooled trapped ions and neutral atoms. 
The latter are further divided into two settings: Rydberg atoms in optical tweezers and ultracold atoms in optical lattices. 
We explain the details and differences of these two platforms in Box~\ref{Box:Rydberg} and Box~\ref{Box:AtomsLattice}. 
While both ions and neutral atoms have the potential for various models of quantum computing, the ion platform is presently considered a mature engine for gate-based model, whereas neutral atoms are mostly used for quantum simulation. 
We also briefly discuss molecular platforms, which are considered promising, but experimentally less advanced.

\paragraph{Trapped ions.} Historically, the ion platform was one of the first in which the two-qubit gates~\cite{Wineland1995,Blatt2003-2} and basic quantum algorithms~\cite{Blatt2003} have been demonstrated. 
The most recent advancements include a 53-ion quantum simulator for the Ising models~\cite{Monroe2017} in 2017, fully-controlled quantum-state engineering in a 20-ion system~\cite{Blatt2018-2} in 2018, 
and demonstration of variational quantum algorithms for chemistry~\cite{Blatt2018,Monroe2020-2} and combinatorial optimization~\cite{Monroe2020} in 2018--2020. 

Progress towards error correction (see Sec.~\ref{sec:errors}) has also been achieved. 
Erhard {\it {\it et al.}}~\cite{Blatt2021-2} demonstrated logical qubits in the framework of the surface code, as well as entanglement and basic operations between them in 2020. 
Egan {\it {\it et al.}}~\cite{Monroe2021}. implemented single-qubit error correction using the Bacon-Shor code~\cite{Bacon2006} (an extension of the Shor code discussed above). 
The coherence time of the qubit has been increased by the factor of 2.5. 
If errors are detected, but not corrected and error-free events are instead post-selected, then the qubit lifetime is increased by more than 10 times. 

Most of existing experiments have been performed in single 1D traps --- such that the ions form a single straight line. 
While attempts to implement 2D traps have been made (with 4 ions in Holz {\it {\it et al.}}~\cite{Blatt2020-2}), this is technologically difficult. 
Not less challenging is scaling 1D traps beyond a few dozen qubits. 
There are a number of ways to address the scaling issue. 
The mainstream idea is to contain ions in multiple traps with the possibility to join or divide traps on demand~\cite{Foss-Feig2021} or move individual ions within the system~\cite{Foss-Feig2021} to enable interactions between arbitrary pair two ions. 
Another approach is to communicate quantum information between ions in different traps via an optical interface~\cite{Lanyon2021}. 

A modern ion trap is a complex microstructure complete with trap electrodes, dielectric insulators, optical waveguides, modulators and detectors integrated together (Fig.~\ref{fig:iontrap}). 
A key factor in fabricating these traps is the materials used. In particular, it is important that the surfaces do not produce significant electric field noise, which could cause decoherence of the ionic states~\cite{Brown2021}.

Recently, the ion platform has been spun off industrially with three ventures emerging as leaders: 
IonQ (USA), Quantinuum (previously Honeywell, USA), and AQT (Austria). 
These companies made progress in different aspects of ion trap quantum computing. 
In particular, AQT demonstrated a compact unit fitting within a standard 19-inch rack capable of operating with 24 qubits~\cite{Blatt2021-3}. 
Quantinuum/Honeywell developed a method for transporting and swapping ions within a trap for quantum gates with all-to-all connectivity~\cite{Foss-Feig2021} and the processor with the highest quantum volume. 
IonQ demonstrated various applications of their devices ranging from chemistry~\cite{Monroe2020-2} to machine learning~\cite{Demler2021,Kerenidis2020}. 

The main mechanism behind the formation of an ordered array of ions in a trap is their electrostatic repulsion (see Box~\ref{Box:TrappedIons}). 
This repulsion leads to the emergence of mechanical modes, which enable full connectivity of quantum units. 
On the other hand, it gives rise to scalability issues as discussed above. 
The situation with neutral atom platforms is opposite. 
They are trapped in optical fields, which enables better scalability, however their interactions are relatively short-ranged limiting the connectivity and the quality of pairwise operations. 

\paragraph{Neutral atoms.} The two leading neutral atom platforms are Rydberg atoms\footnote{The term ``Rydberg atom" is jargon; a more accurate term would be ``Rydberg state of an atom", see Box \ref{Box:Rydberg}.} in optical tweezers (Box~\ref{Box:Rydberg}) and (ground-state) atoms in optical lattices (Box~\ref{Box:AtomsLattice}). 
The key difference between them is the interaction mechanism: 
in the former case, the Rydberg blockade enables digital gates between atomic qubits, whereas in the latter case, 
the quantum states are carried by the atomic motional degrees of freedom with the interaction leading to multiparticle entanglement, 
but not of digital nature. 
A further difference is that optical lattice setups require cooling the atoms to extremely low temperatures (tens of nanokelvins), 
which complicates the setup and requires a long preparation stage (tens of seconds) before the ``payload'' quantum process (a fraction of a second) can be launched. 
In the optical tweezer setting, in contrast, the atomic temperatures are on a scale of hundreds of millikelvins, which is easier to obtain experimentally, and needs much shorter preparation (about 100 milliseconds). 
Moreover, optical tweezers allow one  to craft ordered arrays of atoms with various geometric configurations and interaction schemes~\cite{Browaeys2016,Lukin2016}. 
Therefore, the Rydberg platform is suitable for both analog quantum simulation and digital quantum computing, whereas optical lattices are primary seen as a platform for analog simulation.

\begin{figure}[b]
\center{\includegraphics[width=0.7\linewidth]{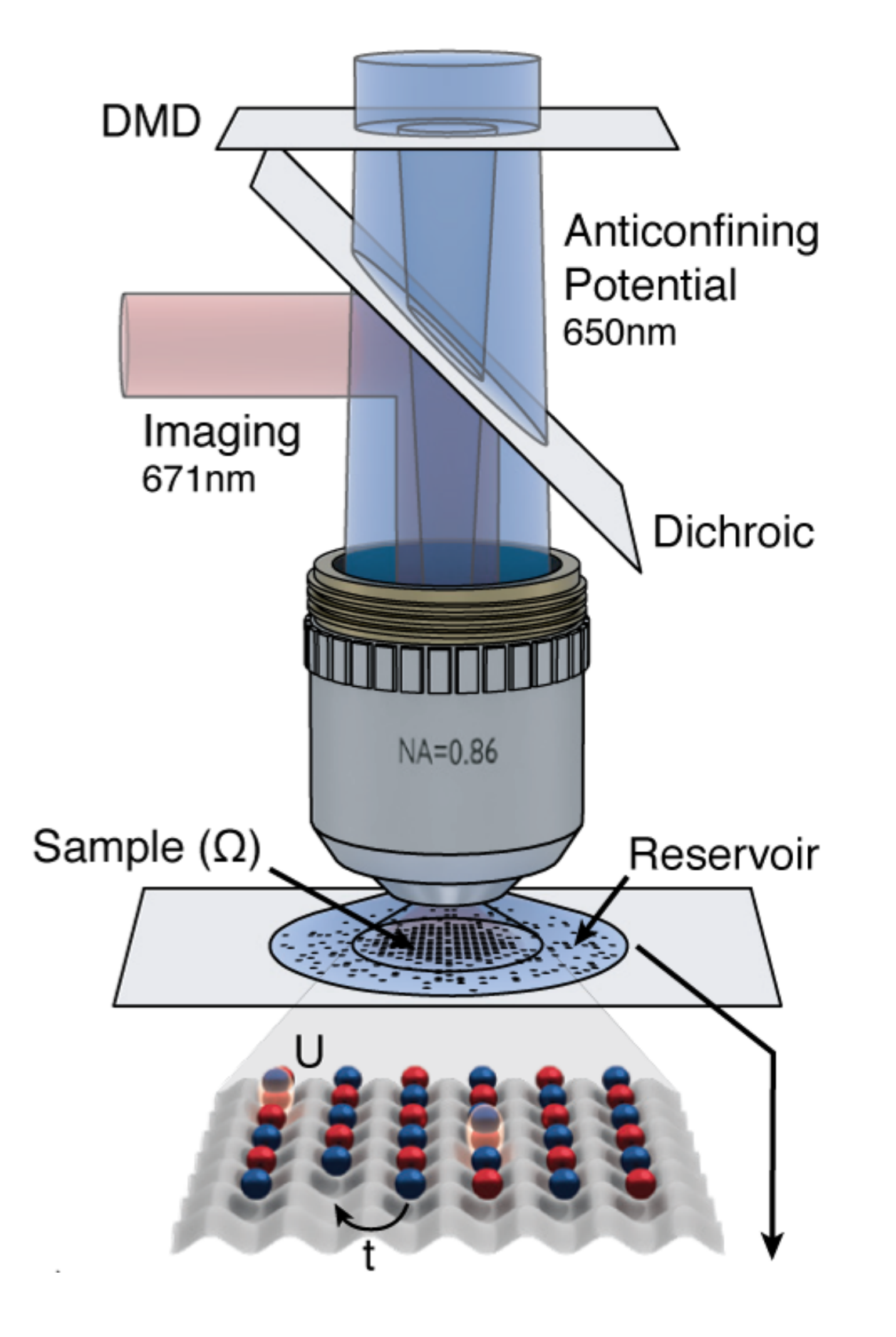}}
\vskip-3mm
\caption{Quantum simulator based on atoms in optical lattices (reproduced from Mazurenko {\it et al.}~\cite{Demler2017}): Lithium atoms are trapped in a two-dimensional square optical lattice, quantum gas microscope is used for detecting the state of the systems with the single-site resolution.}
\label{fig:atom-sim}
\end{figure}

\wideboxbegin{box:AtomsLattice}{Ultracold atoms in optical lattices.}

Two interfering counterpropagating laser beams form a standing wave --- an array of alternating zones of high and low light intensity. Atoms can be trapped in the high-intensity zones (antinodes of the standing wave) via the same mechanism as optical tweezers (see Box~\ref{Box:Rydberg}). As a result, we obtain a periodic array of traps, in which atoms behave akin to electrons in a crystal lattice. 
This system, known as the \emph{optical lattice}, therefore constitutes a simulator for lattice many-body models in physics whose applications range from condensed-matter~\cite{Bloch2008,Bloch2012,Zoller2005} to high-energy physics~\cite{Cirac2015}. 
The quantum degree of freedom in such a simulator can be the motional state of the atom or their spins. 
In contrast to the Rydberg platform, the atoms' electrons rarely leave their ground states. 

A great advantage of this simulator is the tunability of its main parameters, such as dimensionality (1D, 2D or 3D), lattice geometry (rectangular, honeycomb, etc), 
interaction range, directionality (isotropic vs anisotropic), sign (attraction vs repulsion) and strength. 
A further important degree of freedom is the choice of atomic species for the simulator, in particular, the fermionic or bosonic nature of the atoms. 
In many cases, individual lattice sites can be resolved through a regular microscopic objective, which enables one to observe individual atoms and estimate their quantum states (Fig.~\ref{fig:atom-sim}). 
This {\it quantum gas microscope} is a great asset in experimental studies~\cite{Greiner2009,Bloch2010}. 

\label{Box:AtomsLattice}
\wideboxend

The optical lattice platform has existed for about two decades~\cite{Bloch2008,Bloch2012}. 
Starting with the seminal experiments on simulating the phase transition in the Hubbard model in 2002~\cite{Bloch2002}, 
it developed into highly controllable simulators for various condensed matter phenomena~\cite{Bloch2008,Bloch2012} including high-temperature superconductivity~\cite{Jaksch2006}. 

The Rydberg setting has emerged over the last five years. 
Arrays of atoms of different spatial dimensionalities have been demonstrated~\cite{Browaeys2016-2,Lukin2017,Lukin2019,Lukin2019-2,Browaeys2020,Browaeys2021,Lukin2021}.
They have been used to simulate phase transitions in quantum many-body systems with the number of atoms growing from 51 in 2017~\cite{Lukin2017} to 100~\cite{Lukin2021-8}, 196~\cite{Browaeys2021}, and 256~\cite{Lukin2021} in 2020-2021. 
An important result of 2021 is the demonstration of spin liquid phase~\cite{Lukin2021-2}, a quantum phase of matter predicted in 1970s~\cite{Anderson1973}, 
but previously not conclusively observed experimentally. 
This state is characterized by long-range spin entanglement combined with disorder, which is maintained even at very low temperatures. 
These features make spin liquids interesting for topological error correction (see Sec.~\ref{sec:error-topology}).
A further important application of the Rydberg atom platform is combinatorial optimization, specifically, the maximum independent set problem (Sec.~\ref{sec:application-optimization}).

Aside from the analog regime, digital two-qubit gates with Rydberg atom arrays have been reported achieving fidelities in excess of 0.97~\cite{Lukin2018-3}. 
The factors currently preventing even higher fidelities include the Doppler effect, spontaneous emission and laser phase noise \cite{Browaeys2018-2}.

Although Rydberg blockade gates are only possible between nearest neighbors, the atoms can be transported within the array without loss of entanglement thereby enabling a gate sequence for arbitrary pairs of atoms. 
The potential of this approach has been demonstrated in 2021 by Bluvstein {\it {\it et al.}}~\cite{Lukin2021-9}, who used it to implement a variety of error correction schemes involving up to 24 physical qubits.

The Rydberg platform is being commercialized by QuEra (USA) \cite{Lukin2021-7}, Pasqal (France) \cite{Browaeys2020-2}, and ColdQuanta (USA) \cite{Saffman2022}. 

In spite of its great promise, the limitations of the Rydberg platform are associated with atomic states being sensitive to external electric fields and a relatively small blockade radius. 
A way to overcome these shortcomings could be to couple atoms via an artificial optical interface~\cite{Lukin2013}, 
e.g., by placing them in the vicinity of a waveguide to let light communicate quantum information between qubits. 
In this setting, the atoms can be still manipulated individually by means of optical tweezers, but the Rydberg states are no longer necessary. 
Instead, the qubit can be encoded in sublevels of the atomic ground states, which have much longer coherence lifetime. 
A basic experiment in this setup has been realized in 2021~\cite{Lukin2021-7}. 
This setting could become a next step in atomic quantum computation.  

\paragraph{Cold molecules.} A promising platform for quantum computation is based on ultracold molecules. 
The idea is to utilize the dependence of the molecule's dipole moment on its state~\cite{Ni2018,Ni2021,Ni2021-2,Cornish2021}.
As a result, the internal state of a molecule can strongly affect the interaction of neighboring molecules, enabling two-qubit gates. 
The challenge is associated with a rich space of molecular quantum states making them difficult to control. 
To our knowledge, two-qubit gates with individually controlled molecules have not yet been demonstrated. 
An important recent achievement is handling cold molecules using optical tweezers~\cite{Ni2021,Ni2021-2,Ni2021-3}.

\subsection{Optical quantum computing}\label{sec:optical}

The final physical platform with a large footprint in the current quantum computing landscape is based on light. 
Quantum information is encoded in light waves propagating in certain channels (modes), which can be implemented either in free space (on an optical table) or via waveguides in integrated chips. 
The optical platform is special because of the transient nature of optical waves, meaning that the computation has to proceed ``on the fly''. 
Furthermore, light waves under normal conditions do not interact strongly either with each other or with the environment. 
The latter is the reason that the decoherence in the optical platform is strongly reduced, making it appealing for quantum computing. 
On the other hand, the lack of mutual interaction between optical fields makes it a challenge to design two-qubit quantum computational gates. 

In principle, conditions, in which light waves influence each other, do exist. 
They are the subject of a vast field of physics known as {\it nonlinear optics}~\cite{Lukin2014}. 
However, nonlinear optical phenomena typically emerge at light energies on the scale of at least billions of photons. 
It is much more challenging to achieve sizable nonlinear effects at the single-photon level as required for quantum computation. 
Several avenues towards this end are being pursued, mostly based on nonlinear properties of individual atoms, atom-like objects or their ensembles. 
One approach is the aforementioned phenomenon of Rydberg blockade: 
when an atom absorbs a photon and transitions to a Rydberg state, it will affect neighboring atoms, thereby preventing absorption of further photons~\cite{Lukin2013-2}. 
This ``single-photon transistor'' can serve as a quantum gate. 
Other possibilities to enhance optical nonlinearities include tight focusing of an optical beam onto a single atom~\cite{Tey2008}, 
placing an atom in an optical resonator~\cite{Lukin2013}, or using novel nonlinear materials, such as graphene~\cite{Lukin2013-3}. 
All these approaches are, however, difficult to implement and scale up. 
Furthermore, they introduce losses associated with the light-matter interface.

An important alternative is to use much more common linear optical phenomena, such as refraction, reflection, and interference, combined with {\it conditional measurements}. 
After the computational modes undergo linear optical transformations, which entangles them, one performs measurements on some of the modes. The nonlocal effect of this partial measurement (see Box \ref{Box:BasicQ}) on the entangled multimode state causes the state of the remaining modes to change in a way that is similar to a result of nonlinear interaction occurred among them. 
This is known as {\it linear optical quantum computing}. 
It might appear that the probabilistic nature of linear-optical gates, combined with the transient nature of light, 
would preclude viable quantum computing as any undesired measurement results or loss of photons would be fatal for the entire computational process. 
However, in a breakthrough publication of 2001, Knill, Laflamme, and Millburn (KLM) showed this not to be the case~\cite{Knill2001}. 
They proposed to implement probabilistic two-qubit gates on “auxiliary” modes separately from the main quantum computational stream. 
This gate can be tried many times without disrupting that stream. In the event of success, the auxiliary modes will be entangled in a certain manner. 
One can then apply the quantum teleportation protocol to teleport this entangled state onto the computational stream. 
The quantum state of the modes emerging after teleportation will be equivalent to that expected as the output of a two-qubit gate.

While KLM provided a head start to optical quantum computation, the practical implementation of the specific scheme of Ref.~\cite{Knill2001} is prohibitive due to tremendous recourse overhead. 
Multiple ideas for its improvement have been proposed. 
A currently popular paradigm of discrete-variable quantum optics was introduced by Kieling {\it et al.}~\cite{Kieling2007}, 
further improved by Gimeno-Segovia {\it et al.}~\cite{Rudolph2015}, and consists in creating a cluster state (see Sec.~\ref{sec:one-way}) 
by joint measurements on a large number of primitives, each of which is a three-photon entangled state. 
Each of these measurements has a limited probability of success and hence the resulting cluster state will contain ``holes''. 
However, the remaining connectivity is sufficient for meaningful one-way quantum computing. 

Generating the entangled photon primitives is a challenge. 
One approach is to produce these states from single photons by means of an additional preliminary layer of probabilistic linear optical circuitry. 
A scheme to that effect has been proposed by Varnava {\it et al.}~\cite{Rudolph2008} with the success probability of 1/32, albeit requiring feedforward operations. 
Gubarev {\it et al.}~\cite{Straupe2020} proposed a way to eliminate feedforward, however with a lower success probability of 1/54. 
Both these schemes require multiple on-demand single-photon sources. 
The currently leading method for this task is based on quantum dots (previously briefly discussed in Sec.~\ref{sec:solid-state}), in which transitions of single electrons can be used to generate photons~\cite{White2017}. 
However, these sources are imperfect, with the best achieved efficiencies on a scale of 50\%~\cite{Tomm2021}. 
An alternative method is to prepare entangled photon triplets directly from quantum dots~\cite{Rudolph2009} as demonstrated experimentally by Schwartz {\it et al.}~\cite{Schwartz2016} in 2016, albeit, again, with imperfect efficiency and fidelity.

The optical platform was one of the first to be explored in the context of quantum information processing in the late 1990s because its tools have been readily available and relatively inexpensive. 
The initial experimental work was done in free space, with the elements realizing preparation, manipulation, and detection of states of light positioned on an optical table. 
This research, conducted in the discrete, continuous, and hybrid settings, produced many results that are of value for the entire field of quantum science. 
However, the free-space implementation is not practical for large-scale quantum computing due to the lack of scalability, complexity of industrial production, and need for regular alignment. 
An approach with the potential to overcome these shortcomings is based on integrated optics, 
where quantum light is carried in waveguides on the surface of a chip. 

Quantum optical computation faces many challenges and it is not clear presently how to overcome some of them. 
Complications emerge almost at all stages of quantum processing. 
Specifically, the preparation efficiencies of the primitive states (single photons for the discrete-variable and squeezed states for the continuous-variable settings) are significantly below the fault-tolerant quantum computation requirement. 
An additional challenge for the discrete-variable quantum computing is the preparation of the entangled-photon resource for cluster states. 
A subsequent difficulty is associated with losses that are carried by any optical element either in integrated or free-space settings. 
While losses below some threshold can be compensated by means of error correction, the current technology 
does not reach this threshold. Furthermore, cluster state quantum computing requires high-speed electronic processing of measurement results for feedforward onto the circuit elements. 
Characteristic times of such processing are on the scale of nanoseconds. Given that the speed of light is a foot per nanosecond, direct feedforward, especially in an integrated chip setting, appears problematic. 
Facilities like delay lines or quantum optical memory~\cite{Lvovsky2009} may be required, which themselves pose technological challenges. 

Nevertheless, many researchers are ``optimistic about the silicon-photonic route to quantum computing" \cite{Rudolph2017} and this approach is being commercialized by several startups. 
PsiQuantum (USA) is developing a discrete-variable device in an integrated chip setting. 
They announced an ambitious goal of developing a 1 million qubit quantum computer by 2025\footnote{https://psiquantum.com/news/psiquantum-and-globalfoundries-to-build-the-worlds-first-full-scale-quantum-computer}. PsiQuantum is valued at \$3.1 Billion\footnote{https://www.wsj.com/articles/psiquantum-raises-450-million-to-build-its-quantum-computer-11627387321}. 
Their nanophotonic chip technology is expected to be useful not only in quantum, but also in optical computing, particularly for optical neural networks. 

Encoding quantum information in states of photons as particles of light --- such as in the research described above --- is referred to as \emph{discrete-variable} quantum optical computation. 
This however constitutes only one side of light as a dual wave-particle entity. Information can also be carried, and used for quantum computing, by the wave properties of light, such as its amplitude and phase. This is known as \emph{continuous-variable} encoding. 
The main primitive of continuous-variable quantum optical computing is the so-called squeezed state of light \cite{Lvovsky2015}, in which the quantum noise of the light wave amplitude at certain phases is reduced below the standard noise level, which corresponds to light with no nonclassical properties. 
In contrast to the single photon, squeezed light has been possible to reliably produce on-demand for many years. Interfering in linear optical arrangements, squeezed states can produce complex clusters. 
1D~\cite{Furusawa2016,Pfister2014,Fabre2017} and 2D~\cite{Furusawa2019,Andersen2019} clusters as well as measurement-induced feedforward processing~\cite{Andersen2021} have been experimentally demonstrated. 
However, continuous-variables cluster schemes require a certain level of squeezing to enable fault-tolerant quantum computing. 
Current theoretical research sets the threshold in the range of 10--17 decibels (noise power reduction by a factor of 10--50)~\cite{Fukui2018,Menicucci2019,Andersen2021-2}, which compares favorably with the current record of 15 decibels~\cite{Schnabel2016} (2016). 
However, the level of squeezing in ongoing experiments with cluster states is significantly lower: on the scale of 3--5 decibels (noise power reduction factor of 2--3)~\cite{Furusawa2019,Andersen2019}. 
Furthermore, the mentioned theoretical models disregard losses, which are inevitable in optical arrangements. 

Outside the digital model, impressive progress in continuous-variable quantum optical computing was achieved by a Canadian startup Xanadu. 
They presented a chip-based Gaussian boson sampling~\cite{Schuld2021} (discussed in detail in Sec~\ref{sec:special-purpose}) 
circuit with 8 modes with applications in quantum chemistry~\cite{Aspuru-Guzik2015} 
and  mathematics~\cite{Schuld2020}. 
This device, including all the necessary classical infrastructure was remarkably compact, fitting in a standard 19-inch rack.

\section{Quantum computers have diverse applications}\label{sec:applications}

It is expected that quantum computers can be useful in various applications ranging from scientific research to cybersecurity, financial optimization, and drug discovery. 
Here we present an extended but not complete list. Before proceeding, we would like to reiterate a disclaimer. 
Existing publicity hype may mislead the reader into thinking that a computer capable of achieving quantum advantage in application to real-world problems has already been developed. In fact, the state of the art can be described as follows.
\begin{itemize}
    \item First, experiments have been performed to demonstrate quantum advantage on computational problems that are of no practical value, but are widely accepted by the community as difficult to solve.
    \item Second, new scientific insights are obtained by applying quantum simulators to model various physical systems, particularly condensed-matter. While these insights are specifically associated with quantum machines, there are no rigorous proofs of quantum advantage: the classical computational complexity of these  problems has not been thoroughly investigated. 
    \item Finally, there are many efforts to apply quantum computers to various classical problems of practical relevance. In these experiments, the problem sizes are typically far less than what can be solved with a regular classical machine. 
Existing reports are limited to proof-of-concept demonstrations rather than ready-to-use technology. That being said, these demonstrations are of importance for exploring the range of applications, in which the practical quantum advantage can eventually be achieved. 
\end{itemize}
 
We anticipate the progress timeline of quantum computing applications to resemble that of lasers in the second half of the 20th century. 
First applications of the laser immediately after its invention have been in scientific research: 
first in studies of the laser itself, then as a tool for adjacent fields of science, such as atomic, molecular, and optical physics, and finally as a source of light with special properties that are useful a broad range of sciences including chemistry and biology. 
This latter stage coincided with the emergence of narrowly specialized applications, such as holography, spectroscopy, communications, and material processing. 
Finally, lasers developed into a ubiquitous instrument present in virtually every industry and every household, e.g., as an element of a printer or a music player. 
In our discussion of quantum computing applications, we will follow the same general scheme, i.e., separately discuss scientific, specialized, and potentially broad use cases. 

\subsection{Basic science applications}\label{sec:sciapplications}

Applications of quantum computing in scientific research~\cite{Alexeev2021} can in turn be classified into two main categories: 
(1) calculating the properties and simulating complex physical systems and (2) exploring capabilities of quantum computers to achieve computational advantages. 

\paragraph{Simulating physics.} The task of modeling quantum many-body systems has been a subject of close scientific scrutiny for many decades and has reached major milestones with classical computers. 
However, this modeling requires continuously increasing computational power, driving rampant employment of supercomputers in quantum physics research. 
As discussed above, quantum computation offers a qualitatively new capability for quantum modeling. 

It is important to understand that a quantum computer built on a particular platform can study physical phenomena beyond this platform. 
We have already seen several examples to this effect: 
a trapped ion quantum computer was used to study high-energy physics models~\cite{Banuls2020,Blatt2016-2,Blatt2021} 
and a neutral atom machine was employed for simulating condensed matter physics~\cite{Bloch2008,Bloch2012,Browaeys2016-2,Lukin2017}. 
Another example that has not yet been discussed is the simulation of the physical mechanism giving rise to the Higgs boson using atoms in an optical lattice~\cite{Bloch2012-2}. 
Remarkably, this observation has been made at the same time as the Higgs boson has been discovered at the Large Hadron Collider.
As quantum computing develops further capabilities emerge, including 
modelling effects in nuclear physics~\cite{Muschik2021},
simulating dark matter in the Universe~\cite{Mocz2021}, 
studying general relativity and black holes~\cite{Borsten2012,Ekert2020}.

These results have been of primary significance for physics, as they enabled simulating exciting physical phenomena that had been predicted theoretically, 
but were extremely hard to produce in their ``native'' high-energy or condensed-matter physics settings. 
Moreover, some of these experiments, such as those of Refs.~\cite{Monroe2017,Lukin2017,Browaeys2021,Lukin2021} have been at the limit of classical computation accessibility. 
Most recently, a claim to have reached beyond this limit in a physically relevant context has been made~\cite{Amin2021}. 
We shall discuss this result at the end of this subsection. 

\paragraph{Advantage demonstrations.} In 2019, this was demonstrated by the Google team for the random circuit simulation problem~\cite{Martinis2019}. 
This problem consists in predicting the statistics of measurement results for the output of a particular quantum circuit containing a random sequence of single- and two-qubit gates. 
If the circuit is complex enough, the problem becomes intractable for classical computers~\cite{Neven2018,Martinis2018}, but is easily solvable by simply running this quantum circuit multiple times. 

The measured output of a quantum computer running a random circuit is an arbitrary bitstring, whose probability depends on the circuit. 
Therefore, the proof of quantum advantage consists in verifying that the set of random bitstrings produced by the quantum device is consistent with the theoretically expected probability distribution for the given circuit. 
This distribution must be calculated on a classical computer, which is possible for quantum circuits of reduced complexities, for which quantum advantage is not yet present. 
Importantly, each bitstring is likely to occur in the output set only once, making the hypothesis verification a nontrivial mathematical challenge. 
It is exacerbated by the noise and imperfections, which result in high probability for the quantum device to produce completely random outputs. 

The Google team has verified the quantum advantage by first solving this problem for intermediate-size circuits with both their quantum processor 
Sycamore (53 qubits) and the classical supercomputer IBM Summit and observing consistency between the two solutions. 
After that, the complexity of the circuit was increased, so that the estimated time of obtaining the solution classically would be 10,000 years, whereas Sycamore needed only 200 seconds. 

In the conclusion of their work, the authors wrote: 
``We expect that lower simulation costs than reported here will eventually be achieved, but we also expect that they will be consistently outpaced by hardware improvements on larger quantum processors''. 
This prediction turned out to be precisely correct. 
Immediately after the publication of Ref.~\cite{Martinis2019}, 
IBM proposed a technique that would enable reducing the classical calculation to 2.5 days using so-called tensor networks\footnote{\url{https://www.ibm.com/blogs/research/2019/10/on-quantum-supremacy/}}. 
A series of classical numerical experiments followed~\cite{Pednault2019,Huang2020,PanPan2021} with the shortest achieved computation time reaching 5 days on 60 NVIDIA GPUs~\cite{PanPan2021}. 
On the other hand, quantum hardware has also progressed. 
In 2021, a group led by Pan demonstrated two quantum processors, Zuchongzhi 2.0~\cite{Pan2021-4} and 2.1~\cite{Pan2021-5} with 56 and 60 qubits, respectively. 
Zuchongzhi 2.1 performed a random circuit simulation that lasted 4.2 hours, 
but would take 48,000 years on a classical supercomputer even if  tensor networks~\cite{Orus2019-3} are used~\cite{Pan2021-5}. 

A related class of experiments is on quantum advantage with boson sampling (see Secs.~\ref{sec:special-purpose} and \ref{sec:optical}). 
Two groups in China in 2020 realized boson sampling with over 50 photons: 
Zhong {\it et al.}~\cite{Pan2020} in a table-top experiment and Gao {\it et al.}~\cite{Gao2020} in an integrated setting (on a waveguide chip, see Sec.~\ref{sec:optical}) with an additional temporal degree of freedom. 
Results of Ref.~\cite{Pan2020} have later been extended to enable the programmability of the interferometer matrix~\cite{Pan2021}. 
These claims were challenged by Popova and Rubstov~\cite{Rubtsov2021} as well as the Goolge team~\cite{Smelyanskiy2021} arguing that the output photon statistics of the boson sampling circuit can be reproduced using a consumer CPU. 

Both the random quantum circuit and boson sampling demonstrations of quantum advantage are similar in that the problem being solved by the quantum device is the prediction of its own output. 
This may raise a question whether such that a self-serving problem setting is a valid benchmark of computational advantage. 
One can argue, for example, that fluid dynamics equations describing turbulent water flow in a pipe are also beyond the modeling capabilities of a classical computer, 
however are easily ``solved'' by a direct experiment – thereby also offering an ``advantage”. 
The difference between these cases is that the complexity of fluid dynamics is largely associated with the lack of knowledge of precise parameters of the system combined with extreme sensitivity of the solution to these parameters (known as chaos). 
For NISQ devices, in contrast, the parameters are known precisely and the system is not chaotic; 
the complexity is merely a consequence of the exponential scaling of the computational space with the system size. 
An important further difference is the programmability of the quantum computers, i.e., our ability to arbitrarily change the parameters of the systems. 
Finally, there exists a roadmap towards developing the NISQ computation state of the art into a technology capable of solving practical computational problems, which is not the case for the aforementioned fluid dynamics setting.

That being said, the above demonstrations have been in application to ``toy” problems of little practical value. 
In 2021, Google and D-Wave reached the next major benchmark~\cite{Amin2021}: demonstrating quantum advantage for a physically relevant problem. 
They took advantage of the fact that many problems in condensed matter physics can be reduced to the quantum version of the Ising problem (Box~\ref{Box:QUBO}), 
which is the subject of the D-Wave solver. 
They have chosen one such problem (simulating geometrically frustrated magnets) and compared the solution for up to 1440 qubits generated via a D-Wave annealer with a state-of-the-art classical computing technique (path-integral Monte-Carlo). 

\subsection{Specialized applications}

\subsubsection{Certified random number generation} 

Random number generation is important for many applications, such as cryptography and numerical simulations. 
The crucial task in this context is convincing the user of a random number generator in high quality of the output randomness. 
In principle, there exist randomness test that are used to check random sequences\footnote{\url{https://csrc.nist.gov/projects/random-bit-generation/documentation-and-software}}, 
however no mathematical test can guarantee absence of any (intended or inadvertent) hidden regularities therein. 

Quantum technologies offer a unique opportunity to generate random numbers with guaranteed randomness. 
A simplest example is a single photon polarized at 45 degrees incident on a polarizing beamsplitter, which will randomly transmit or reflect this photon with the probability of $1/2$. 
This technology has been the basis for commercial random number generators, with the first product of this kind introduced by ID Quantique in 2001\footnote{\url{https://www.idquantique.com/random-number-generation/products/quantis-random-number-generator/}}. 
However, in order to trust such a unit, the user must be fully aware of and understand its physical design, which is not always possible. 

A much more attractive setting is when the user can convince themselves of the nature of generated randomness by means of information exchange with the device. 
This requirement, known as device-independent random-number generation, can be satisfied by using a NISQ computer as a random generator. 
The idea is to still use the fundamental randomness of a quantum measurement, but replace a single photon by an output multiqubit entangled state of a quantum register.
The ability of the random number generator to rapidly produce multiple samples resulting from preparing and measuring such a state will indicate the quantum origin of these samples to the user. 

Specifically, the protocol proposed by Aaronson\footnote{Scott Aaronson. Certified randomness from quantum supremacy. Talk at CRYPTO 2018, October 2018.} is as follows. 
A user with a classical computer first obtains a few random bits from some trusted source and uses this ``seed randomness'' to design a quantum circuit. 
The user then sends this design to the operator of the random number generator. 
The operator implements this circuit with their quantum computer and runs it multiple times, supplying many random bit strings in a short time. 

How can the user be convinced that the samples they receive indeed originate from measuring a quantum state? 
The idea is to use the quantum circuit of a complexity at the borderline of quantum advantage~\cite{Aaronson2016}, i.e., such that the user is able to simulate it and compute the output state using classical resources, 
but it is prohibitive to classically {\it sample} measurement results from these states. 
Knowing the state, the user is able to statistically test whether the samples they received are consistent with the probability distribution associated with the state they calculated. This consistency will prove the quantum nature of the data.

If the outcome of the test is positive, the user can then use standard techniques to amplify the randomness and remove all the correlations that may present in the set~\cite{Aaronson2016,Guruswami2009}. 
Certifiable quantum randomness generation was announced by Google as one of the first commercial applications of their NISQ devices. 

Ironically, it is essential for this protocol to use quantum computers at the borderline of the quantum advantage threshold, but not far above it. 
That is, developing quantum computing with larger quantum volume is not beneficial for quantum random number generation. 

\begin{table*}[]
\begin{tabular}{|l|l|l|l|}
\hline
\textbf{Cryptographic algorithm} & \textbf{Type} & \textbf{Purpose}            & \textbf{Quantum security} \\ \hline
AES                              & Symmetric     & Encryption                  & Larger key sizes needed   \\ \hline
SHA-2, SHA-3                     & --            & Hash functions              & Larger output needed      \\ \hline
RSA                              & Public key    & Signature, key distribution & No longer secure          \\ \hline
ECDSA, ECDH                      & Public key    & Signature, key distribution & No longer secure          \\ \hline
DSA                              & Public key    & Signature, key distribution & No longer secure          \\ \hline
\end{tabular}
\caption{Security of cryptographic algorithms in the post-quantum era.}
\label{tab:security}
\end{table*}

\wideboxbegin{box:PKC}{Public-key cryptography.}

Cryptography is easily implemented if the communication parties, 
which we call Alice and Bob, share a prearranged, secret data set (a sequence of 0's and 1's) known as \emph{secret key} or \emph{one-time pad}. 
Alice can use the secret key to encrypt her message. 
She can then send it to Bob via an insecure channel. 
While anyone can read this encrypted message, nobody can decrypt it except Bob who possesses another copy of the secret key. 
Such a family of protocols is known as \emph{private-key cryptography}. 
They are secure, simple, and existed for hundreds of years. 

However, the secret key is a high-cost recourse because sharing it would in turn require secure communication between Alice and Bob. 
Therefore a great majority of applications uses an approach known as \emph{public-key cryptography}. 
This ingenious technique is based on the existence of so-called ``one-way" functions that are straightforward to run on a conventional computer, but difficult to calculate in reverse. 
For example, multiplying two large prime numbers is easy, but factorizing a given product is exponentially hard. Such public-key protocols enable secure communication between parties who have never had an opportunity to exchange a secret key.  

Public-key cryptography is jeopardized by the arrival of quantum computation. 
This is because some one-way functions --- unfortunately those that are deployed in currently popular protocols 
--- are reversible in polynomial time with quantum computers. 
Examples include RSA and Diffie-Hellman protocols for key exchange and digital signatures. 
Fortunately, not all one-way functions are vulnerable to quantum cryptanalysis. 
In the next few years, public-key cryptography protocols world-wide are expected to transition to such {\it quantum-safe} or {\it post-quantum} primitives. 
For example, NIST has initiated Post-Quantum Cryptography Standardization Program in 2016, which is expected to be completed by 2024 with a set of standards for quantum-safe algorithms. 

These solutions, however, do not provide ultimate security as they are still based on computational complexity assumptions and their potential vulnerabilities are a subject of ongoing research~\cite{Bernstein2017,Fedorov2021-2}. 
An ultimate solution to the communication security problem is offered by the quantum key distribution~\cite{Gisin2001}. 
However, broad deployment of this technology is facing many challenges, such as cost, speed, 
losses in communication lines, and practical security~\cite{Diamanti2016}.   

\label{Box:PKC}
\wideboxend

\subsubsection{Sampling probability distributions}\label{sec:sampling}

The classically difficult task of sampling from a known probability distribution is useful beyond random number generation. 
This is of particular significance for generative neural networks in machine learning, such as variational autoencoders and generative adversarial networks. 
An important task in this context is obtaining sample sets that are not necessarily \emph{better}, but \emph{different} from those obtained by classical algorithms. 
NISQ technology is often able to satisfy this criterion because of a fundamentally different process of sample generation. 

Dumoulin {\it et al.}~\cite{Dumoulin2014} and Benedetti {\it et al.}~\cite{Benedetti2016} pointed out that quantum annealers that  strongly interact with the environment freeze out the dynamics of a spin system before the termination of the annealing process. 
As a result, such annealers sample from a thermal distribution with some finite temperature. 
The proposed method was experimentally implemented using the D-Wave 2X quantum annealer~\cite{Benedetti2016,Benedetti2017} for the training of a Boltzmann machine. 
However, shortcomings of existing quantum annealers (see Box~\ref{Box:DWave}) limited the study to low-dimensional datasets (see also Refs.~\cite{Neven2008,Neven2012,Henderson2015,Coffrin2021}). 

A more successful application of the D-Wave machine for sampling was reported by Gircha {\it et al.}~\cite{Fedorov2021-5} 
to train a restricted Boltzmann machine as a layer in a discrete variational autoencoder for generative chemistry and drug design. 
A few thousand novel chemical structures with potential medicinal properties have been generated. Gibbs sampling was further studied in 2020 by researchers from Harvard University and QuEra~\cite{Lukin2021-6} using the programmable Rydberg simulator (see Box~\ref{Box:Rydberg}). 
The proposed method potentially leads to a speedup over a classical Markov chain (state-of-the-art sampling technique) for several examples. 

\subsubsection{Cryptanalysis}\label{sec:applicationcrypto}

Modern public-key cryptography is based on the concept of one-way functions (see Box~\ref{Box:PKC}). 
The aforementioned Shor’s algorithm can be used in cryptanalysis (deciphering) of currently deployed public-key cryptography algorithms, such as RSA and Diffie-Hellman (see Table~\ref{tab:security}), 
which is reducible to the task of prime factorization. 
Proof-of-concept experimental factoring of 15, 21, and 35 have been demonstrated on superconducting~\cite{Martinis2012-2}, trapped ion~\cite{Blatt2016}, and photonic~\cite{Pan2007,White2007,OBrien2012} quantum computers. 
Shor’s algorithm for practically relevant key sizes (2048- or 4096-bit), however, requires capabilities far beyond those of NISQ devices. 
We have already mentioned calculations by C. Gidney and M. Ekera on factoring 2048-bit RSA key, 
which would require 8 hours using 20 million physical qubits~\cite{Gidney2021}. 
Another very recent proposal~\cite{Gouzien2021} suggests a way to factor 2048 RSA integers in 177 days with 13436 physcial qubits and a multimode memory. 
A recent forecast review~\cite{Sevilla2020} estimates the likelihood for quantum devices capable of factoring RSA-2048 to exist before 2039 as less than 5$\%$. 

Special-purpose quantum machines can be used for factorization as well~\cite{Jiang2018,Peng2019}. 
For example, the D-Wave annealer was used to factor $1,005,973=1009\times1019$~\cite{Wang2020}. 
Variational quantum algorithms have also been studied in this context~\cite{Aspuru-Guzik2019}. 
However, it seems that the complexity of this task for special purpose quantum computers grows much faster than for universal ones. 
Thus, these studies are of conceptual interest in the NISQ era, but are not expected to lead to quantum advantage.  

The secrecy of key distribution is not the only vulnerability of modern cryptography. 
Another problem is that the secret key is typically much shorter than the dataset it is used to encrypt. 
There do exist encryption algorithms, which solve this problem so that it is hard for classical computers to decrypt messages encrypted with relatively short keys; the difficulty grows exponentially with the key size. 
For example, widely deployed AES protocol uses keys of length as little as 256 bits to securely encrypt terabytes of data. 

Quantum computers have less of an effect on this matter, since Shor's algorithm does not apply, and exponential speedups are not expected (see Table~\ref{tab:security}). 
However, Grover's algorithm~\cite{Grover1996} enables quadratic speedup in brute force search, which means that the key length should be doubled to enable the same level of protection~\cite{Kim2018}. 
The same scaling applies to cryptographic hash functions, for which the primary attack method is also brute-force search~\cite{Kim2018}. 
 
An area of particular concern in the context of quantum security is blockchains and cryptocurrencies~\cite{Tomamichel2018,Fedorov2018,Fedorov2018-2}, which are argued to be the ``blueprint for a new economy''~\cite{Swan2015}. 
Typical blockchain and cryptocurrency protocols use several cryptographic schemes, such as digital signatures and hash functions for achieving a consensus (proof-of-work) between users in the absence of trust. 
The quantum vulnerability of hash functions is similar to that of AES: that is, the primary known method of attack is brute-force search~\cite{Kim2018} and hence no more than quadratic advantage can be expected. 
However, private keys can be extracted from digitally signed messages by means of Shor’s algorithm allowing parties in possession of a quantum computer to impersonate any other party, 
which will obviously collapse any blockchain relying on current protocols. 
For example, Bitcoin (a cryptocurrency blockchain), which uses the elliptic curve signature scheme, could be completely broken by a quantum computer as early as 2027~\cite{Tomamichel2018}. 

Attacks with quantum computers have become a subject of many studies that proposed solutions for quantum-resistant blockchains~\cite{Fedorov2018,Fedorov2018-2}: 
blockchains that use quantum key distribution~\cite{Gisin2001} or post-quantum digital signatures and consensus schemes. 
A quantum-secured blockchain protocol was experimentally demonstrated in 2018~\cite{Fedorov2018}. 

In summary, the detrimental effect of quantum computing on information security can be thwarted by upgrading information exchange protocols to quantum or post-quantum technologies. 
Importantly, this transition must be implemented long before the emergence of practical quantum computation~\cite{Mosca2017,Fedorov2018,Wallden2019}. 
Otherwise a variety of potential risk scenarios can be envisioned. For example, a present-day hacker might intercept and store encrypted messages with the hope to decrypt them with a quantum computer a few years later. 
If the information is long-term sensitive (medical records, genetic data, strategic plans, etc.), this attack may result in damages.  

\subsection{Economically impactful application}

There are two ways one can classify general-purpose applications of quantum computing. 
On the one hand, one can consider the classes of problems that  quantum computers are able to solve. Roughly, three such classes can be identified: 
\begin{itemize}
	\item simulation, that is predicting the behaviour of a certain complex system based on an existing mathematical model;
	\item optimization, i.e., finding the best setting of a large combination of discrete parameters according to some criterion; 
	\item machine learning, that is constructing a mathematical model that would fit the properties of a certain dataset. 
\end{itemize}
These classes form the ``supply'' of services that quantum computation can provide. 
These services, on the other hand, can provide a variety of ``demands'' from various areas of technologies, such as chemistry, material science, life science, finance, etc. 
The significance of such demands is evidenced, for example, by the ``quantum challenges" 
announced by several industry leaders such as Airbus\footnote{\url{https://www.airbus.com/en/innovation/disruptive-concepts/quantum-technologies/airbus-quantum-computing-challenge}} 
and BMW\footnote{\url{https://aws.amazon.com/blogs/quantum-computing/winners-announced-in-the-bmw-group-quantum-computing-challenge/}}, 
in which quantum scientists and technologists are invited to provide solutions to a variety of problems faced by these companies. 


\subsubsection{Simulation}\label{sec:application-simulation} 

Before we proceed to a specific discussion, we caution the reader to distinguish the notions of ``simulation'' as a problem class and ``quantum simulators'' as a type of a quantum computer. 
Many simulation tasks can be solved with other types of quantum computers and quantum simulators can be applied to other types of problems beyond simulation.

Simulation problems can be further classified into quantum and classical according to the object of study. 
We begin with the former, specifically, with the application in chemistry, life science, and materials science. 
Many subjects of these fields, such as fuels, drugs, biologically active compounds, and fertilizers, are quantum systems consisting of a large number of interacting components. 
As discussed above, such systems are hard to model classically, but naturally amenable to quantum computation and simulation. 
This constitutes a major component of the expected landscape of quantum technology applications. 

\paragraph{Quantum chemistry.} A case in point is the calculation of energies and electronic structures of molecular ground and exited states, which is important for theoretical understanding of chemical reactions. In order to encode electronic states of molecules into qubits, one uses a basis of predefined ``spin orbitals'', i.e., quantum states that can be occupied by individual electrons\footnote{\label{SpinOrbitsFootnote}The number of spin orbitals associated with each atom in a molecule equals the maximum electron capacity of the corresponding period in the periodic table: for example, H and He are represented by 2 orbitals, all atoms from Li to F by 10 orbitals, and so on.}. 
Each qubit in the register of the quantum computer represents one such spin orbital, with its value --- 0 or 1 --- determining whether this spin orbital is occupied. The molecular state is then a superposition of multiple occupancy configurations. The quantum computer is used to simulate the molecular Hamiltonian and find the lowest energy state. 

Currently, the primary tools of the trade are variational quantum algorithms (such as VQE; see Sec.~\ref{sec:variational}) thanks to their modest hardware requirements. 
They were used to analyze small molecules, such as hydrogen (H$_2$)~\cite{Gambetta2017-2,Martinis2016-2,Blatt2018-3}, lithium hydride (LiH)~\cite{Gambetta2017-2,Blatt2018-3}, beryllium hydride (BeH$_2$)~\cite{Gambetta2017-2}, 
and water (H$_2$O)~\cite{Monroe2020-2}, 
as well as to simulate diazene isomerizations~\cite{Babbush2020} and carbon monoxide oxidation~\cite{Fedorov2021-3}. 

We are observing a surge of interest to quantum computing applications in chemistry from automobile industry, 
including BMW~\cite{Luckow2021}, Daimler-Benz~\cite{Kim2019}, Nissan~\cite{Fedorov2021-3},
Ford\footnote{https://medium.com/@ford/why-ford-is-taking-a-quantum-leap-into-the-future-of-computing-453128a2ea9f}, 
and Toyota~\cite{Hirai2021} (and also Volkswagen~\cite{Lieb2019}, see below). 
Objects of interest include a new generation of electric batteries, combustion efficiency optimization and fuel cells. 
For example, Sapova {\it et al.} in collaboration with Nissan improved VQE to simulate the molecules involved in carbon monoxide oxidation~\cite{Fedorov2021-3}. 
Kim {\it et al.}~with Daimler-Benz~\cite{Kim2019} estimated the cost of simulating electrolyte molecules in Li-ion batteries on a fault-tolerant quantum computer.

However, existing simulations have limitations arising not only from available hardware capabilities, 
but also of conceptual nature, such as the quality of the initial ansatz, convergence speed, presence of local minima, and the large number of measurements required in each iteration~\cite{Aspuru-Guzik2020,Aspuru-Guzik2021,Babbush2021-4}. 
Active research is underway to overcome some of these challenges via software and hardware improvements of variational quantum computing. 
However, most significant expectations are associated with the deployment of error correction. 
This may open doors to simulating various chemical systems of practical relevance, including medium-sized inorganic catalysts, biomimetics, metalorganic molecules, and homogeneous catalysts for C-H bond activation~\cite{Elfving2020}. 
This latter process is relevant for the production of methanol, which can replace coal and petroleum as a cleaner source of energy and a primary product for synthetic materials. 

Longer-term perspectives, requiring scalable fault-tolerant quantum computers, are the chemistry of enzyme active sites since they can involve multiple coupled transition metals~\cite{Aspuru-Guzik2019-2,Bravyi2020}, 
famous examples being the four manganese ions in the oxygen evolving complex~\cite{Cady2008} or the eight transition metals in the iron-sulfur clusters of nitrogenase~\cite{Lowdin1963,Cha1989,Troyer2017,Berry2019,Troyer2021,Babbush2021-3}. 
The latter task is crucial for understanding nitrogen fixation by the enzyme nitrogenase, which allows obtaining ammonia at room temperature and standard pressure (so-called FeMoco problem). 
Solving it would be a major breakthrough in comparison with the state-of-the-art industry Haber-Bosch process, which requires high temperature and high pressure and is therefore energy intensive. 
The FeMoco problem corresponds to finding the lowest energy state of 108 spin orbital qubits occupied by 54 electrons~\cite{Troyer2017}. 
A concrete guide and the corresponding quantum circuit for such calculations using a fault-tolerant quantum computer have been presented for the first time by Reiher {\it et al.}~\cite{Troyer2017}
and improved in Refs~\cite{Berry2019,Troyer2021,Babbush2021-3}.
The most recent conclusion from 2020 shows that FeMoco can be simulated using about four million physical qubits in four days of runtime, 
assuming 1 $\mu$s cycle times and physical gate error rates no worse than 0.1\%~\cite{Babbush2021-3}.

\paragraph{Life science and drug discovery.} Quantum simulation of molecules and chemical reactions has direct application in life science, specifically, in drug discovery~\cite{Fedorov2021-4,Aspuru-Guzik2018-3}. 
Over 99\% of the approved drug molecules in DrugBank 5.0~\cite{Wishart2017} have molecular weights between up to 1800 atomic mass units~\cite{Yamazaki2018,Aspuru-Guzik2020}. 
This allows us to estimate the number of spin orbitals, and therefore the number of qubits, required for modelling these molecules (as per footnote \ref{SpinOrbitsFootnote}) to be on the order of $10^2$ to $10^3$. 
Although this is outside the capacities of current NISQ devices, collaborations between pharmaceutical and quantum computing businesses are starting to emerge. 
For example, in 2021 Google established a partnership with Boehringer Ingelheim \cite{Babbush2022}. 

Quantum simulation tasks in life sciences extend beyond modeling individual molecules to analysis of macroscopic quantum phenomena in molecular clusters. 
These phenomena are particularly manifest in the photochemistry of conjugated organic molecules interacting with light. 
Examples include light harvesting in plants and vision in animals~\cite{Nori2013,Nori2018,Benjamin2011}.

\paragraph{Materials science.} The frontier that follows quantum chemistry in the order of complexity is materials science. 
Unlike a molecule, a crystal is described by an infinite lattice containing infinitely many electrons. 
For this reason, simulation of real materials can be done only approximately and is hard even for quantum computers. 
A case in point is high-temperature superconductivity~\cite{Bravyi2020}. 
Originally discovered in 1986, this phenomenon has no comprehensive theoretical explanation to date~\cite{Lee2006}. 
Understanding the physics behind high-temperature superconductors is extremely rewarding as it would pave the way toward the ``holy grail” --- 
 materials with the critical temperature above the room level, which would enable lossless transmission of electrical energy. 

Typically, simulating a solid-state lattice is performed within a framework of a certain theoretical model that simplifies the interaction between electrons in this lattice, yet accurately predicts salient properties of the material at hand. 
Many solid-state phenomena including superconductivity are described by the so-called Hubbard model. 
This model accounts for only two types of behavior exhibited by electrons: single-electron hopping between lattice sites and two electrons (of different spins) interacting within a single site. In spite of its apparent simplicity, 
Hubbard model is classically intractable. On the other hand, NISQ computers are capable of analyzing lattices of a few periods in size. 

Two approaches are being pursued in this context. The first one is simulation via a digital (gate-based) quantum computer. In this setting, each lattice site is represented by two qubits. 
For example, an 8-site 1D Fermi-Hubbard model was simulated by the Google Sycamore quantum processor~\cite{Babbush2020-2}. 
This quantum analysis is not yet of practical interest since such lattice sizes are also amenable to classical simulation. 
The quantum advantage threshold is expected at around physical 200 qubits~\cite{Montanaro2020}, which enables modelling, e.g., a 2D square lattice of $10\times10$ sites. 

The second approach involves analog quantum simulators (Sec.~\ref{sec:quantumsimulators}), currently based primarily on ultracold atoms. 
These experiments are already bringing new insights into many-body  systems that were not known before. 
For example, experiments with cold atoms in lattices (see Sec.~\ref{sec:ionsatoms} and Box~\ref{Box:AtomsLattice}), 
involving up to 80 lattice sites, enabled detailed reproduction of the primary physical phenomena defining phase transitions in the Fermi-Hubbard model~\cite{Demler2017,Demler2019,Demler2021-2}.

In addition to superconductivity, near-term quantum computing can be useful for simulating 2D materials (such as graphene and heterostructures), frustrated spin systems, and materials’ dynamical effects~\cite{Bravyi2020,Vashishta2020,Babbush2022-2}. 
On the long-term horizon, one may expect that quantum computing will become a main tool for designing bespoke materials with the required properties. 

So far in this section we focused on applying quantum computers to solving quantum many-body problems as a whole and found that achieving quantum advantage is beyond the reach of existing technologies. 
This is particularly because there exist many classical techniques that model even very large quantum systems remarkably well. 
However, these methods can be further accelerated or made more precise by making a part of calculations quantum. 

The current workhorse for the classical simulation of quantum chemistry and materials is density functional theory (DFT). 
The crux of this method is explained in Ref.~\cite{Troyer2016}: 
``DFT circumvents the exponential scaling of resources required to directly solve the electronic quantum many-body Hamiltonian by mapping the problem 
of finding the total energy and particle density of a system to that of finding the energy and particle density of 
noninteracting electrons in a potential that is a functional only of the electron density, and requiring self-consistency between the density and potential”. 
That is, the many-body problem is replaced by solving the motion of a single particle in the field created by other particles. 
However, this approach is insufficient in those settings for which the entanglement is essential for describing the system state. 
This includes the aforementioned high-temperature superconductivity~\cite{Lee2006} as well as molecular complexes involving transition metals~\cite{Kovaleva2008} and actinides~\cite{Albers2001}.

To overcome this restriction, DFT can be supplemented by a quantum treatment of those spin orbitals that are relevant at a particular geometric location. 
These spin orbitals are considered as an entangled ``cluster'', whose quantum state is analyzed in the potential that depends on the density of other electrons. 
The treatment of the cluster can be implemented using either classical or quantum tools. 
To our knowledge, such an approach has not yet been tried with a \emph{bona fide} quantum computer, 
however the simulations by Bauer {\it et al.}~\cite{Troyer2016} led to the conclusion that quantum advantage can be reached with 100 logical qubits. 

\paragraph{Simulation of classical processes.} We now proceed to discussing classical simulation problems that can be solved faster using quantum computers. 
An important class of such problems is solving systems of equations, which is important for a large variety of applications including aerodynamics, hydrodynamics, market dynamics in finance, and disease spreading in epidemiology. 

In 2009 Harrow, Hassidim, and Lloyd (HHL) proposed a quantum algorithm~\cite{Lloyd2009} 
(later improved by Ambainis~\cite{Ambainis2010} and Childs {\it et al.}~\cite{Childs2017}) that enables solving systems of linear equations with a gate-based quantum computer. 
The time to solution required by the HHL algorithm scales as the logarithm of the total number of equations, thereby providing exponential speedup with respect to classical algorithms, whose time to solution scales polynomially. 
However, the HHL algorithm has a limitation in that the output is represented as a multiqubit state with the solution encoded in the amplitudes of qubit configurations~\cite{Kyriienko2021}. 
This is an example of the problem mentioned in the Introduction: because the parallelism of quantum computers hinges on their ability to solve problems in a superposition state, the resulting solution will also be in a superposition state. 
To extract the classical solution (in this case the amplitude of each qubit configuration), one would need to run the quantum algorithm multiple times, each time performing the measurement of the output. The number of such measurements scales exponentially with the number of qubits, thereby negating quantum advantage if HHL is used in a straightforward manner to replace the classical linear equation solver. 
A further complication associated with HHL is the need to use sophisticated techniques, such as ``quantum random access memory'' to prepare the input state~\cite{Lloyd2008,Lloyd2008-2,Hong2012}.
Such memory, in contrast to its classical counterpart, allows one to query memory cells in superposition. 
This technology has not been implemented in practice, although some setups have been proposed and tested~\cite{Duan2019}.

Although the HHL algorithm has initially been proposed only for solving systems of linear equations, 
it can be readily extended to a broader range of equations including nonlinear, ordinary differential, nonlinear differential, and partial differential. 
Indeed, under certain conditions, any such system can be reduced to linear by means of finite element method, which discretizes the parameter space via a finite mesh. 
In this setting, HHL might be able to achieve quantum advantage in spite of aforementioned challenges~\cite{Kyriienko2021}. 
This is because the resulting linear equations in the finite element method are produced algorithmically rather than input as classical dataset. 
Furthermore, the resulting system of linear equations is typically sparse. 
For example, this can be the case for the electromagnetic scattering cross-section problem. 
Clader {\it et al.}~\cite{Clader2013} studied the application of the HHL algorithm in this setting and argued that the exponential speedup can be achieved. 
However, later the speedup was shown to be polynomial~\cite{Montanaro2016-3}. 

The HHL algorithm is based on the gate-based quantum computing model and is hence beyond the current level of technology. 
However, related algorithms, realizable on current NISQ computers, are being proposed in the framework of the variational model for solving linear and nonlinear equation systems. 
An experimental realization for systems containing up to 1024 equations has been presented in 2020 on a Rigetti 16Q Aspen-4 superconducting quantum computer~\cite{Coles2020}.  

The applicability of the finite element method to nonlinear differential equations is however limited~\cite{Lloyd2020,Lloyd2020-2}. 
For example, chaotic systems, such as fluid dynamics (governed by the Navier-Stokes equation), cannot be solved via this approach. 
This motivated the development of a suite of quantum algorithms specially designed for this purpose (Gaitan~\cite{Gaitan2020}, Lloyd {\it et al.}~\cite{Lloyd2020-2}, and Kyriienko {\it et al.}~\cite{Kyriienko2021}). 

A further application of quantum computing to classical simulation is as an alternative to Monte-Carlo methods. 
The latter is a large family of methods for estimating the properties (e.g., mean and variance) of a statistical distribution by taking multiple samples from that distribution.
The quantum alternative is to reduce this problem to estimating the amplitude of a certain state vector in the Hilbert space. 
This can be done efficiently using a variant of Grover’s algorithm. 
While this approach is applicable in a variety of fields, the current interest appears to be focused on applications in 
finance~\cite{Alexeev2022}, particularly for derivative pricing and risk analysis~\cite{Woerner2020,Zeng2021}. 
In 2020 researchers from Goldman Sachs and IBM  found a quantum advantage for derivative pricing achievable with 7.5k logical (ideal) qubits~\cite{Woerner2019-2}. 
This is beyond the capabilities of the existing and upcoming generation of quantum computing devices. 

\subsubsection{Optimization}\label{sec:application-optimization} 

Problems of discrete optimization, that is, finding the best solution among a countable set, 
are ubiquitous in human civilization: 
from single individuals attempting to choose the best route to work in the morning traffic or the best portfolio for their retirement savings to transnational retailers aiming to find the best schedule for their delivery trucks. 
The characteristic feature of these problems is the exponential growth of the complexity with the problem size. 
Many classes of optimization problems might be amenable to quantum speedup. 

Of particular relevance here is quadratic unconstrained binary optimization (QUBO, Box~\ref{Box:QUBO}). 
The primary quantum approach to solving QUBO problems is via quantum annealing. 
Experiments to this effect have been attempted on D-Wave System machines for a variety of applications:
\begin{itemize}
	\item chemistry, specifically, finding ground states of molecules~\cite{Lieb2019};
	\item life science, including lattice protein folding~\cite{Aspuru-Guzik2012-2,Fingerhuth2018}\footnote{Protein folding, i.e. the prediction of the three-dimensional protein structure given a specific amino acid sequence, is a grand challenge in biology.} 
	and genome assembly~\cite{Fedorov2021,Sarkar2021};
	\item solving polynomial systems of equations for engineering applications~\cite{Sota2019} and linear equations for regression~\cite{Sota2019};
	\item materials science, in particular, designing metamaterials~\cite{Kitai2020,Boltasseva2021};
	\item likelihood-based regularized unfolding for processing high-energy physics data~\cite{Wittek2019-2};
	\item finance, such as portfolio optimization~\cite{Orus2019,Orus2020,Grant2021,Alexeev2022}, forecasting crashes~\cite{Orus2019-2}, 
	finding optimal trading trajectories~\cite{Rosenberg2016-2}, optimal arbitrage opportunities~\cite{Rosenberg2016},  optimal feature selection in credit scoring~\cite{Rounds2017},
	and, foreign exchange reserves management~\cite{Vesely2022};
	\item logistics, including traffic optimization~\cite{Neukart2017,Inoue2021,Hussain2020}, 
	and scheduling~\cite{Venturelli2016,Ikeda2019,Sadhu2020,Botter2020,Domino2021,Domino2021-2} including railway conflict management~\cite{Domino2021,Domino2021-2}.
\end{itemize}
Further examples are listed on the website of D-Wave Systems\footnote{https://www.dwavesys.com/learn/featured-applications/}. 

An approach alternative to quantum annealing is the quantum approximate optimization algorithm (QAOA), which falls within the framework of variational quantum computing (see Sec.~\ref{sec:variational}). 
Initial proposals on QAQA considered applications to graph optimization, in particular, to the MaxCut problem (see Box~\ref{Box:QUBO}). 
Experimentally, this was demonstrated by Google with up to 23 qubits, achieving  ``an advantage over random guessing but not over some efficient classical algorithms''~\cite{Babbush2021}. 
Recently, an international collaboration led by Volkswagen applied QAOA to the paint shop problem~\cite{Lieb2021}, 
i.e., minimizing the number of changes of color when painting a certain sequence of cars. 
This NP-hard optimization problem reduces to QUBO. 
Solutions for instances of small sizes have been obtained via the IonQ trapped-ion quantum computer~\cite{Monroe2020}. 
Another use case of a quantum variational algorithm is flight schedule optimization recently presented by Delta Airlines~\cite{Stojkovic2021}. 

Finally, a particular case of QUBO, the maximum independent set problem (see Box~\ref{Box:QUBO}), 
which has direct applications in network design~\cite{Hale1980} and finance~\cite{Boginski2005} and is furthermore important for interval scheduling, 
can be tackled using programmable Rydberg atom simulators (see Sec.~\ref{sec:ionsatoms} and Box~\ref{Box:Rydberg}). 
This system allows implementation of either quantum annealing or QAOA. 
As an example, a collaboration of academics led by \'Electricit\'e de France and a quantum startup Pasqal have applied QAOA in this system to optimize smart-charging of electric vehicles~\cite{Henriet2021}. 
Quantum advantage in this setting can be expected with as few as 1000-1200 atoms provided that the coherence time is substantially increased~\cite{Ayral2020}.

This latter conclusion is supported by a 2021 theoretical study of limitations of optimization algorithms on noisy quantum devices~\cite{Stilck-Franca2021}, which argues that ``substantial quantum advantages
are unlikely for classical optimization unless noise rates are decreased by orders of magnitude or the topology of the problem matches that of the device.” 
In some cases, hopes to achieve quantum advantage by reducing an optimization problem to QUBO appear unviable altogether because of an exponential overhead associated with such reduction. 
Examples include quantum chemistry~\cite{Lieb2019,Chermoshentsev2021} and lattice protein folding~\cite{Chermoshentsev2021}. Note that for pretein folding, the issue can be resolved by means of the variational algorithm without reducing the problem to QUBO. An experiment to this effect was performed in 2021 on 9 qubits of an IBM 20-qubit quantum computer~\cite{Woerner2021}.

\subsubsection{Machine learning}\label{sec:application-ML} 

\begin{table*}[]
\begin{tabular}{|l|c|c|c|l|}
\hline
\multicolumn{1}{|c|}{\textbf{Algorithm}} & \textbf{Classical}                         & \textbf{Quantum}                                                   & \textbf{QRAM} & \textbf{Reference}                                                                                                                                                                                                  \\ \hline
Linear regression                        & $\mathcal{O}(N)$                           & $\mathcal{O}(\log(N))^{*}$                    & \textbf{Yes}  & \cite{Schuld2016,Wang2017,Panigrahi2020,Li2019}                                                                                                                                                    \\ \hline
Gaussian process regression              & $\mathcal{O}(N)$                           & $\mathcal{O}(\log(N))^{\dagger}$ & Yes           & \cite{Zhao2019,Zhao2019-2}                                                                                                                                                                         \\ \hline
Decision trees                           & $\mathcal{O}(N\log N)$                     & Unclear                                                            & No            & \cite{Lu2014}                                                                                                                                                                                      \\ \hline
Ensemble methods                         & $\mathcal{O}(N)$                           & $\mathcal{O}(\sqrt{N})$                                            & No            & \cite{Schuld2018,Wang2019,Arunachalam2020}                                                                                                                                                         \\ \hline
Support vector machines                  & $\approx\mathcal{O}(N^2)-\mathcal{O}(N^3)$ & $\mathcal{O}(\log{N})$                                             & Yes           & \cite{Lloyd2014,Chatterjee2017,Schuld2019}                                                                                                                                                         \\ \hline
Hidden Markov models                     & $\mathcal{O}(N)$                           & Unclear                                                            & No            & \cite{Monras2012}                                                                                                                                                                                  \\ \hline
Bayesian networks                        & $\mathcal{O}(N)$                           & $\mathcal{O}(\sqrt{N})$                                            & No            & \cite{Chuang2014,Wiebe2015}                                                                                                                                                                        \\ \hline
Graphical models                         & $\mathcal{O}(N)$                           & Unclear                                                            & No            & \cite{Benedetti2017}                                                                                                                                                                               \\ \hline
k‐Means clustering                       & $\mathcal{O}(kN)$                          & $\mathcal{O}(k\log{N})$                                            & Yes           & \cite{Lloyd2013,Wiebe2015-2,Kerenidis2018}                                                                                                                                                         \\ \hline
Principal component analysys             & $\mathcal{O}(N)$                           & $\mathcal{O}(\log{N})$                                             & No            & \cite{Lloyd2014-2}                                                                                                                                                                                 \\ \hline
Persistent homology                      & $\mathcal{O}(\exp(N))$                     & $\mathcal{O}(N^2)$                                                 & No            & \cite{Lloyd2016}                                                                                                                                                                                   \\ \hline
Gaussian mixture models                  & $\mathcal{O}(\log(N))$                     & $\mathcal{O}({\rm polylog}(N))$                                        & Yes           & \cite{Kerenidis2020-2,Miyahara2020}                                                                                                                                                                \\ \hline
Variational autoencoder                  & $\mathcal{O}(\exp(N))$                     & Unclear                                                            & No            & \cite{Khoshaman2018}                                                                                                                                                                               \\ \hline
Multilayer perceptrons                   & $\mathcal{O}(N)$                           & Unclear                                                            & No            & \cite{Kak1995,Zak1998,Aspuru-Guzik2017,Kim2017,Lloyd2019}                                                                                                                                          \\ \hline
Convolutional neural networks            & $\mathcal{O}(N)$                           & $\mathcal{O}(\log{N})$                                             & No            & \cite{Lukin2019-3}                                                                                                                                                                                 \\ \hline
Bayesian deep learning                   & $\mathcal{O}(N)$                           & $\mathcal{O}(\sqrt{N})$                                            & No            & \cite{Wittek2019}                                                                                                                                                                                  \\ \hline
Generative adversarial networks          & $\mathcal{O}(N)$                           & $\mathcal{O}({\rm polylog}(N))$                                        & No            & \cite{Lloyd2018,Dallaire-Demers2018,Gao2017}                                                                                                                                                       \\ \hline
Boltzmann machines                       & $\mathcal{O}(N)$                           & $\mathcal{O}(\sqrt{N})$                                            & No            & \cite{Dumoulin2014,Anschuetz2019,Benedetti2016,Amin2018,Wiebe2015-3}\footnote{a)} 
\\ \hline
Long short-term memory                   & $\mathcal{O}(N)$                           & Unclear                                                            & No            & \cite{Chen2020}                                                                                                                                                                                    \\ \hline
Reinforcement learning                   & Unclear                                    & Unclear                                                            & No            & \cite{Dunjko2021}                                                                                                                                                                                  \\ \hline
\end{tabular}
\caption{Overview of quantum machine learning algorithms (based on Refs.~\cite{Biamonte2017,Benjamin2021}; $N$ is the characteristic layer size). $^{a)}$ see also: https://www.cs.ubc.ca/~nando/papers/quantumrbm.pdf}
\label{tab:ML}
\end{table*}

Machine learning techniques are powerful for finding patterns in data. 
Quantum technology and machine learning are developing rapidly and overlap each other in several contexts, comprising a new field known as quantum machine learning~\cite{Biamonte2017}. 
One can identify three primary directions of these field:
\begin{enumerate}
	\item classical neural networks for obtaining variational solutions for many-body quantum-mechanical problems;
	\item fully-quantum neural networks operating with quantum data, possibly augmented with classical neural networks;
	\item quantum algorithms that could act as building blocks of classical machine learning programs.
\end{enumerate}
Item 1 in this list does not typically involve quantum computers~\cite{Dunjko2018}, therefore it is outside the scope of our review. 

Item 2 refers to quantum neural networks processing quantum or classical data using a circuit with gates described by continuous parameters~\cite{Verdon2018,Benedetti2019,Schuld2019,Schuld2019-2,Wolf2020,Woerner2021}. 
This is related to variational quantum computing (see Sec.~\ref{sec:variational}) with the difference being that a variational circuit aims to generate a quantum state optimizing a certain cost function, 
whereas a quantum neural network is trained to process a more general dataset. 
For example, quantum convolutional neural networks~\cite{Henderson2020-2,Lukin2019-3} were proposed and used to recognize complex many-body quantum states~\cite{Lukin2019-3}. 
Another example is generative adversarial networks~\cite{Lloyd2018,Dallaire-Demers2018,Woerner2019} aimed to produce a state whose statistical properties are consistent with an input sample set. 
This approach has a potential application to facilitate the financial derivative pricing~\cite{Woerner2019} and learning the financial dataset~\cite{Coyle2021} using the Rigetti quantum processor.
Aside from the context of variational quantum algorithms, 
quantum neural networks have not yet been extensively studied and the scope of their practical quantum advantage is undetermined. 

In classical machine learning (item 3), linear algebraic calculations, such as Fourier transforms, 
matrix-vector multiplication, diagonalizing matrices, and solving linear systems of equations, constitute the computationally heaviest part. 
Quantum processing enables polynomial advantage in many of these calculations~\cite{Biamonte2017,Dunjko2018,Lloyd2009,Aspuru-Guzik2018-3}.
A further application of quantum computation in machine learning is sampling from a given probability distribution (see Sec.~\ref{sec:sampling}), which is an important component of generative neural networks. 

Table~\ref{tab:ML} lists the known possibilities for applying quantum computing in machine learning as well as associated advantages~\cite{Biamonte2017,Benjamin2021}. 
At the same time, Ref.~\cite{Biamonte2017}, which is often seen as the ``manifest'' of quantum machine learning, raises four challenges that must be addressed to achieve these advantages in practice: 
\begin{enumerate}
	\item The input problem. The computational cost of loading classical input data into a quantum register is significant and may exceed that of the quantum computation \emph{per se}. 
	\item The output problem. Multiple samples of the output quantum register are required in order to obtain a specific solution of interest. 
	\item The costing problem. Little is known about the true number of gates required by quantum machine learning algorithms. 
	\item The benchmarking problem. The expected degree of polynomial quantum advantage may change because classical algorithms keep developing and improving.
\end{enumerate}

Recent use cases in quantum machine learning~\cite{Gambetta2017-3} include classifiers for handwritten digits datasets (D-Wave  \cite{Benedetti2018} and IonQ \cite{Kerenidis2020}),
analyzing NMR readings (IonQ~\cite{Demler2020,Demler2021}), 
learning for the classification of lung cancer patients (D-Wave~\cite{Jain2020}), 
classifying and ranking DNA to RNA transcription factors (D-Wave~\cite{Lidar2018-2}), 
satellite imagery analysis~ (Rigetti~\cite{Henderson2020}) and weather forecasting (Rigetti~\cite{Rigetti2021}), and many others~\cite{Perdomo-Ortiz2018}. 
As previously, we caution the reader that these demonstrations are of proof-of-principle nature and do not yet present quantum advantage.

\section{Quantum computing requires software}\label{sec:software}

Although the implementation of the quantum hardware --- long-lived qubits and their interaction mechanisms --- constitutes the heart and grand challenge of quantum computational technology, 
its practical application is impossible without means of its classical control and the software that would enable its programming by a human (i.e., classical) user~\cite{Chong2017}. 

We identify three levels of quantum software. 
The highest level is that of applications, i.e., programs that solve computational problems for the end-user --- for example, traffic optimization in a given city or simulating a particular material. 
This level is both platform- and model-agnostic, that is, the end user need not know what is under the hood of the quantum computational service. 

This level interacts with the second-level quantum software, which deals with quantum algorithms --- 
sequences of instructions that implement quantum computation in the language of abstract information carriers (e.g., qubits) within a certain computational model. 
This level is \emph{platform-agnostic}, but \emph{model-specific}. An algorithm that invokes a gate-based quantum calculation will be the same for a trapped-ion or superconducting quantum computer. 
However, an algorithm for the digital model would not be suitable, e.g., for a quantum annealer. 

The final third level operates directly with qubits in a particular machine and is both platform- and model-specific. 
This level transforms an abstract quantum algorithm into a sequence of signals that control a physical implementation of qubits, e.g., superconducting junctions, atoms, ions, or photons. 
Some authors divide this third level into two sublevels: 
the upper sublevel transforms logical qubits into physical qubits taking into account the specific design of the quantum circuit and the associated error rate; the lower sublevel interacts with physical qubits. 
Because the lowest-level software is strongly hardware-specific, its development is implemented by quantum hardware manufacturers and is typically hidden from external programmers and users. 
These manufacturers, however, provide cloud interfaces, programming languages, and software development kits (SDKs), using which an external programmer can write algorithms. 
These interfaces enable one to address individual qubits, initialize them, perform single-, two-, and multiqubit operations, and measure them. In this way, one can compose, manipulate, and optimize quantum circuits. 
Examples of such SDKs for the gate-based model include:
\begin{itemize}
	\item graphical user interface, Python package Qiskit\footnote{\url{https://qiskit.org}}, and quantum assembly language (QASM)\footnote{\url{https://github.com/Qiskit/openqasm}} by IBM; 
	\item Python library Cirq by Google\footnote{\url{https://github.com/quantumlib/Cirq}}, which has an important added capability of supporting calculations with qudits;
	\item assembly-type language Quil by Rigetti\footnote{\url{https://www.rigetti.com}};
	\item programming language Q\#\footnote{\url{https://azure.microsoft.com/en-us/resources/development-kit/quantum-computing/}} by Microsoft tailored for compatibility with other Microsoft products, such as Visual studio, .NET, and Azure. 
\end{itemize}
In addition, bespoke SDKs are being developed to program special-purpose quantum machines, for example:
\begin{itemize}
	\item Strawberry Fields by Xanadu\footnote{\url{https://strawberryfields.ai}}, a cross-platform Python library for simulating and developing quantum optical circuitry.
	\item Ocean SDK\footnote{\url{https://docs.ocean.dwavesys.com/en/stable/}} developed by D-Wave Systems for solving QUBO and related problems on D-Wave hardware or compatible tools including simulators.
\end{itemize}
Software of the second (algorithmic) level has been developing for many years, some examples listed in Sec.~\ref{sec:applications}. 
However, most of existing quantum algorithms have been designed with a perfect multiqubit quantum computer in mind and are, therefore, inapplicable to current NISQ machines. 
Bridging the desired and available levels of technology constitutes a major challenge in this field. 
An example of an SDK that attempts to address this challenge is TensorFlow Quantum~\cite{Babbush2021-2} --- a quantum machine learning library for rapid prototyping of hybrid quantum-classical machine learning models. 

Finally, the first (application) level is only starting to emerge. Interestingly, this level is currently driven not only by hardware giants, 
but also by startups, which use a business model known as quantum-computing-as-a-service model (QCaaS). 
Examples include AWS Braket by Amazon\footnote{\url{https:https://aws.amazon.com/braket/}}, QC Ware\footnote{\url{https:qcware.com}}, Zapata\footnote{\url{https://www.zapatacomputing.com}}, and QBoard\footnote{\url{https://qboard.tech}}. 
In the framework of this model, the interaction between a service provider and a client is in the context of a computational problem, which is formulated in the language familiar to the client. 
The provider works with the client to advise whether and to which extend a quantum computer can be helpful in solving this problem, pick the most suitable quantum computer model(s), 
and then develops a program that utilizes the second (algorithmic) level quantum software to solve this problem. 

We note that in addition to the above discussed software for quantum computers, special software is also required for designing and optimizing quantum hardware~\cite{Pan2021-6}; 
however, we leave this topic outside the scope of the present review. 

\section{Leading countries have announced national quantum programs}\label{sec:national}

The field of quantum science and technology is considered a strategic priority for many countries across the globe. 
National programs on quantum technologies have been announced by several countries~\cite{Thew2019}: 
\begin{enumerate}	
	\item Europe (EU), around 3 billion EUR and special programs in France (1.8 billion) and Germany~\cite{Thew2019,Riedel2019};
	\item Japan, around 1 billion USD~\cite{Yamamoto2019-2}; 
	\item Canada (exact amount is not known; 1 billion USD already invested)~\cite{Sussman2019}; 
	\item USA (US\$1.2 billion over five years in a national quantum initiative)~\cite{Raymer2019}; 
	\item Australia (exact amount is not known)~\cite{White2019}; 
	\item Russia, around 0.5 billion USD~\cite{Fedorov2019};
	\item UK (1 billion pounds over ten years)~\cite{Walmsley2019}; 
	\item China (exact amount is not known; around 1 billion USD for the past 10 years)~\cite{Zhang2019}; 
	\item India (80 billion rupees (US\$1.12 billion) over five years)~\cite{Padma2020};
	\item Taiwan\footnote{\url{https://web.phys.ntu.edu.tw/qcip/}}, Sweden, and Singapore\footnote{https://www.hpcwire.com/off-the-wire/singapores-quantum-engineering-programme-teams-up-with-aws-to-boost-quantum-technologies/} 
	(exact amount is not known).
\end{enumerate}	
One may expect this trend to continue, with additional countries publishing their national plans and funding objectives. 

\section{The quantum computing market is growing}\label{sec:market}

The growth of the quantum computing as field of science, technology, and economy manifests itself in a variety of ways. 
\begin{itemize}
	\item The number of scientific publications increases\footnote{Trends in Quantum Computing by Jacob Farinholt, 2019 (published in 2019), no comprehensive data for more recent years are not yet available.}~\cite{Dhawan2018,Scheidsteger2021}. 
	Quantum computing research accumulated 4703 publications in 10 years, with a 3.39\% growth per annum and averaging 14.30 citations per paper during this period~\cite{Dhawan2018}.
	\item The number of patent applications has also grown for the last two decades, from tens in 2000 up to hundreds in 2020\footnote{\url{ttps://patinformatics.com/wp-content/uploads/2018/01/Quantum-Applications-Patent-Landscape-Report-Opt.pdf}}. 
	\item Over the last 10 years~\cite{MacQuarrie2020}, approximately 140 of quantum computing enterprises emerged, of which 43\% are hardware and the rest are software. 
\end{itemize}

These factors give rise to the formation of a quantum computing market~\cite{Gibney2019,MacQuarrie2020,Martinis2017,Galitski2021}. 
Because quantum computing does not yet surpass classical in practical tasks, the market is currently dominated by investments rather than direct sales of hardware or services. 
According to Gibney~\cite{Gibney2019}, 
by the start of 2019 private investors had funded at least 52 quantum-technology companies globally since 2012 --- many of them spin-offs from universities. 
Although the value of some of the cash infusions remains secret, this study captures the scale of this activity. 
It finds that, in 2017 and 2018, companies received at least \$450 million in private funding --- more than four times the \$104 million disclosed over the previous two years. 
Because presently it is universally believed that quantum advantage is within reach, the growth estimates are supremely optimistic. To argue  this point, we highlight several forecasts.
\begin{itemize}
	\item According to Research and Markets\footnote{\url{https://www.researchandmarkets.com/reports/5010716}}, the quantum computing market was valued at \$507.1 million in 2019, and is projected to grow at a compound annual growth rate of 56.0\% 
	\item	According to Market Insights Reports\footnote{\url{https://www.quantaneo.com/}}, 
	the Global Quantum Computing Market is expected to witness a compound annual growth rate of 34\% during the forecast period 2019-2025, reaching a size of USD 2.82 billion.
	\item An Inside Quantum Technology report\footnote{\url{https://www.insidequantumtechnology.com/}} 
	estimates revenues from quantum computing at \$1.9 billion USD in 2023, increasing to 8.0 billion USD by 2027.
\end{itemize}

A recent BCG analysis\footnote{\url{https://www.bcg.com/en-ca/publications/2019/quantum-computers-create-value-when}} predicts  three phases of the quantum computing market progress: 
\begin{itemize}
	\item the NISQ era, lasting 3-5 years and focusing on  scientific and specialized applications, with the estimated market impact of \$2-5 billion;
	\item Broad quantum advantage era, lasting for 10+ years with  scientific, specialized, and some general-purpose applications and an estimated impact of \$25-50 billion
	\item Era of full-scale fault-tolerant quantum computers, lasting for the next 20+ years with scientific, specialized, and various general-purpose applications and an estimated impact of \$450-850 billion.  
\end{itemize}

\section{Conclusion and outlook}\label{sec:outlook}

Ten years ago, when we were asked about the time horizon, at which we expect to see practical quantum computing, our answer was ``20-30 years''. 
This estimate reflected the lack of realistic roadmap from a conceptual understanding and basic demonstrations, 
which existed at that time, to a viable product. 
This situation drastically changed over the last decade. 
Currently, this roadmap does exist: leading research groups and computing companies confidently plan the development of quantum computers of 1000-qubit size by 2023\footnote{\url{https://research.ibm.com/blog/ibm-quantum-roadmap}} 
and error corrected devices by 2029\footnote{\url{https://quantumai.google/learn/map}}. 

Remarkably, this change appears to be not due to any specific scientific discovery or technological breakthrough, 
but thanks to multiple achievements in various fields leading to the emergence of NISQ devices based on a variety of physical principles. 
Although these devices are not yet useful for many practical tasks, 
they serve as playground for testing various quantum computational concepts and models, and therefore as fulcrum for further progress. 
By building NISQ devices, we learn how build  even better NISQ devices and better algorithms, resulting in exponential progress. 
Perhaps equally important is a psychological side: 
NISQ technology made people believe that quantum computation is no longer a matter of science fiction, but the reality of humankind's immediate future. 
Therefore, we are currently at an inflection point heading towards explosive growth of quantum computational technology and the market associated therewith.

Claims of quantum advantage motivate scientists to develop classical algorithms that challenge these claims. 
Moreover, attempts to simulate quantum computation classically resulted in a new class of algorithms and techniques know as {\it quantum-inspired}. 
This leads in an important side effect of NISQ technology: new  classical algorithms for simulating quantum systems~\cite{Orus2019-3}, 
optimization~\cite{Lvovsky2019,Oshiyama2022}, and data processing~\cite{Tang2019,Lloyd2020-3}. 

Quantum computing is sometimes considered a niche solution for specific problems. 
But historically the same belief was widely held for both classical computers and the internet. 
As we now know, both these technologies have progressed far beyond expectations, changing not only the technological or economic landscape, 
but the entire fabric of our society. 
These developments were made possible thanks to, first, rapid growth of the capabilities, such as the computational power and communication rate, and, 
second, wide availability of these technologies not only to a narrow circle of specialists, but to the general population. 
Until these developments took place, the impact of computation and communication technologies was impossible to predict. 

We can extrapolate these expectations to quantum computing. 
We expect quantum technology to change our society to the same extent as semiconductor technology changed it over past seventy years. 
This hope however hinges on the same conditions of steadily growing capability and availability. 
We need to develop the quantum analog of Moore’s law, i.e., 
the situation, in which each new generation of quantum processors surpasses the previous generation by a significant factor. 
It is not possible to tell at present whether this would be the case. 
However, we did observe this trend over the past decade and expect this to continue at the same zeal for the next ten or so years. 

As more researchers and businesses begin adopting and adapting quantum computing technology, the network effect will start to play a role, 
allowing for the development and testing of new quantum algorithms and applications, facilitating education, 
and fostering the understanding of the vectors for further development. 
This would require wide availability of quantum computers as cloud offerings to a broad range of users for experimentation, 
tinkering, and even a bit of play.

\section*{Acknowledgments}
We thank M.D. Lukin, E. Demler, T. Calarco, A. Kitaev, W. Ketterle, D. Abanin, and A. Ekert for valuable discussions 
as well as M. Atature, S. Simon, S. Markoff, B. Mayer, F. Pammolli, Y. Tsybrovskyy, A. Mastiukova, 
I. Semerikov, S. Straupe, I. Besedin, and E. Kiktenko for useful comments and information. 
We specially thank D. Deutsch for insights into the history of quantum computing. 

\bibliographystyle{naturemag}
\bibliography{bibliography.bib}

\end{document}